\documentclass[twocolumn,showpacs,preprintnumbers,amsmath,amssymb]{revtex4}

\usepackage{graphicx}
\usepackage{dcolumn}

\usepackage{amsmath,amssymb}

\usepackage[mathscr]{euscript}

\usepackage{calc}

\usepackage{nicefrac}

\usepackage{color}

\usepackage{mathptmx, courier, pifont}
\usepackage[scaled=0.92]{helvet}
\usepackage[T1]{fontenc}
\usepackage{textcomp} 

\usepackage{bm}

\catcode`\@=11
\renewcommand{\ps@myheadings}{\let\@mkboth\@gobbletwo
\def\@oddhead{\hbox{}\hfil {\normalsize\sl\rightmark}\hfil 
{\bf\arabic{page}}\hbox{}}
\def\@oddfoot{}\def\@evenhead{{\bf\arabic{page}}\hfil  
{\normalsize\sl\leftmark} \hfil\hbox                  
{}}\def\@evenfoot{}\def\sectionmark##1{}\def\subsectionmark##1{}}

\newcommand{\yes}{yes}
     
\newcommand{\labelshow}{no}        

\newcommand{\mwgtag}[1]{\label{eq:#1}}

\newcommand{\eq}[1]{%
\ifx\labelshow\yes
Eq.~{\{#1\}}(\ref{eq:#1})%
\else%
Eq.\ (\ref{eq:#1})
\fi}

\newcommand{\eqnoeq}[1]{%
\ifx\labelshow\yes
{\{#1\}}(\ref{eq:#1})%
\else%
(\ref{eq:#1})
\fi}

\newcommand{\rangeref}[2]{%
\ifx\labelshow\yes
Eqs.\ ({\{#1\}}%
\ref{eq:#1})--({\{#2\}}%
\ref{eq:#2})
\else%
Eqs.\ (\ref{eq:#1})--(\ref{eq:#2})\nobreak   
\fi}

\newcommand{\reference}[1]{%
\ifx\labelshow\yes
{\{#1\}}\ \cite{#1}%
\else%
\cite{#1}
\fi}

\newcommand{\figgy}[4]{%
\ifx\labelshow\yes
\begin{figure}[tbp]
\vspace{#2}%
\begin{center}\parbox{5 true in}{%
\caption[]{\{#1\} \protect\label{fg:#1} #4}%
}\end{center}%
\end{figure}%
\else
\begin{figure}[tbp]
\vspace{#2}%
\begin{center}\parbox{5 true in}{%
\caption[]{\small \protect\label{fg:#1} #4}%
}\end{center}%
\end{figure}%
\fi}

\newcommand{\twofiggy}[8]{%
\ifx\labelshow\yes
\begin{figure}[tbp]
\vspace{#2}%
\begin{center}\parbox{5 true in}{%
\caption[]{\small \{#1\} \protect\label{fg:#1} #4}%
}\end{center}%
\vspace{#6}%
\begin{center}\parbox{5 true in}{%
\caption[]{ \small \{#5\} \protect\label{fg:#5} #8}%
}\end{center}%
\end{figure}%
\else
\begin{figure}[tbp]
\vspace{#2}%
\begin{center}\parbox{5 true in}{%
\caption[]{ \small \protect\label{fg:#1} #4}%
}\end{center}%
\vspace{#6}%
\begin{center}\parbox{5 true in}{%
\caption[]{\small \protect\label{fg:#5} #8}%
}\end{center}%
\end{figure}%
\fi}

\def\picture #1 by #2 (#3){ 
  \vbox to #2{
    \hrule width #1 height 0pt depth 0pt
    \vfill
  \special{picture #3} } }

\def\centerpicture #1 by #2 (#3 scaled #4){
  \dimen0=#1 \dimen1=#2 
  \divide\dimen0 by 1000 \multiply\dimen0 by #4
  \divide\dimen1 by 1000 \multiply\dimen1 by #4
	\ashift{\dimen0}
	\picture \dimen0 by \dimen1 (#3 scaled #4)}

\newlength{\alen}
\newcommand{\ashift}[1]{
\setlength{\alen}{\textwidth}
\addtolength{\alen}{-1#1}
\addtolength{\alen}{-0.5\alen}
\addtolength{\alen}{-8 pt} \hspace{\alen}}

\newcommand{\alignc}{c}
\newcommand{\isok}{c}

\newcommand{\fgcaption}[3]{\renewcommand{\isok}{#1}%
 \ifx  \alignc \isok
\begin{center}%
\parbox{4 in}{\begin{center}%
{\small{\bf\noindent Fig.\  \ref{fg:#2}\hspace{1em}} #3}  \end{center}}%
\end{center}%
\else%
\begin{center}%
\parbox{4 in}{%
{\small{\bf \noindent Fig.\  \ref{fg:#2}\hspace{0.5em}} #3}}%
\end{center}%
\fi}

\newcommand{\fig}[1]{%
\ifx\labelshow\yes
Fig.~{ \{#1\}} \ref{fg:#1}%
\else%
Fig.~\ref{fg:#1}
\fi}

\newcommand{\figs}[2]{%
\ifx\labelshow\yes
Figs.~{ \{#1\}} \ref{fg:#1}--{ \{#2\}} \ref{fg:#2}%
\else%
Figs.~\ref{fg:#1}--\ref{fg:#2}
\fi}

\newcommand{\tableinsert}[5]{
\ifx\labelshow\yes
\begin{table}
\begin{center}\parbox{5 true in}{%
\caption[]{\{#1\} \protect\label{tb:#1}#4}%
}\end{center}%
\par\vspace {#2}%
\noindent{\footnotesize #5}%
\end{table}%
\else%
\begin{table}
\begin{center}\parbox{5 true in}{%
\caption[]{\protect\label{tb:#1}#4}%
}\end{center}%
\par\vspace {#2}%
\noindent{\small #5}%
\end{table}%
\fi}

\newcommand{\tableref}[1]{%
\ifx\labelshow\yes
Table~{\{#1\}} \ref{tb:#1}%
\else%
Table~\ref{tb:#1}
\fi}%

\newcommand{\heading}[2]{%
\ifx\labelshow\yes
\section{\{#1\} \protect\label{h:#1}#2}%
\else%
\section{\protect\label{h:#1}#2}%
\fi}%

\newcommand{\subheading}[2]{%
\ifx\labelshow\yes
\subsection{\{#1\} \protect\label{sh:#1}#2}%
\else%
\subsection{\protect\label{sh:#1}#2}%
\fi}%

\newcommand{\subsubheading}[2]{%
\ifx\labelshow\yes%
\subsubsection{\{#1\} \protect\label{ss:#1}#2}
\else%
\subsubsection{\protect\label{ss:#1}#2}
\fi}

\newcommand{\tightdot}{\cdot}

\newcommand{\units}[1]{\mbox{\,#1}}

\newcommand{\halfthin}{\kern 0.0834em}
\newcommand{\neghalfthin}{\kern -0.0834em}

\newcommand{\quarterthin}{\kern 0.0417em}
\newcommand{\negquarterthin}{\kern - 0.0417em}

\newcommand{\boldsym}[1]{\mbox{\boldmath$#1$}}

\newcommand{\onematrix}[2]{\begin{pmatrix}#1\\#2\end{pmatrix}}
\newcommand{\twomatrix}[4]{\begin{pmatrix}#1&#2\\#3&#4\end{pmatrix}}

\newcommand{\paulix}{\begin{pmatrix}0&1\\1&0\end{pmatrix}}
\newcommand{\pauliy}{\begin{pmatrix}0&-i\\i&\phantom{-}0\end{pmatrix}}

\newcommand\deriv[2]{\frac{d#1}{d#2}}

\newcommand{\pardiv}[2]{\frac{\partial #1}{\partial #2}}

\newcommand{\ttfrac}[2]{\scriptscriptstyle\frac{#1}{#2}}

\newcommand{\bra}[1]{\left\langle#1\right|}
 \newcommand{\ket}[1]{\left|#1\right\rangle}
\newcommand{\bracket}[2]{\langle#1|#2\rangle}

\newcommand{\ev}[1]{\langle#1\rangle}
\newcommand{\mel}[3]{\bra{#1}#2\ket{#3}}

\newcommand{\clebsch}[6]{\bracket{#1 \,#2 \,#3 \,#4}{#5 \,#6}}

\newcommand{\comm}[2]{[ \, #1 , #2 \, ]}

\newcommand{\su}[1]{\mbox{SU($#1$)}}

\newcommand{\so}[1]{\mbox{SO($#1$)}}
\newcommand{\unitary}[1]{\mbox{U($#1$)}}

\newlength{\dynkwid}
\newlength{\dynklength}

\newlength{\dynkbase}
\newcommand\filledcirc{{\color{black}\bullet}\mathllap{\circ}}

\newcommand{\dynkinZero}{
 \begin{Large}
        \hbox{
            $\circ$\kern-0.1em%
            \hspace{1.5em}%
            \kern-0.1em$\circ$
        }
    \end{Large}
}

\newcommand{\dynkinOne}{
    \begin{Large}
        \hbox{
            $\circ$\kern-0.1em%
            \raisebox{0.52ex}{\hbox{\vbox{\hrule width 1.5em}}}%
            \kern-0.1em$\circ$
        }
    \end{Large}
}

\newcommand{\dynkinTwo}{
    \begin{Large}
        \hbox{
             $\circ$\kern-0.1em%
             \raisebox{0.32ex}{\hbox{\vbox{\hrule width 1.5em}}}%
             \kern-1.5em
             \raisebox{0.72ex}{\hbox{\vbox{\hrule width 1.5em}}}%
             \kern-0.1em$\circ$
        }
    \end{Large}
}

\newcommand{\dynkinThree}{
    \begin{Large}
        \hbox{
            $\circle$\kern-0.1em%
            \raisebox{0.55ex}{\hbox{\vbox{\hrule width 1.55em}}}%
            \kern-1.58em
            \raisebox{0.83ex}{\hbox{\vbox{\hrule width 1.58em}}}%
            \kern-1.60em
            \raisebox{0.27ex}{\hbox{\vbox{\hrule width 1.70em}}}%
            \kern-0.18em$\circ$
        }
    \end{Large}
}

\newcommand{\dynkinZeroH}{
 \begin{huge}
        \hbox{
            $\circ$\kern-0.1em%
            \hspace{1.3em}%
            \kern-0.1em$\circ$
        }
    \end{huge}
}

\newcommand{\dynkinOneH}{
    \begin{huge}
        \hbox{
            $\circ$\kern-0.1em%
            \raisebox{0.54ex}{\hbox{\vbox{\hrule width 1.30em height 
\dynkwid}}}%
            \kern-0.1em$\circ$
        }
    \end{huge}
}

\newcommand{\dynkinTwoH}{
    \begin{huge}
        \hbox{
            $\circ$\kern-0.1em%
            \raisebox{0.33ex}{\hbox{\vbox{\hrule width 1.3em height \dynkwid}}}%
            \kern-1.3em
            \raisebox{0.74ex}{\hbox{\vbox{\hrule width 1.3em height \dynkwid}}}%
            \kern-0.1em$\circ$
        }
    \end{huge}
}

\newcommand{\dynkinThreeH}{
    \begin{huge}
        \hbox{
$\circ$\kern-0.1em%
\raisebox{0.55ex}{\hbox{\vbox{\hrule width 1.29em height\dynkwid}}}%
\kern-1.30em
\raisebox{0.80ex}{\hbox{\vbox{\hrule width 1.36em height\dynkwid}}}%
\kern-1.36em
\raisebox{0.30ex}{\hbox{\vbox{\hrule width 1.38em height\dynkwid}}}%
\kern-0.14em$\circ$
}
    \end{huge}
}

\newcommand{\dynkinThreeHfilled}{
    \begin{huge}
        \hbox{
$\circ$\kern-0.1em%
\raisebox{0.55ex}{\hbox{\vbox{\hrule width 1.29em height\dynkwid}}}%
\kern-1.30em
\raisebox{0.80ex}{\hbox{\vbox{\hrule width 1.36em height\dynkwid}}}%
\kern-1.36em
\raisebox{0.30ex}{\hbox{\vbox{\hrule width 1.38em height\dynkwid}}}%
\kern-0.14em$\filledcirc$
}
    \end{huge}
}

\newcommand{\tsub}[1]{_{\mbox{\scriptsize#1}}}
\newcommand{\tsup}[1]{^{\mbox{\scriptsize#1}}}

\newcommand{\runningheads}[2]{\markboth{\hfill #1\hfill}{\hfill #2\hfill}}

\newcount\hours
\newcount\minutes
\newcount\tmp
\hours=\time
\divide\hours by 60
\tmp=\hours
\minutes=\time
\multiply\tmp by 60
\advance\minutes by -\tmp
\newcommand{\timenow}{\number\hours:\ifnum\number\minutes>9%
{\number\minutes}\else%
0\number\minutes\fi}

\newcommand{\datefile}{\date{\vspace{10pt}
            \small\sc  
            ---\thinspace Printed from File: {\small
           \lowercase{\jobname}.tex at \timenow}
            \thinspace---\vphantom{\bigg[} 
            \\ 
             \small ---\today---}}

\newcommand{\mnote}[1]
{\setlength{\marginparwidth}{50pt}%
\setlength{\marginparsep}{10pt}%
\marginpar{\scriptsize\em#1}}

\newcounter{exercisenumber}

\newcommand{\newexercise}{\addtocounter{exercisenumber}{1}
            \vspace{0.5em}\noindent{\small\bf\thechapter.\theexercisenumber
\hspace{1.0em}}}

\def\pointsize@{10}
\def\fivepoint{\def\pointsize@{5}%
 \normalbaselineskip7\p@
 \abovedisplayskip7\p@ plus1.8\p@ minus5.4\p@
 \belowdisplayskip7\p@ plus1.8\p@ minus5.4\p@
 \abovedisplayshortskip\z@ plus1.8\p@
 \belowdisplayshortskip2.6\p@ plus1.8\p@ minus1.0\p@
 \textfont@\rm\fiverm
 \textfont@\it\fivei
 \textfont@\bf\fivebf
 \ifsyntax@\def\big##1{{\hbox{$\left##1\right.$}}}\else
 \let\big\eightbig@
 \textfont\z@\fiverm \scriptfont\z@\fiverm \scriptscriptfont\z@\fiverm
 \textfont\@ne\fivei \scriptfont\@ne\fivei \scriptscriptfont\@ne\fivei
 \textfont\tw@\fivesy \scriptfont\tw@\fivesy \scriptscriptfont\tw@\fivesy
 \textfont\thr@@\tenex \scriptfont\thr@@\tenex \scriptscriptfont\thr@@\tenex
 \textfont\itfam\fivei
 \textfont\bffam\fivebf \scriptfont\bffam\fivebf \scriptscriptfont\bffam\fivebf
 \fi
 \setbox\strutbox\hbox{\vrule height 5\p@ depth2\p@ width\z@}%
 \setbox\strutbox@\hbox{\vrule height 4\p@ depth1\p@ width\z@}%
 \normalbaselines\fiverm\ex@=.2326ex}

\def\boxit#1{\vbox{\hrule\hbox{\vrule\kern3pt
     \vbox{\kern3pt#1\kern3pt}\kern3pt\vrule}\hrule}}

\def\yspacef{\kern 0.6\pointsize@\p@}
\def\yspaceo{\kern 0.3em }

\def\cstok#1{
\if5\pointsize@ \let\yspace\yspacef \else
 \let\yspace\yspaceo \fi
  \leavevmode\hbox{\kern-.4pt\vrule\vtop{\vbox{\hrule
      \kern 1.8\pointsize@\p@
        \hbox{\vphantom{\rm/}\yspace{#1}\yspace}}
      \kern 1.8\pointsize@\p@\hrule}\vrule}}
 
\newcommand{\tcstok}[1]{\tiny
\if5\pointsize@ \let\yspace\yspacef \else
 \let\yspace\yspaceo \fi
  \leavevmode\hbox{\kern-.4pt\vrule\vtop{\vbox{\hrule
      \kern 1.8\pointsize@\p@ 
        \hbox{\vphantom{\rm/}\yspace{#1}\yspace}}
      \kern 1.8\pointsize@\p@\hrule}\vrule}}  

\newcommand{\scstok}[1]{\small
\if5\pointsize@ \let\yspace\yspacef \else
 \let\yspace\yspaceo \fi
  \leavevmode\hbox{\kern-.4pt\vrule\vtop{\vbox{\hrule
      \kern 1.8\pointsize@\p@ 
        \hbox{\vphantom{\rm/}\yspace{#1}\yspace}}
      \kern 1.8\pointsize@\p@\hrule}\vrule}}  

\def\Young#1{\null\,\vcenter{\normalbaselines\m@th
     \if8\pointsize@ \baselineskip=6pt \lineskip=-0.4pt
     \else \baselineskip=7.5pt \lineskip=-0.4pt \fi
     \if5\pointsize@ \baselineskip=3pt \lineskip=-0.4pt\fi
    \ialign{$##$&&$##$\crcr
      \mathstrut\crcr\noalign{\kern-\baselineskip}
      #1\crcr\mathstrut\crcr\noalign{\kern-\baselineskip}}}\,}

\makeatletter
\newcommand\boxpargfont{\@setfontsize\boxpargfont{9.5pt}{11.5}}
\makeatother

\renewcommand{\heading}[2]{%
\setcounter{exampleNumber}{0}%
\setcounter{boxCount}{0}%
\chapter{\protect\label{h:#1}#2}%
}%

\renewcommand{\subheading}[2]{%
\section{\protect\label{sh:#1}#2}%
}%

\renewcommand{\subsubheading}[2]{%
\subsection{\protect\label{ss:#1}#2}%
}

\renewcommand{\paragraph}{\subsubsection}

\renewcommand{\boldsym}[1]{\mbox{\kern  -0.0em ${\bm #1}$}}

\newlength{\itemsepnow}
\newlength{\figup}
\newlength{\tabup}
\newlength{\tabtitlesep}
\newlength{\afterTableLineOne}
\newlength{\afterTableLineTwo}
\newlength{\beforeTableLineThree}
\newlength{\tablerowMore}
\newlength{\tableLineShift}
\newlength{\aboveTableFootnote}
\newcommand{\angstrom}{A\kern -1.2ex \raise0.8em\hbox{$\scriptstyle\circ$\ }}
\newcommand{\redebro}{\lambda\kern-0.9ex \raise.3em\hbox{-}}

\newcommand{\sprod}[2]{
\boldsym#1\tightdot\boldsym#2}

\renewcommand{\tightdot}{\cdot}  

\newcommand{\diffelement}[1]{d\kern-0.00em#1}

\renewcommand{\deriv}[2]{\frac{\diffelement#1}{\diffelement#2}}

\renewcommand{\pardiv}[2]{\frac{\partial {#1}}{\partial {#2}}}

\newcommand{\spin}{\mathscr{S}}   

\newcommand{\reducedI}{I\raisebox{.13em}{\kern-.35em-}{\kern.1em}}
\newcommand{\tripleDotReducedI}{\reducedI\raisebox{0.85em}{\kern-.48em
  {.} {\kern-0.35em .} {\kern-0.35em .} } \kern-.45em}
\newcommand{\tripleDotI}{I\raisebox{0.85em}{\kern-.40em{.} {\kern-0.35em .}
            {\kern-0.35em .} } \kern-.45em}

\newcommand{\ntoone}[1]{#1\kern 0.05em:\kern 0.05em1}
\newcommand{\oneton}[1]{1\kern 0.05em:\kern 0.05em#1}

\def\slashchar#1{#1\kern-0.60em/}
\def\itslashchar#1{#1\kern-0.5em/}

\newcommand{\Adagcoupled}[4]{A_{\scriptscriptstyle 
#1#2}^{\raisebox{0.2em}{$\scriptstyle\dagger$} {\scriptscriptstyle 
#3 #4}}}
\newcommand{\phantomdagger}{^{\vphantom{\dagger}}}
\newcommand{\phd}{\phantomdagger}

\newcommand{\casimir}[1]{C_{\scriptscriptstyle#1}}

\newlength{\howmuch}
\newlength{\parindentnow}
\newcounter{exampleNumber}
\newcounter{boxCount}

{%
\begin{examplelist}
\item 
#2
\label{example:#1}
\end{examplelist}
}

{%
\begin{small}

\addtocounter{exampleNumber}{1}%
\setlength{\parindentnow}{\parindent}%
\par\vspace{2.5ex}\hrule\par\vspace{1.8ex}%
\par\setlength{\parindent}{6pt}%
\noindent{\bf Example \arabic{chapter}.\arabic{exampleNumber}}%
\par\vspace{0.7ex}\par\noindent%
}%
{\par\vspace{1.4ex}\hrule\vspace{2.6ex}%
\parindent=\parindentnow%
\end{small}
}

\newlength{\boxup}
\newlength{\boxparindent}
\newlength{\boxparl}
\newlength{\boxparskip}
\newcommand{\makeBoxedPronouncement}[1]{
\begin{center}
\begin{tabular}{c}
\begin{minipage}{33em}
\vspace{1ex}
\begin{shadedbox}
{#1}
\end{shadedbox}
\end{minipage}
\end{tabular}
\vspace{1ex}
\end{center}
}

\newcommand{\graybox}[1]{
\medskip
\makeBoxedPronouncement{#1}
\medskip
}

\renewcommand{\newexercise}[1]{
\addtocounter{exercisenumber}{1}
\vspace{0.5em}\noindent{\small\bf\thechapter.\theexercisenumber\hspace{0.25em}}
}

\renewcommand{\onematrix}[2]{\begin{pmatrix}#1\\#2\end{pmatrix}}
\renewcommand{\twomatrix}[5][0ex]{\begin{pmatrix}#2&#3\\[#1]#4&#5\end{pmatrix}}

\newcommand{\parslash}{\partial \kern-0.60em/}
\newcommand{\Dslash}{D \kern-0.60em/}
\newcommand{\Aslash}{A \kern-0.55em/}

\newcommand{\crossprod}[2]{\bm#1 \times \bm#2}

\newcommand{\bmk}{{\bm k}}
\newcommand{\bmq}{{\bm q}}

\renewcommand{\makeBoxedPronouncement}[1]{
\begin{center}
\begin{tabular}{c}
\begin{minipage}{22em}
\vspace{0.0ex}
{#1}
\end{minipage}
\end{tabular}
\vspace{0.0ex}
\end{center}
}

\newcommand{\SO}[1]{\mbox{SO($#1$)}}

\newcommand{\Adag}[2]{A^\dagger_{#1#2}}
\newcommand{\AdagPower}[2]{(\Adag{#1}{#2})^{N_{#1#2}}}
\newcommand{\hws}{\ket{\rm HW}}
\newcommand{\iqhe}{integer quantum Hall effect}
\newcommand{\fqhe}{fractional quantum Hall effect}

\newcommand{\Piop}[2]{\Pi_{#1#2}}

\newcommand{\singlefig}[6]{%
\begin{figure}[tb]\vspace{#3}%
\centering
\includegraphics*[scale=#5]{#2}%
\caption{\label{fg:#1} #6}%
\vspace{#4}%
\end{figure}}

\newcommand{\singlefigbottom}[6]{%
\begin{figure}[b]\vspace{#3}%
\centering
\includegraphics*[scale=#5]{#2}%
\caption{\label{fg:#1} #6}%
\vspace{#4}%
\end{figure}}

\newcommand{\doublefig}[6]{%
\begin{figure*}[tb] \vspace{#3}%
\includegraphics*[scale=#5]{#2}%
\caption{\label{fg:#1} #6}
\vspace{#4}
\end{figure*}}

\begin{document}

\title{Emergent Fermion Dynamical Symmetry for Monolayer Graphene in a Strong 
Magnetic Field}

\author{Mike Guidry$^{(1)}$}
\email{guidry@utk.edu}
\author{Lian-Ao Wu$^{(2)}$}
\email{lianaowu@gmail.com}
\author{Fletcher Williams$^{(1)}$}
\email{fletcher.williams@knoxvillecatholic.com}

\affiliation{
$^{(1)}$Department of Physics and Astronomy, University of
Tennessee, Knoxville, Tennessee 37996, USA \\
 $^{(2)}$Department of Physics, University of the Basque Country UPV/EHU, 48080 Bilbao, IKERBASQUE Basque Foundation for Science, 48013 Bilbao,
EHU Quantum Center, University of the Basque Country UPV/EHU, Leioa, Biscay 48940, Spain
}

\date{\today}

\begin{abstract}
We review the physics of monolayer graphene in a strong magnetic  field, with  emphasis on  highly collective states that emerge from the weakly interacting system because of correlations (emergent states).  After reviewing the  general properties of graphene and of electrons in a magnetic field, we  give a brief introduction to the \iqhe\ (IQHE) and the  \fqhe\ (FQHE) in a 2D electron gas as foundation to  show that monolayer graphene in a magnetic field exhibits both effects, but with properties modified by the influence of  the graphene crystal.  After giving an introduction to standard methods  of dealing with emergent states for this system, we show that an \SO8  fermion dynamical symmetry governs the emergent degrees of freedom and that the  algebraic and group properties of the dynamical symmetry provide a new  view of strongly correlated states observed in monolayer graphene subject to a strong magnetic field.
\end{abstract}

\pacs{71.10.-w, 71.27.+a, 74.72.-h}

\maketitle

\runningheads{}{{\em Emergent Fermion Dynamical Symmetry for Monolayer Graphene \ldots} M. W. Guidry, L.-A. Wu, and F. Williams}

\clearpage
\tableofcontents

\section{Introduction}

\noindent
Graphene is a crystalline material composed of a single one-atom thick layer  of carbon; thus it represents   the ideal two-dimensional material.  It was first isolated by   Geim and Novoselov  in 2004, by exfoliating graphite using strips of tape. The common pencil was invented in 1564; it functions because the  graphite in its ``lead'' stacks layers of graphene having strong bonding within layers but weak van der Waals bonding between layers.  In writing with a pencil, layers of graphene are sloughed  off by breaking the weak bonds between layers and are transferred to the  paper; presumably some parts of a line drawn with a pencil consist of a  single layer of graphene.  Thus, it is likely that graphene has been produced  since 1564 by the ordinary practice of writing with pencils, but until 2004 it  was thought that the single layer allotrope of carbon atoms perhaps did not  exist in its free state.  Hence the common view that graphene is easy to produce but  difficult to detect, which delayed its discovery for more than 400  years  after it was first unknowingly produced.

This paper provides an overview of graphene that emphasizes its role as a laboratory for studying emergent states in pesudo-relativistic matter.  There is considerable current interest in samples having two or more layers of graphene, but this review will confine itself to a single layer (\textit{monolayer graphene}).  We will begin by giving a concise overview of graphene as a material, its basic structure at the atomic and crystalline level, and of conventional approaches to studying its emergent states when placed in a strong magnetic field.  Then we shall introduce an alternative view of this many-body system that is the primary emphasis of this review: emergent states for monolayer graphene in a strong magnetic field in terms of Lie algebras, Lie groups, and associated fermion dynamical symmetries of the Hamiltonian. Since the Lie algebra and Lie group methodology employed in this discussion may be less familiar than more conventional condensed matter methods for some readers, we will include important details and proofs of various assertions in a supplemental web document \cite{supplementMaterials}.

\section{Some material properties of graphene}

\noindent
Graphene has many unusual material properties; let us summarize a few of the more important ones.

\begin{itemize}

\item
It is planar, with a hexagonal unit cell  of area 0.052 nm$^2$ containing two carbon atoms.  Taking the mass of the unit cell to be twice the mass  of  carbon,  the mass density is $\rho = 0.77 \units{mg\,m}^{-2}$.

\item
A single layer of graphene is almost, but not quite, transparent, absorbing  $\sim 2.3$\% of the light passing through it at optical wavelengths.  Absorption is additive for multiple layers, so the number of layers in a  sample may be inferred by how dark  it appears in normal light. This method can be used to distinguish regions of a  sample having one, two, three, \ldots, layers of graphene after exfoliation  with  tape, for example. 

\item
The breaking strength of graphene is 42 N\,m$^{-1}$, which is $\sim$100 times 
larger than that of a hypothetical film of steel having the same thickness.

\item
The thermal conductivity of graphene is dominated by  phonons and is large, 
 with a  measured value of 5000 W/mK that is ten times higher than that of copper.

\item
The mobility of electrons in graphene is very high.  Electric fields can be used to tune 
charge carriers  continuously between electrons and holes, with carrier
densities as high as $10^{13} \units{cm}^{-2}$ and carrier mobilities that can exceed 
15,000 cm$^2$V$^{-1}$s$^{-1}$, even at room temperature and pressure.

\end{itemize}

\noindent
As a consequence of such material properties, graphene is of large 
 interest as a basis for new devices and other practical applications.  
But because of its ideal 2D geometry, non-trivial crystal structure, and unusual 
electronic properties, it also represents a novel laboratory for 
fundamental physics, as elaborated in Section \ref{fundPropGraphene}.

\section{\label{fundPropGraphene}Fundamental properties of graphene}

\noindent
Because graphene exhibits linear dispersion near the Fermi surface (see \fig{dispersionGraphene}), its charge-carrying electrons behave in a manner  analogous to that of relativistic electrons described by a Dirac equation, with  the fermi velocity $v\tsub F$ (maximum velocity of a fermion in a system) playing a role analogous to that of light speed,  where the actual speed of light is $c/v\tsub F\sim$ 300 times larger than the graphene fermi  velocity.  Just as an actual relativistic electron moving very near the speed of light typically has a small rest mass, an electron in graphene  near  the fermi surface typically displays an effective mass near zero. Thus we may view graphene as \textit{pseudo-relativistic matter} that behaves mathematically as actual Lorentz-invariant, relativistic matter would be expected to behave.

Strongly-correlated, \textit{non-relativistic matter} has been studied  extensively in phenomena such as superconductivity.  However, before  the isolation of graphene it was not easy to observe highly-correlated states for \textit{relativistic fermions}  because it was difficult to find  them under conditions that could produce the strong particle--particle interactions necessary to form such states.  It has been proposed that neutrinos in the hot, dense environs of a  core-collapse supernova, or in the early Universe before about one second after  the Big Bang, could undergo sufficient interactions to produce a correlated,  relativistic, many-body state but no  observations  yet support this conjecture. As we now discuss, graphene  affords a benchtop laboratory for studying strongly-correlated,  pseudo-relativistic matter.

In the presence  of a magnetic field applied transverse to the sample, a single layer of graphene  exhibits both an \textit{integer quantum Hall effect} (IQHE) and a \textit{fractional quantum Hall effect} (FQHE). The integer quantum Hall effect can occur for  weakly interacting electrons but  FQHE states can exist only  because of strong correlations, so observation of a \fqhe\ for graphene placed in a magnetic field is evidence of  strongly-correlated, pseudo-relativistic electronic states.  A description of these emergent states  will be the 
overall focus of this review.

\section{Structure of Graphene}

\noindent
Let's consider  in more detail the electronic and crystal structure of graphene. Since graphene is pure carbon, the obvious point of departure is the chemical bonding tendencies of carbon.  In  this section we adapt extensively from the presentation of Ref.\  \cite{goer2011}.

\subsection {Electronic structure}

\noindent
Carbon has six electrons in a $1s^22s^22p^2$ configuration, with the $2s$ and  $2p$ orbitals playing the dominant role in chemical bonding.  A signature characteristic of  carbon is its tendency to hybridize the $2s$ and $2p$ orbitals and form covalent bonds  with other atoms.  A common hybridization is $sp^n$, in which the $\ket{2s}$ orbital  and $n$ of the $\ket {2p_i}$ orbitals mix.  Because graphene is a planar  crystal of carbon atoms bonded to each other, the dominant hybridization is  $sp^2$, in which the spherically symmetric $2s$ orbital mixes with two of the  $2p$ orbitals lying in the same plane (typically chosen to be $2p_x$ and  $2p_y$).  This gives the orbital geometry illustrated in \fig{sp2-orbitals}(a),%
%
%
 \doublefig
          {sp2-orbitals} 
          {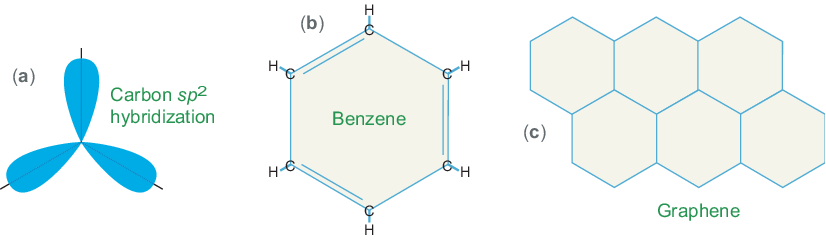}
          {0pt} 
          {0pt}
          {1.0}
          {(\textbf{a}) Carbon $sp^2$ hybridization. (\textbf{b}) Schematic structure of the ring compound benzene;  the actual ground state is a quantum superposition of this configuration and one with  the positions of the double and single bonds exchanged.  (\textbf{c}) Schematic picture of  graphene as an extended set of benzene rings joined together, with each benzene C--H  bond replaced by a C--C bond between adjacent rings.}
with three planar bonding orbitals separated by 120$^\circ$ angles.   In \fig{sp2-orbitals}(b) the schematic structure of benzene is indicated, with  the covalent bonding at each vertex between a carbon and two other carbons and a  hydrogen corresponding to $sp^2$ hybridization, with the in-plane bonds termed  {\em $\sigma$-bonds.}  The double lines indicate double bonds, with the second  bond created by overlap of the non-hybridized $2p_z$ orbitals on adjacent  carbons directed perpendicular to the plane of the ring ({\em $\pi$-bonds}).   The $\pi$-bonding electrons are delocalized over the ring and the benzene  ground state corresponds to a quantum superposition of configurations having the double  bonds in alternating positions around the ring.  

In \fig{sp2-orbitals}(c) we indicate that the  planar structure of graphene may be viewed as similar to fitting many benzene  rings together geometrically, with the hydrogen stripped off and the bond to  hydrogen at each benzene vertex replaced by a covalent bond to a carbon in the  adjacent ring. The in-plane $\sigma$ bonds between the carbon atoms are responsible for the  robustness of the lattice structure in all allotropes of carbon;  they lead to a  $\sigma$-band that is completely filled.  As in benzene, the $p$ orbital that does not participate in the $sp^2$ hybridization (typically chosen to be $p_z$)  is perpendicular to the carbon-atom plane and can bind covalently with  neighboring carbon atoms, delocalizing and leading to a $\pi$-band.  In pure graphene each $p$ orbital has one extra electron, so the $\pi$-band is  half full.

We shall refer to the resulting graphene structure  as the {\em honeycomb lattice,} for obvious geometrical reasons.  It is  believed that all graphitic compounds (graphite, carbon nanotubes, graphene,  and fullerenes) have as their basic building block this honeycomb graphene  lattice: {\em graphite} corresponds to stacks of graphene sheets with strong  covalent bonds within the sheets and weak van der Waals forces between the  sheets, a {\em carbon nanotube} corresponds to a graphene sheet rolled into a tube,  and a {\em fullerene} is wrapped-up graphene with pentagons inserted to allow an average spherical shape.

\subsection{\label{latticeStructure}Lattice structure}

\noindent
To address the direct (real-space) and reciprocal (momentum-space)  graphene 
lattices it is useful to define the direct lattice vectors illustrated in 
\fig{grapheneLatticeVectors}(a).%
%
%
 \doublefig
          {grapheneLatticeVectors} 
          {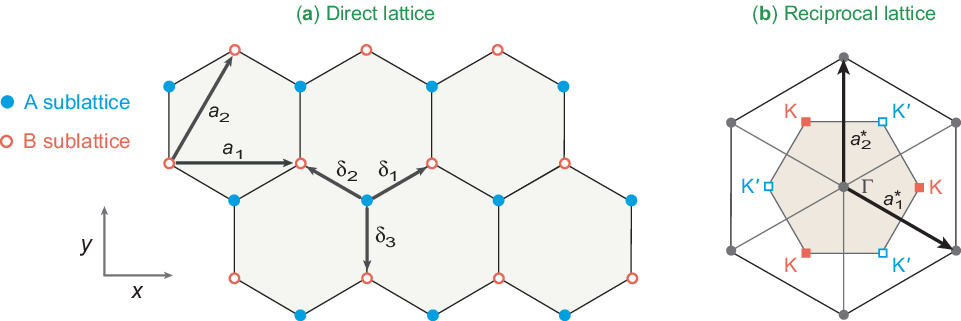}
          {0pt} 
          {0pt}
          {0.90}
          {(\textbf{a}) The direct lattice for graphene. The graphene honeycomb  lattice is not a Brevais lattice since from the crystallographic point of  view two neighboring sites are inequivalent.  It may be viewed as a  triangular Brevais lattice with a two-atom basis, A and B, or as  bipartite, with interlocking triangular lattices A and B.  The vectors  $\bm a_1$ and $\bm a_2$ are the basis vectors of the triangular  Brevais lattice.  The vectors $\bm\delta_1$ , $\bm  \delta_2$, and $\bm\delta_3$ connect nearest-neighbor (NN)  carbon atoms, with a measured separation of $a=0.142$ nm.  (\textbf{b}) Reciprocal  lattice of triangular lattice and first Brillouin zone (shaded).  The primitive lattice vectors are  $\bm a_1^*$ and $\bm a_2^*$.  The center $\Gamma$ and inequivalent corners K and  K$^\prime$ are marked. Adapted from Ref.\ \cite{goer2011}.}
As described in the figure caption, the graphene lattice may be 
viewed as bipartite, consisting of two distinct triangular sublattices 
(labeled A and B in the figure, with the A sublattice indicated by solid blue circles and the B sublattice by open red circles).  Alternatively, the lattice may be viewed as triangular, with a two-atom basis.  It is common to define a quantum number 
associated with this distinction between sublattices
A and B that takes two values and thus may be 
described by an \su2 symmetry that is termed the {\em sublattice pseudospin}. The 
lattice vectors are
\begin{equation}
     \bm a_1 = \sqrt 3 a\bm e_x
     \qquad
     \bm a_2 = \frac{\sqrt 3  a}{2} \big( \bm e_x + \sqrt 3 \, 
    \bm e_y \big),
     \mwgtag{latticeVectorEq}
\end{equation}
where the spacing between carbon atoms is $a \simeq 0.142 \units{nm}$. The corresponding reciprocal lattice and  Brillouin zone are illustrated in \fig{grapheneLatticeVectors}(b). The lattice vectors of the reciprocal lattice are given by
\begin{equation}
     \bm a_1^* = \frac{2\pi}{\sqrt 3 a} \left( \bm e_x - \frac{\bm 
    e_y}{\sqrt 3}\right)
    \qquad
    \bm a_2^* = \frac{4\pi}{3a} \bm e_y ,
     \mwgtag{reciprocalLatticeVectorEq}
\end{equation}
and are illustrated in \fig{grapheneLatticeVectors}(b).

\subsection{Band structure}

\noindent
The carbon--carbon bonding in graphene involves  in-plane $\sigma$-bonds and  out-of-plane $\pi$-bonds.  The $\sigma$ electrons typically are associated with energy bands far from the Fermi surface while the low-energy electronic  properties are primarily associated with the $\pi$ electrons.  Since our primary interest is low-energy states,  our focus will be on the $\pi$ electrons.  It is conventional to  address the energy bands of the $\pi$ electrons in graphene using the  \textit{tight-binding approximation}, as formulated originally by Wallace in 1947 \cite{wall47} to study the structure of graphite (then of large practical interest because of the importance of graphite in early  nuclear reactors). Our discussion  will follow the presentation by Goerbig \cite{goer2011}.

\subsubsection{Tight binding approximation}

%
%
\noindent
The honeycomb lattice may be viewed as triangular with two  atoms per unit cell, suggesting a trial wavefunction \cite{goer2011}
\begin{equation}
     \psi_\bmk(\bm r) = a_\bmk \psi_\bmk^{\rm\scriptstyle (A)} (\bm r)
     + b_\bmk \psi_\bmk^{\rm\scriptstyle (B)} (\bm r),
     \mwgtag{tbind1.1}
\end{equation}
where the $\psi_\bmk^{\rm\scriptstyle (A)}$ and $\psi_\bmk^{\rm\scriptstyle 
(B)}$ are Bloch functions
\begin{subequations}
\begin{align}
     \psi_\bmk^{\rm(A)} (\bm r) &= 
     \sum_{\small\bm R_\ell}
      \phi ^{\rm (A)}
     (\bm r + \bm\delta_ {\rm A} - \bm R_\ell)
     e^{i\,\sprod {k}{R_\ell}} ,
      \mwgtag{tbind1.1a}
     \\
     \psi_\bmk^{\rm\scriptstyle (B)} (\bm r) &= \sum_{\small\bm R_\ell}
      \phi \tsup{(B)}
     (\bm r + \bm\delta\tsub B - \bm R_\ell)
     e^{i\,\sprod kR_\ell} ,
     \mwgtag{tbind1.1b}
\end{align}
\mwgtag{tbind1.1tot}%
\end{subequations}
for the A and B lattices, respectively, $\bm \delta\tsub A$ and  $\bm \delta\tsub B$ are the vectors that connect the sites of the Brevais  lattice with the site of the A or B atom, respectively, in the unit cell, and
$
    \phi ^{\rm (A)}
     (\bm r + \bm\delta\tsub A - \bm R_\ell)
$
and
$
    \phi ^{\rm (B)}
     (\bm r + \bm\delta\tsub B - \bm R_\ell)
 $
 are atomic orbital wavefunctions defined near the A or B atoms, respectively, at the  Brevais lattice site $\bm R_\ell$. Utilizing \eq{tbind1.1}, the Shr\"odinger equation $H\psi_\bmk = \epsilon_\bmk  \psi_\bmk$ may be written as
\begin{equation}
     (a^*_\bmk, b^*_\bmk) \, H_\bmk 
     \begin{pmatrix}
      a_\bmk \\ b_\bmk
     \end{pmatrix}
     =
     \epsilon_\bmk (a^*_\bmk, b^*_\bmk) \, S_\bmk
     \begin{pmatrix}
      a_\bmk \\ b_\bmk
     \end{pmatrix}
      ,
     \mwgtag{tbind1.2}
\end{equation}
where the \textit{Hamiltonian matrix} $H_\bmk$ is
\begin{equation}
    H_\bmk = H_\bmk^\dagger = 
    \begin{pmatrix}
     \psi_\bmk^{\rm\scriptstyle (A)^*} H \psi_\bmk^{\rm\scriptstyle (A)}
     &
      \psi_\bmk^{\rm\scriptstyle (A)^*} H \psi_\bmk^{\rm\scriptstyle (B)}
      \\[4pt]
       \psi_\bmk^{\rm\scriptstyle (B)^*} H \psi_\bmk^{\rm\scriptstyle (A)}
     &
      \psi_\bmk^{\rm\scriptstyle (B)^*} H \psi_\bmk^{\rm\scriptstyle (B)}
    \end{pmatrix}
     \mwgtag{tbind1.3}
\end{equation}
and the \textit{overlap matrix} $S_\bmk $  reflecting non-orthogonality of the trial wavefunctions
is
\begin{equation}
    S_\bmk = S_\bmk^\dagger = 
    \begin{pmatrix}
     \psi_\bmk^{\rm\scriptstyle (A)^*} \psi_\bmk^{\rm\scriptstyle (A)}
     &
      \psi_\bmk^{\rm\scriptstyle (A)^*}  \psi_\bmk^{\rm\scriptstyle (B)}
      \\[4pt]
       \psi_\bmk^{\rm\scriptstyle (B)^*}  \psi_\bmk^{\rm\scriptstyle (A)}
     &
      \psi_\bmk^{\rm\scriptstyle (B)^*}  \psi_\bmk^{\rm\scriptstyle (B)}
    \end{pmatrix}  .
     \mwgtag{tbind1.4}
\end{equation}
The Hamiltonian includes an atomic orbital part $H^a$ satisfying
\begin{equation}
     H^a \phi^{(j)} (\bm r + \bm\delta_j - \bm R_\ell) = \epsilon^{(j)}
     \phi^{(j)} (\bm r + \bm\delta_j - \bm R_\ell)
     \mwgtag{tbind1.4a}
\end{equation}
and a perturbative part $\Delta V$ accounting for all other terms not contained 
in the atomic-orbital Hamiltonian.  Then the matrix components of \eq{tbind1.3} for $N$ unit cells may be written as
%
%
\begin{equation}
     H_\bmk^{ij} = N \big(\epsilon^{(j)} s_\bmk^{(ij)} + t_\bmk^{(ij)}\big) ,
     \mwgtag{tbind1.6}
\end{equation}
where $i$ and $j$ denote  A or B,  we have utilized Eqs.\ \eqnoeq{tbind1.1tot} and defined
\begin{equation}
    s_\bmk^{ij} \equiv \sum_{\bm R_\ell} e^{i\,\sprod{k}{R_\ell}}
    \int d^2 \negquarterthin r \, \phi^{(i)^*}(\bm r) \phi^{(j)}(\bm r + 
    \bm\delta_{ij} - \bm R_\ell))
    = \frac{S^{ij}_\bmk}{N}
     \mwgtag{tbind1.7}
\end{equation}
with $\bm\delta_{ij} \equiv \bm \delta_j - \bm \delta_i$, and the {\em 
hopping matrix} $t_\bmk^{ij}$ is defined by
\begin{equation}
     t_\bmk^{ij} \equiv \sum_{\bm R_\ell} e^{i\,\sprod{k}{R_\ell}}
     \int d^2 \negquarterthin r \, \phi^{(i)^*}(\bm r) \,\Delta V \, \phi^{(j)}  
    (\bm r + \bm\delta_{ij} - \bm R_\ell) .
     \mwgtag{tbind1.8}
\end{equation}
The electronic bands corresponding to eigenvalues of the Schr\"odinger 
equation follow from  the secular equation
%
%
\begin{equation}
    \det\, [H_\bmk -\epsilon^\lambda_\bmk S_\bmk] = 
    \det \left[t^{ij}_\bmk -(\epsilon_\bmk^\lambda - \epsilon^{(j)})\, s_\bmk^{ij} \right]
     = 0,
     \mwgtag{tbind1.9}
\end{equation}
which has two solutions labeled by the band index $\lambda$ for the case of two  atoms per unit cell.  For the special case where the atoms on the different  sublattices have the same electronic configurations, the onsite  energy $\epsilon^{(i)}$ is a physically irrelevant constant for all $i$ that may be omitted and for graphene 
\begin{equation}
    \det \left[t^{ij}_\bmk -\epsilon_\bmk^\lambda  s_\bmk^{ij} \right] = 0 ,
     \mwgtag{tbind1.10}
\end{equation}
determines the bands in the tight-binding approximation. 

To solve this equation we may choose the Brevais lattice vectors to correspond  to the A sublattice, with the equivalent site on the B sublattice obtained from $\bm\delta\tsub B = \bm\delta\tsub{AB} = \bm\delta_3$, as illustrated in \fig{tightBindingVectors},%
 \singlefig
          {tightBindingVectors} 
          {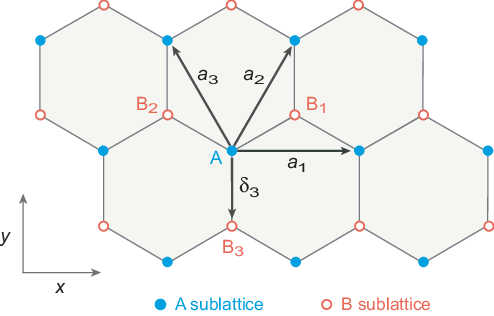}
          {0pt} 
          {0pt}
          {1.0}
          {Vectors for a tight-binding model with nearest-neighbor (NN) and  next-nearest-neighbor (NNN) interactions in graphene \cite{goer2011}.}
and include nearest-neighbor (NN) and next-nearest-neighbor (NNN) terms.  From \fig{tightBindingVectors}, the NN hopping amplitude $t$  connects points on \textit{different sublattices} and is given by
\begin{equation}
     t \equiv 
     \int d^2 \negquarterthin r \, \phi^{A*}(\bm r) \, \Delta V \, \phi^{B}  (\bm r 
    + 
    \bm\delta_{3}),
     \mwgtag{tbind1.11}
\end{equation}
while the NNN hopping amplitude connects points on the \textit{same sublattice} and is given by
\begin{equation}
     t \tsub{NNN} \equiv 
     \int d^2 \negquarterthin r \, \phi^{A*}(\bm r) \, \Delta V \, \phi^{A}  (\bm r 
    + 
    \bm a_1) .
     \mwgtag{tbind1.12}
\end{equation}
The atomic orbitals are assumed normalized
\begin{equation}
     \int d^2 \negquarterthin r \, \phi^{(i)*}(\bm r)  \phi^{(i)}  (\bm r) = 1 ,
     \mwgtag{tbind1.13}
\end{equation}
the overlap correction for NN site orbitals is
%
%
\begin{equation}
    s \equiv 
    \int d^2 \negquarterthin r \, \phi^{A^*}(\bm r)  \phi^{B}  (\bm r + 
    \bm\delta_{3}),
     \mwgtag{tbind1.14}
\end{equation}
and the overlap corrections for sites that are not nearest-neighbor are assumed to be small and will be neglected. 

For an arbitrary site on the A sublattice the off-diagonal terms of the hopping  matrix [$i \ne j$ in \eq{tbind1.8}, meaning that the hopping is between the A  and B sublattices] consist of three terms corresponding to the nearest  neighbors $B_1$, $B_2$, and $B_3$, as illustrated in \fig{tightBindingVectors}  (since we are neglecting higher than NNN couplings).  All of these have the  same hopping amplitude but different phases. From  \fig{tightBindingVectors} the site $B_3$ is described by the same lattice  vector as the site A (shifted by $\bm\delta_3$), so the phase factor in the  hopping matrix \eq{tbind1.8} is just 1.  But the site $B_1$ is shifted by  the vector $\bm a_2$ relative to $B_3$ and so contributes a phase factor $\exp(i\sprod{k}{a_2})$, while the site $B_3$ is shifted by the vector $\bm  a_3$ relative to $B_3$ and so contributes a phase factor  $\exp(i\sprod{k}{a_3})$.  Defining the sum of the NN phase factors at $\bm k$ by
%
%
\begin{equation}
    \gamma_\bmk \equiv 1 + e^{i\,\sprod{k}{a_2}} + e^{i\,\sprod{k}{a_3}} ,
     \mwgtag{tbind1.15}
\end{equation}
the off-diagonal elements of the hopping matrix may be written as $t_\bmk^{AB} = (t_\bmk^{BA})^* = t\gamma_\bmk^*$ (with the convention that $t_\bmk ^{AB}$ corresponds to hopping from B to A), and the overlap matrix  as
\begin{equation}
     s^{AB}_\bmk = (s^{BA}_\bmk)^* = s\gamma_\bmk^*
     \qquad s_\bmk^{AA} = s_\bmk^{BB} = 1 ,
     \mwgtag{tbind1.17}
\end{equation}
where the values of the diagonal terms follow from the normalization  \eqnoeq{tbind1.13}. The NNN hopping amplitudes yield the diagonal elements of the hopping matrix in  our approximation, since they connect sites on the same sublattice,
\begin{align}
     t_\bmk^{\rm AA} &= t_\bmk^{\rm BB} = 2\,t\tsub{NNN} \sum_{i=1}^3 \cos (\sprod{k}{a_i})
     = t\tsub{NNN}\, \left(|\gamma_\bmk|^2 -3\right) .
     \mwgtag{tbind1.18}
\end{align}
Thus, upon inserting these results in \eq{tbind1.10} the secular equation takes the form
\begin{equation}
    \det
    \left[
    \begin{matrix}
    t_\bmk^{\rm\scriptstyle AA} - \epsilon_\bmk & (t-s\epsilon_\bmk)\gamma_\bmk^*
      \\[2pt]
      (t-s\epsilon_\bmk)\gamma_\bmk & t_\bmk^{\rm\scriptstyle AA} - \epsilon_\bmk
    \end{matrix}
    \right] = 0,
    \mwgtag{tbind1.19}
\end{equation}
which has two solutions labeled by the band index $\lambda = \pm 1$,
\begin{equation}
     \epsilon_\bmk^\lambda = \frac{t_\bmk^{\rm\scriptstyle AA} + \lambda t \,|\gamma_\bmk|}
     {1+\lambda s \,|\gamma_\bmk|} .
     \mwgtag{tbind1.20}
\end{equation}
 As a first approximation, it may be assumed  that the overlap $s$ is much less than one and that the NN  interaction dominates the NNN interaction so that  $t\tsub{NNN}  \ll t$. Expanding \eq{tbind1.20} with these assumptions gives
\begin{align}
     \epsilon_\bmk^\lambda  &\simeq t_\bmk^{\rm\scriptstyle AA} +\lambda t 
    |\gamma_\bmk|
     -st |\gamma_\bmk|^2
     \nonumber
     \\
    &=
     \lambda t |\gamma_\bmk| + (t_{\rm\scriptstyle NNN} - st) |\gamma_\bmk|^2 ,
     \mwgtag{tbind1.21}
\end{align}
where a constant $-3 t\tsub{NNN}$ has been omitted. (Thus the effect of the  overlap $s$ at the present level of approximation is to renormalize the strength of the  next-nearest-neighbor hopping.) Finally, inserting \eq{tbind1.15} into \eq{tbind1.21} yields the dispersion relation
\begin{equation}
\begin{gathered}
     \epsilon_\bmk^\lambda =
     \lambda t \sqrt{ f_\bmk}
    \ + \ (t\tsub{NNN} -st) 
    f_\bmk ,
    \\
     f_\bmk \equiv 3 + 2\sum_{i=1}^3 \cos\, (\sprod{k}{a_i}).
     \mwgtag{tbind1.23}
    \end{gathered}
\end{equation}
Comparison with more sophisticated numerical calculations or  spectroscopic measurements suggests that reasonable physical choices for the  parameters are
$$
    t \sim 3 \units{meV} \qquad (t\tsub{NNN} -st) \sim 0.1 t \sim 0.3 
    \units{meV},
$$ 
which justifies the expansion used in going from \eq{tbind1.20}  to \eq{tbind1.21}. 
From the vectors in \fig{tightBindingVectors},  the scalar  products $\bm k\cdot \bm a_i$ in \eq{tbind1.23} may be evaluated to give
\begin{equation}
\begin{gathered}
    \sprod{k}{a_1} = \sqrt 3 a k_x
    \qquad
    \sprod{k}{a_2} = \frac{\sqrt 3 a}{2} \left(
    k_x + \sqrt 3 k_y
    \right)
    \\
    \sprod{k}{a_3} = \frac{\sqrt 3 a}{2} \left(
    -k_x + \sqrt 3 k_y
    \right)
     \mwgtag{tbind_sprod}
 \end{gathered}
\end{equation}
and  $f_\bmk$ is given explicitly by
\begin{align}
     f_\bmk &= 3 
     +2\cos \left(\sqrt 3 \,a k_x \right)
     +2\cos\left( \frac{\sqrt 3\, a}{2} \big(k_x +\sqrt 3\, k_y\big)\right)
      \nonumber
      \\
      &\quad+2\cos\left( \frac{\sqrt 3 \,a}{2} \big(-k_x +\sqrt 3 \,k_y \big)\right)
      \nonumber
      \\
       &= 3 +
       2\cos(\sqrt 3\, a k_x)
       \nonumber
       \\
       &\quad
       +4 \cos\left(\frac{\sqrt 3 \,a}{2} \, k_x\right) 
    \cos\left(\frac{\vphantom{\sqrt 3} 3a}{2}\,  k_y \right),
     \mwgtag{tbind1.23c}
\end{align}
where $2\cos\theta \cos\phi =  \cos(\theta-\phi) + \cos(\theta+\phi)$ was used in the last step.
Figure \ref{fg:dispersionGraphene}%
 \doublefig
          {dispersionGraphene} 
          {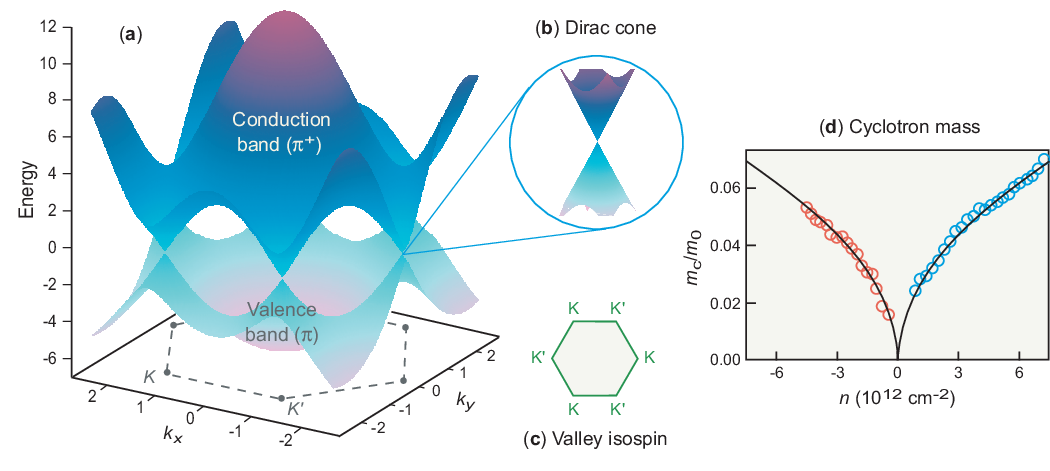}
          {0pt} 
          {0pt}
          {1.0}
          {(\textbf{a}) Electronic dispersion [energy versus momentum $(k_x, k_y)$] of  graphene calculated from Eqs.\ \eqnoeq{tbind1.23} and \eqnoeq{tbind1.23c} in a tight-binding model with no magnetic field; figure adapted  from Ref.\ \cite{guid2022}. Two  inequivalent points  in the Brillouin zone (points not connected by reciprocal  lattice vectors) are labeled $K$ and $K'$.  (\textbf{b}) Near these {\em K-points} the dispersion is approximately {\em linear,} leading to six  {\em Dirac cones,} three labeled by $K$ and three labeled by $K^\prime$.  The Fermi surface  for undoped graphene lies at zero energy in  this diagram (in the plane where the cones just touch), and the level density  vanishes there for undoped graphene. This implies that the electrons near the  Fermi surface are described by a Dirac equation.  (\textbf{c}) The six minima in the  conduction band at the Dirac cones are  called {\em valleys}, which are labeled by $K$ or $K'$.  The two  possible valley labels $K$ or $K'$ for an electron are termed  {\em valley isospin.}  (\textbf{d}) The cyclotron mass of electrons in graphene, adapted from  Ref.\ \cite{novo2005}. Near the Dirac cones of \fig{dispersionGraphene}(a) the  electronic number density vanishes $(n=0)$, since the level density tends to  zero; the data indicate that the electrons become essentially massless there  and hence obey a massless Dirac equation.
}
illustrates the electronic dispersion for graphene calculated using  Eqs.\ \eqnoeq{tbind1.23} and \eqnoeq{tbind1.23c}. The valley isospin labels  in  \fig{dispersionGraphene}(c) will be discussed further below.

\subsubsection{\label{dispersionRelation} Dispersion in graphene}

\noindent
In \fig{dispersionGraphene} the valence band $\pi$, corresponding to $\lambda =  -1$, and the conduction band $\pi^*$, corresponding to $\lambda = +1$, meet at  six discrete points that are termed {\em K points}.  As shown in   \fig{dispersionGraphene}(b), near the K points the dispersion is approximately  linear.  Since a linear dispersion is characteristic of the Dirac equation for  ultrarelativistic massless electrons, these points are also termed the {\em  Dirac points,}  and the dispersion in the vicinity of a Dirac point is called a  {\em Dirac cone.} In \fig{dispersionGraphene}(d) we show experimental evidence that the effective mass of  graphene electrons approaches zero near the Dirac points (where the electron  density vanishes). As we shall elaborate further below,   the low-energy excitations of graphene are expected to look formally like those of  {\em relativistic, massless fermions.} 

Each carbon atom contributes one $\pi$  electron and in the ground state the lower band is completely full and the  upper  band is completely empty.  Thus, the Fermi surface lies at the Dirac points  where the $\pi$ band and $\pi^*$ bands just touch,  corresponding to  $\epsilon_\bmk^\lambda = 0$.  From  \eq{tbind1.21}, this condition requires the real and imaginary  parts of $\gamma_k$ to vanish. From Eqs.\ \eqnoeq{tbind1.15} and \eqnoeq{tbind_sprod} we require the simultaneous conditions
\begin{align*}
    {\rm Re\,\, } \gamma_\bmk &=1 + \cos\left( \frac{\sqrt 3 a}{2}(k_x + 
    \sqrt 3 k_y)\right)
    \\
    &\quad+
    \cos\left( \frac{\sqrt 3 a}{2}(-k_x + 
    \sqrt 3 k_y)\right)  = 0,
    \\
    {\rm Im\,\, } \gamma_\bmk &= \sin\left( \frac{\sqrt 3 a}{2}(k_x + 
    \sqrt 3 k_y)\right)
    \\
    &\quad +
    \sin\left( \frac{\sqrt 3 a}{2}(-k_x + 
    \sqrt 3 k_y)\right) =0,
\end{align*} 
where $e^{ix} = \cos 
x + i\sin x$ has been used. Since  sine is an odd function the second of  these equations is satisfied if $k_y = 0$. Inserting that into the first of the  above equations gives $k_x = \pm 4\pi/(3\sqrt 3 a)$. Thus there are two solutions, K and K$^\prime$ , corresponding to
%
%
\begin{equation}
     \bm K = + \frac{4\pi}{3\sqrt 3 a} \, \bm e_x
      \qquad
      \bm K^\prime = - \frac{4\pi}{3\sqrt 3 a} \, \bm e_x .
     \mwgtag{KKprimeSolutions}
\end{equation}
As illustrated in \fig{Kequivalence},%
 \singlefigbottom
          {Kequivalence}  
          {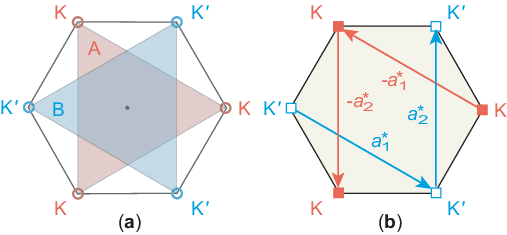}
          {0pt} 
          {0pt}
          {1.0}
          {(\textbf{a}) The graphene A sublattice (red triangle connecting sites labeled by K) and B sublattice (blue triangle connecting points labeled by K$^\prime$)  are related by inversion. (\textbf{b}) The points K and K$^\prime$ form inequivalent sets of three connected by reciprocal lattice vectors.}
there are two other equivalent points that are connected to the K solution 
by the reciprocal lattice vectors $\bm a_1^*$ and $\bm a_2^*$ defined in 
\eq{reciprocalLatticeVectorEq} and \fig{grapheneLatticeVectors}. Likewise, 
there are two other equivalent points connected to the K$^\prime$ solution by 
reciprocal lattice vectors.  The sets K and K$^\prime$  are distinct, 
representing the two independent solutions of \eq{KKprimeSolutions}. 
The solutions come in pairs because the band Hamiltonian 
\eqnoeq{tbind1.3} is time-reversal invariant ($H_k = H_{-k}^*$).  Hence 
if $\bm k$ is a solution  of $\epsilon_\bmk = 0$, so is $-\bm k$. Because of the 
lattice symmetry there is only one such independent $({\rm K},{\rm K'})$ pair in the Brillouin zone.

Thus, the six Dirac points can be divided into two  sets of three,  corresponding to the points K and K$^\prime$ in \fig{grapheneLatticeVectors}(b).  The points K are all equivalent because they are connected by reciprocal  lattice vectors; likewise for the points K$^\prime$. The inequivalent points K  and K$^\prime$ constitute a 2-dimensional degree of freedom called the {\em  valley isospin} (or just \textit{isospin} for brevity) $\xi = \pm 1$, so-called because a two-component wavefunction can be  treated formally as a ``spin'' [fundamental representation of SU(2)],  similar to the isotopic spin formalism of nuclear  physics].  One says that there is a {\em two-fold valley degeneracy} in graphene  labeled by the quantum number $\xi$. This  degeneracy is expected to survive approximately for excitations having energy  much smaller than the bandwidth $\sim |t|$, since they are restricted to the  vicinity of a given K-point.

\subsection{\label{lowEnergyExcitations} Low-energy excitations}

\noindent
Let us now consider the low-energy excitations expected in graphene.  The  Hamiltonian and dispersion relations derived above are complicated by the  non-orthogonality of the atomic wavefunctions, which necessitates the overlap  integrals \eqnoeq{tbind1.7}. As noted in conjunction with \eq{tbind1.21},  the primary effect of the overlaps is to renormalize the strengths of  next-nearest-neighbor hopping.  For low-energy excitations in particular, the  overlap effect should be small and it is convenient to absorb it  into NNN hopping amplitudes and define an {\em effective tight-binding Hamiltonian}  that takes the form \cite{goer2011}
%
%
\begin{equation}
    H_\bmk \simeq
    t 
    \begin{pmatrix}
     0 & \gamma_\bmk^*
     \\[2pt]
     \gamma_\bmk & 0
    \end{pmatrix}
    +
    t\tsub{NNN}
    \begin{pmatrix}
    |\gamma_\bmk|^2 & 0
    \\[2pt]
    0 & |\gamma_\bmk|^2
    \end{pmatrix} .
     \mwgtag{effectiveTB_ham}
\end{equation}
The corresponding eigenvectors are the spinors
\begin{equation}
     \psi_\bmk^\lambda = 
     \begin{pmatrix}
      a_\bmk^\lambda 
      \\[2pt]
      b_\bmk^\lambda
     \end{pmatrix} ,
     \mwgtag{sublatticeSpinor}
\end{equation}
with components that are the amplitudes for the Bloch wavefunctions of Eqs.\
\eqnoeq{tbind1.1a} and \eqnoeq{tbind1.1b} on the two different sublattices A 
and B.


From  \fig{dispersionGraphene}, the low-energy 
excitations in graphene are expected to occur in the vicinity of the Dirac 
points $\pm \bm K$. We decompose the wavevector $\bm k$ as
\begin{equation}
    \bm k = \pm \bm K + \bm q,
     \mwgtag{tbind1.24}
\end{equation}
where we assume that $|\bm q| \ll |\bm K| \sim a^{-1}$ and expand the energy  dispersion around $\pm\bm K$.  The relative phase between the two sublattice components that we used above was  an arbitrary choice.  As a consequence, the dispersion relation  \eqnoeq{tbind1.21} is unaffected by a change $\gamma_\bmk \rightarrow \gamma_\bmk  \exp(ig_\bmk)$ for any real, nonsingular function $g_\bmk$ and it is convenient to  use this freedom to replace \eq{tbind1.15} with the more symmetric expression
%
%
\begin{equation} 
    e^{i\,\sprod{k}{\delta_3}} \gamma_\bmk
    = e^{i\,\sprod{k}{\delta_1}} + e^{i\,\sprod{k}{\delta_2}} + e^{i\,\sprod{k}{\delta_3}}
    = \sum_{i=1}^3 e^{\pm i \,\sprod{k}{\delta_i}} ,
     \mwgtag{tbind1.25}
\end{equation}
where the relationship of the vectors $\bm\delta_i$ and $\bm a_i$  that is illustrated in \fig{tightBindingVectors2}%
 \singlefig
          {tightBindingVectors2} 
          {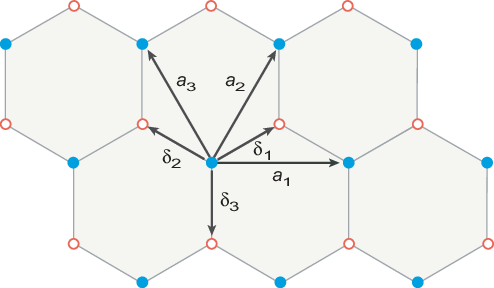}
          {0pt} 
          {0pt}
          {1.0}
          {Relationship among the vectors $\bm\delta_i$ and $\bm a_i$. Adapted from Ref.\ \cite{goer2011}.}
has been used to write $\bm a_2 + \bm \delta_3 = \bm \delta_1$ and $\bm a_3 + \bm \delta_3 = \bm \delta_2$ \cite{goer2011,bena2009}.   Thus
\begin{equation}
    e^{i \,\sprod{k}{\delta_3}} \gamma_k  | _{\bm k = \pm \bm K + \bm q} 
     = \sum_{i=1}^3 e^{i\,(\pm \bm K + \bm q) \cdot \bm \delta_i}
    = \sum_{i=1}^3 e^{\pm i \,\sprod{K}{\delta_i}}  e^{i\, \sprod{q}{\delta_i}}.
     \mwgtag{tbind1.26}
\end{equation}
Assuming $\bm q$ to be a small displacement from $\bm K$ and expanding the 
last exponential in the preceding expression to second order gives
%
%
\begin{align}
\gamma_{\bm q}^{\pm} &\equiv e^{i \,\sprod{k}{\delta_3}} \gamma_{\bm k}  | _{\bm k = \pm \bm K + \bm q}
     \nonumber
     \\
    &=
     \sum_{i=1}^3  e^{\pm i \,\sprod{K}{\delta_i}}
     \left(1 + i\,\sprod{q}{\delta_i} - \frac12 (\sprod{q}{\delta_i})^2 \right)
     \nonumber
     \\
    &= \sum_{i=1}^3  e^{\pm i \,\sprod{K}{\delta_i}}
     +
     i \sum_{i=1}^3  e^{\pm i \,\sprod{K}{\delta_i}} (\sprod{q}{\delta_i})
     \nonumber
     \\
      &\quad-\frac12
      \sum_{i=1}^3  e^{\pm i \,\sprod{K}{\delta_i}} (\sprod{q}{\delta_i})^2
     \nonumber
     \\[2pt]
    &\equiv
     (\gamma_{\bm q}^{\pm}) ^{(0)} + (\gamma_{\bm q}^{\pm})^{ (1)} + (\gamma_{\bm q}^{\pm})^{ (2)} .
     \mwgtag{tbind1.27}
\end{align}
From the geometry in \fig{tightBindingVectors2}, the components of the vectors 
$\delta_i$ are
\begin{equation}
\begin{gathered}
    \delta_1^x = \displaystyle\frac{\sqrt 3 }{2} a
     \qquad
     \delta_2^x = -\frac{\sqrt 3 }{2} a
      \qquad
     \delta_3^x = 0,
     \\
     \delta_1^y = \displaystyle\frac a2
     \qquad
     \delta_2^y = \frac a2
      \qquad
     \delta_3^y = -a .
     \mwgtag{tbind1.28}
\end{gathered}
\end{equation}
The zero-order term in \eq{tbind1.27} vanishes:
$$
     (\gamma_{\bm q}^{\pm})^{ (0)} = \sum_{i=1}^3  e^{\pm i \sprod{K}{\delta_i}}
     = e^{\pm 2\pi i/3} + e^{\mp 2\pi i/3} + 1 = 0,
$$
where Eqs.\ \eqnoeq{KKprimeSolutions} and \eqnoeq{tbind1.28},  and that $e^{ix} 
+ e^{-ix} = 2 \cos x$, were used. Thus, if second and higher order
terms are neglected,
\begin{equation}
    e^{i \sprod{k}{\delta_3}} \gamma_{\bm k}  | _{\bm k = \pm \bm K + \bm q} 
    = 
    (\gamma_{\bm q}^\pm)^{(1)}
    =  i \sum_{i=1}^3  e^{\pm i \sprod{K}{\delta_i}} (\sprod{q}{\delta_i}).
    \mwgtag{tbind1.29}
\end{equation}
From \eq{tbind1.28}
\begin{gather*}
    \sprod{q}{\delta_1} = \frac{\sqrt 3 }{2} a q_x + \frac a2 q_y
    \qquad
    \sprod{q}{\delta_2} = - \frac{\sqrt 3 }{2} a q_x + \frac a2 q_y,
    \\
    \sprod{q}{\delta_3} = -aq_y ,
\end{gather*}
which may be used to evaluate \eq{tbind1.29}; to this order
 \cite{goer2011}
 %
 %
\begin{equation}
     \gamma_{\bm q}^{\pm}
     \simeq \mp \frac{3}{2}a(q_x \pm iq_y ).
     \mwgtag{tbind1.30}
\end{equation}
To deal with the $\pm$ and $\mp$ symbols  that follow from the two-fold degeneracy of solutions in \eq{KKprimeSolutions}, it  is convenient to employ the {\em valley isospin quantum number} $\xi$ introduced earlier, 
with $\xi = +1$ corresponding to the ${\rm K}$ point at $\bm K$ and  $\xi=-1$  to the K$^\prime$ point at $-\bm K$ [modulo translations  by reciprocal lattice vectors; see \eq{KKprimeSolutions} and  \fig{Kequivalence}]. Thus, \eq{tbind1.30} becomes
\begin{equation}
     \gamma_{\bm q}^{\xi}
     =  -\frac{3}{2} \xi a(q_x  + i\xi q_y ).
     \mwgtag{tbind1.32}
\end{equation}
Since we are evaluating only to first order,  the $t\tsub{NNN}$ term in  \eq{effectiveTB_ham} may be dropped and inserting the preceding expression for $\gamma_k$ in \eq{effectiveTB_ham} gives 
$$
     H_k^\xi \simeq 
     t 
    \begin{pmatrix}
     0 & \gamma_k^*
     \\
     \gamma_k & 0
    \end{pmatrix}
    \simeq
     -\frac32 \xi a t 
     \begin{pmatrix}
      0 & q_x - i\xi q_y
      \\
      q_x+i\xi q_y & 0
     \end{pmatrix} 
$$
as the effective low-energy Hamiltonian.
Defining a Fermi velocity 
%
%
\begin{equation}
     v\tsub F \equiv  \frac{-3at}{2\hbar}  = \frac{3a|t|}{2\hbar}
     \mwgtag{fermiVelocity}
\end{equation}
(where $t$ is generally negative), gives for the Hamiltonian
\begin{equation}
     H_k^\xi = \hbar \xi v\tsub F
     \begin{pmatrix}
      0 & q_x -i\xi q_y
      \\
      q_x + i\xi q_y & 0 
     \end{pmatrix} ,
     \mwgtag{tbind1.33}
\end{equation}
which may be expressed as
%
%
\begin{equation}
    H_q^{\xi} = \hbar \xi v\tsub F (q_x\sigma^x + \xi q_y \sigma^y) ,
     \mwgtag{tbind1.34}
\end{equation}
upon utilizing the standard $2\times2$ representations of the Pauli matrices $\sigma^x$ and $\sigma^y$. The dispersion at this level of approximation is obtained by dropping the term proportional to $t\tsub{NNN}$ from \eq{tbind1.21} and inserting $\gamma_\bmk$ from \eq{tbind1.32}, giving
%
%
\begin{equation}
     \epsilon_\bmk = \lambda t |\gamma_\bmq| 
     = \frac{3}{2} a \lambda t|\bm q|
     = \hbar v\tsub F \lambda |\bm q|,
     \mwgtag{tbind1.35}
\end{equation}
where \eq{fermiVelocity} was used. The energy $\epsilon_\bmk$  depends on the band 
index $\lambda$ but not the valley isospin $\xi$.

\subsection{\label{chiralityGraphene} Chirality}

\noindent
The sublattice symmetry of graphene (two interlocking  triangular  sublattices) should be approximately valid for low energy excitations. Since it  corresponds to two degrees of freedom, it is convenient to describe it in terms  of a pseudospin $\boldsym \sigma$ that takes two values.  To the degree that  the  sublattice symmetry is respected, there is an associated conserved quantity.  For  actual relativistic electrons it is convenient to express the spin degree of  freedom in terms of  helicity, which is the projection of the spin on the  direction of motion.  Since the solution near the Dirac points (low-energy  states) behaves effectively as that for massless relativistic  electrons, it  is convenient to define the helicity $\eta_q$ operator associated with  projection of the sub-lattice pseudospin on the wavevector by \cite{goer2011}
%
%
\begin{equation}
      \eta_q \equiv   \frac{\boldsym{\sigma} \cdot \boldsym q}{|\boldsym q|}
     \mwgtag{helicityDefiningEq},
\end{equation}
which has eigenvalues $\eta_q \ket{\eta = \pm 1} = \pm \ket{\eta = \pm 1}$. The helicity for \textit{massless particles} commutes with the Dirac Hamiltonian, so it is  a conserved quantum number. 

For massless particles the helicity is the same as the {\em chirality}  (``handedness''), which is the eigenvalue of the Dirac $\gamma_5$ operator. We  shall assume electrons near the Dirac points to be exactly massless [see  \fig{dispersionGraphene}(d)], so that their helicity and chirality may be identified.  It is common to refer to these fermions as {\em massless chiral  fermions,} and to the sublattice symmetry as {\em conservation of chirality}. (Remember that this is a ``pseudo-chirality'' associated with  the sublattice  pseudospin, not the actual spin $s$ of the electron.) The band index $\lambda$ is the product of the sublattice chirality and the  valley  isospin, $\lambda = \eta  \xi$. Conservation of chirality implies that there is no backscattering of electrons  upon encountering a slowly-varying barrier, since to rotate the wavevector by  $\pi$ would flip the sign of the chirality.  From the tight-binding model, the effective low-energy Hamiltonian at the  corners of the Brillouin zone may be written
\begin{equation}
 H = \hbar v\tsub F\, \boldsym \sigma \cdot \boldsym p,
 \mwgtag{chiralHam}
\end{equation}
which is equivalent to the Dirac equation for massless chiral  fermions (also called the \textit{Weyl equation}), but with the speed of  light $c$ replaced by the Fermi velocity $v\tsub F$.

\section{\label{FQHEgeneral}Classical and Quantum Hall Effects}

\noindent
The integer and fractional quantum Hall effects represent remarkable physics that appears when a strong magnetic field is applied to a  low-density electron gas confined to two dimensions at very low temperature. These effects were first  observed in the early 1980s for 2D electron gases created in semiconductor  devices.  To understand the quantum Hall effect in graphene it is important to  understand the basics of this extensive earlier work. This section gives a general introduction to the quantization of  non-relativistic electrons in a magnetic field that is the basis of the quantum  Hall effect, and the following two sections give a general introduction  to the integer and fractional quantum Hall effects, respectively.   Then we  shall be  prepared to address the issue of quantum Hall effects in  graphene.

\subsection {\label{normalHallEffect}The classical Hall effect}

\noindent
 Let's begin by recalling the basics of the classical Hall effect.%
 \doublefig
          {classicalHallEffect}  
          {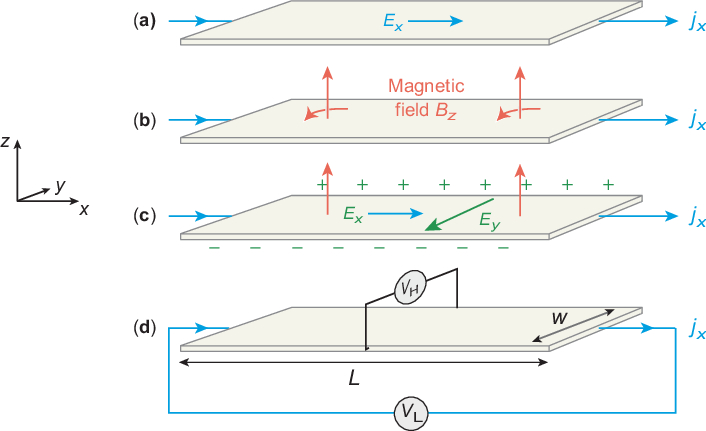}
          {0pt} 
          {0pt}
          {1.0}
          {The classical Hall effect.  (\textbf{a}) An electric field $E_x$ causes a current $j_x$ to flow through a thin rectangular sample in the $x$ direction. (\textbf{b}) A uniform magnetic field $B_z$  is placed on the  sample in the positive $z$  direction.  The curved arrows indicate the Lorentz-force response of the electrons to the  magnetic field, causing a deflection in the $y$ direction. (\textbf{c}) This causes  electrons to accumulate on one edge and  a positive ion excess on the opposite edge, producing a transverse electric field  $E_y$ (the Hall field) that just cancels the force produced by the magnetic  field. Adapted from Ref.\ \cite{guid2022}}
In \fig{classicalHallEffect}(a), an electric field $E_x$ causes a current $j_x$  to flow through a thin rectangular sample in the $x$ direction. In  \fig{classicalHallEffect}(b) a uniform magnetic field $B_z$  is placed on the  sample in the positive $z$  direction, which   causes a deflection of the electrons in the  $y$ direction (Lorentz force).  As indicated in  \fig{classicalHallEffect}(c), electrons accumulate on  one edge and a positive ion excess accumulates on the opposite edge, producing a  transverse electric field $E_y$ (the {\em Hall field}) that in steady state just  cancels the Lorentz force produced by the magnetic field. As a result, the current is entirely  in the $x$ direction and for uniform samples the Hall field is perpendicular to the current, in  the direction $\crossprod jB $. Typically the transverse voltage $V\tsub H$ and the longitudinal voltage $V\tsub L$ are measured in the experiment, as indicated in \fig{classicalHallEffect}(d).

The situation may be analyzed quantitatively  in terms of the classical Lorentz force $\bm F$  acting on the electrons in the magnetic field,
\begin{equation}
    \bm F = -e \left(\bm E + \frac1c \, \crossprod{v}{B} \right),
     \mwgtag{classicalHall1.1}
\end{equation}
where $\bm B$ is a magnetic field  oriented in the  $+z$ direction, $\bm E$ 
is the electric field, and $\bm v$ is the electron velocity (taken to be in 
the $x$ direction). Requiring forces in the $y$ direction to cancel,
$$
    F_y = -e \left( E_y - \frac1c v_xB_z \right) = 0,
$$
implies that the transverse electric field and the 
perpendicular magnetic field are related by
\begin{equation}
    E_y = \frac1c v_xB_z .
     \mwgtag{classicalHall1.2}
\end{equation}
The velocity $\bm v$ can be estimated using a force $\bm$ derived from the Drude model \cite{kitt2004,quii2009,phil2003},
\begin{equation}
    \bm F = m\left(\deriv{\bm v}{t} + \frac{\bm v}{\tau}\right) =  -e 
    \left(\bm E + \frac1c \, \crossprod{v}{B} \right) ,
     \mwgtag{callicalHall1.3}
\end{equation}
where $m$ is the effective mass of an electron and $\tau$ is the mean time  between electron collisions.  In steady state, $d{\bm  v}/dt = 0$ and the equations of motion are
\begin{equation}
\begin{gathered}
   v_x = -\frac{e\tau}{m}\, E_x - \omega\tsub c \tau  v_y
   \qquad
   v_y = -\frac{e\tau}{m}\, E_y + \omega\tsub c \tau v_x,
   \\
   v_z = -\frac{e\tau}{m}\, E_z ,
\mwgtag{classicalHall1.4}
\end{gathered}
\end{equation} 
where $B \equiv B_z = |\bm B|$ and the {\em cyclotron frequency} $\omega\tsub c$ is defined by
\begin{equation}
   \omega\tsub c \equiv \frac{eB}{mc}.
   \mwgtag{classicalHall1.3}
\end{equation}
Inserting the first of Eqs.\ \eqnoeq{classicalHall1.4} into \eq{classicalHall1.2} gives a relationship between the transverse electric field $E_y$ and longitudinal electric field 
$E_x$,
\begin{eqnarray}
    E_y = \frac{B}{c} \, v_x = \frac{B}{c}
    \left(
    -\frac{e\tau}{m}\, E_x - \omega\tsub c \tau v_y
    \right)
    = - \frac{eB\tau}{mc} \, E_x,
    \mwgtag{classicalHall1.5}
\end{eqnarray}
where at equilibrium $v_y = 0$ has been required. If $\tau$ is the mean collision time for an electron, the current density may  be approximated in the Drude model as
\begin{equation}
    \bm j = \frac{e^2 n_e}{m} \, \tau \bm E  = \sigma \bm E
    \qquad
    \sigma \equiv \frac{n_e e^2 \tau}{m},
    \mwgtag{classicalHall1.6}
\end{equation}
where $n_e$ is the electron number density and $\sigma$ is the conductance.

In a Hall effect  experiment, one typically measures the  transverse voltage $V\tsub H$ and the longitudinal voltage $V\tsub L = E_x  \ell$,  where $\ell$ is the distance over which the voltage changes by $V\tsub L$. If  the sample is approximated as 2D with  transverse width $w$ [see \fig{classicalHallEffect}(a)], the relationship  between the Hall voltage $V\tsub H$, Hall field $E_y$, and total current  $I$ is given by
\begin{equation}
     E_y = \frac{V\tsub H}{w}
     \qquad
     I = w j_x.
     \mwgtag{classicalHall1.7}
\end{equation}
The {\em Hall resistance} $R\tsub H$ is then defined by [see Appendix 
\ref{ResistanceAndStuff})]
\begin{equation}
    R\tsub H = \frac{V\tsub H}{I} = \frac{w E_y}{w j_x} = \frac{E_y}{j_x}
    = - \frac{B}{ecn_e},
    \mwgtag{clasicalHall1.8}
\end{equation}
where Eqs.\ \eqnoeq{classicalHall1.5}--\eqnoeq{classicalHall1.7} were used. 
Likewise, the longitudinal resistance $R\tsub L$ is given by
\begin{equation}
    R\tsub L = \frac{V\tsub L}{I} 
    = \frac{LE_x}{w j_x} = \frac{\ell}{w\sigma}, 
    \mwgtag{classicalHall1.9}
\end{equation}
where \eqnoeq{classicalHall1.6} was used. Now let us see how this classical picture is modified by quantum effects, guided by the presentation in Ref.\ \cite{guid2022}.

\subsection{Quantum Hall effects}

\noindent
A much richer set of possibilities   is  found when  Hall effect experiments are performed at low  temperatures and high magnetic fields.  These {\em quantum Hall effects (QHE)}  can be separated into two sets of phenomena: 

\begin{enumerate}

\item 
the  {\em \iqhe\ (IQHE)}, and

\item
the {\em \fqhe\ (FQHE).}

\end{enumerate}

\noindent
To understand these quantum Hall effects we must first examine the quantization of electrons confined to two dimensions in a strong  magnetic field.

\section{\label{landauLevels} Non-Relativistic Landau Levels}

\noindent
Consider the quantum description of non-relativistic electrons in a 2D gas  with a strong magnetic field  transverse to the 2D plane.    The magnetic field will lead to substantial spin polarization, so we  assume   states to have a single electron spin polarization  and  neglect the constant Zeeman energy for each polarized spin.  This is not strictly justified since some spin effects are observable even at high magnetic  field,  but it should not change substantially the general features that  we shall emphasize.  Initially the  Coulomb interaction also will be neglected relative to the effect of the strong  magnetic field, but we will return to its effect later.

\subsection{Schr\"odinger equation and solution}

\noindent
This section follows the presentation by Phillips \cite{phil2003}.  By the usual minimal prescription to ensure gauge invariance, the effect of an  electromagnetic field   may be included by adding to the  momentum operator $\bm p = (p_x, p_y)$  a  term depending on the vector potential and the  Hamiltonian may be written
\begin{equation}
    H = \frac{1}{2m} \left( \frac{\hbar}{i} \,\boldsym\nabla  + \frac{e}{c} \,\,\bm 
    A \right)^2 ,
    \mwgtag{ll1.1}
\end{equation}
where $\bm A$ is a 3-vector potential having a curl equal to the magnetic field $\bm 
B$:
\begin{equation}
     \bm B = \bm\nabla \times \bm A \qquad \bm B \equiv (B_x,\, B_y,\, B_z) = (0, 
    0, B).
    \mwgtag{magfieldDef}
\end{equation}
Various gauge choices give this magnetic field.  Two common  ones are the {\em Landau gauge} and the {\em symmetric gauge},  which are defined by
\begin{subequations}
\begin{gather}
    \bm A = (0,\, Bx,\, 0) \quad \units{(Landau gauge)},
    \mwgtag{ll1.2a}
    \\
    \bm A = \frac B2 (-y,\, x,\, 0) \quad \units{(symmetric gauge)} .
    \mwgtag{ll1.2b}
\end{gather}
\mwgtag{ll1.2}%
\end{subequations}
The electromagnetic field is unchanged by gauge transformations, so {\em no 
physical quantities depend on the gauge choice,} but the forms of the vector 
potential $\bm A$ and  wavefunction $\psi$ do.  That is not an ambiguity because $\bm A$ and $\psi$ are {\em  not observables } in this context.

It will prove convenient for the initial discussion to work in the Landau gauge, $\bm A = (0,\, Bx,\, 0)$, so that $A_y = Bx$ and $A_x = 0$ (with the $z$ dimension 
neglected, since we are concerned with 2D electron gases). The Schr\"odinger 
equation is then
\begin{equation}
    -\frac{\hbar^2}{2m} \left[
    \partial_x^2 + \left(\partial_y - \frac{ieBx}{\hbar c}\right)^2
    \right]
    \psi(x,y) = E\psi(x,y) ,
    \mwgtag{ll1.4}
\end{equation}
with $m$ the effective mass, $\partial_y \equiv \partial/\partial y$, and  $\partial_x^2 \equiv \partial^2/\partial x^2$. Since the  Hamiltonian does not depend on $y$ in the chosen gauge, it is convenient to write the  wavefunction  in the separable form $\psi(x,y) =  \phi(x) e^{iky}$. Inserting this into \eq{ll1.4} indicates that $\phi(x)$ obeys
\begin{equation}
    \frac{\hbar\omega\tsub c}{2}
    \left[
    -\Lambda^2 \partial_x^2 + \left( \frac{x}{\Lambda} - \Lambda\,k\right)^2 \right]
    \phi(x) = E \phi(x),
    \mwgtag{ll1.6}
\end{equation}
where $\omega\tsub c = eB/mc$ is the cyclotron frequency of  \eq{classicalHall1.3}  and the {\em magnetic length} $\Lambda$ is 
\begin{equation}
    \Lambda \equiv \sqrt{\frac{\hbar c}{eB}}.
    \mwgtag{ll1.7}
\end{equation}
But \eq{ll1.6} is just the harmonic-oscillator Schr\"odinger equation,   so
the wavefunction is of harmonic oscillator form
\begin{equation}
    \psi_{nk}(x,y) = H_n \left(\frac x\Lambda -\Lambda k \right) 
    e^{-(x-x_k)^2/2\Lambda^2} e^{iky} ,
    \mwgtag{ll1.8}
\end{equation}
where $H_n$ is a Hermite polynomial and $x_k = \Lambda^2 k$.  The  corresponding energy is
\begin{equation}
    \epsilon_n = \hbar\omega\tsub c \left(n + \frac12 \right),
    \mwgtag{ll1.9}
\end{equation}
which  depends on the principle Landau quantum  number $n= 0, 1, 2, \ldots, $ but not on $k$.

\subsection{Properties of Landau levels}

\noindent
From  \eq{ll1.8}  the wavefunction is extended in  $y$  but localized in $x$ near $x_k = \Lambda^2 k$.  This is  specific to our gauge choice: under a local gauge transformation the vector potential $A(\bm r,t)$ and  the wavefunction  $\psi(\bm r, t)$ are changed simultaneously according to
\begin{subequations}
\begin{gather}
    A(\bm r, t) \ \rightarrow\ A(\bm r,t) + \boldsym\nabla \chi(\bm r, t),
    \mwgtag{gaugexformA}
    \\
    \psi(\bm r, t) \ \rightarrow\ \psi(\bm r, t) e^{i\hbar c\chi(\bm r, t)/e} ,
    \mwgtag{gaugexformB}
\end{gather}
\mwgtag{gaugexform}%
\end{subequations}
where $\bm r \equiv (x,y)$ and $\chi(\bm r, t)$ is some scalar function, so the forms of both $\bm A$ and $\psi$ (but {\em no observable quantities})  depend on  the gauge chosen. If $L$ is the spatial extent in the $y$ direction, $k_m = 2\pi m/L $, with $m$ an integer.  Therefore the spacing in $k$ is given by
\begin{equation}
k_{m+1} - k_m = \frac{2\pi}{L}.
\mwgtag{k-spacing}
\end{equation}
Energy levels labeled by $n$ in \eq{ll1.9} are termed {\em Landau levels.} Semiclassically, they correspond to   electrons moving in circles of quantized radius specified by  the quantum number $n$, with quantization arising from requiring an integer number of electron de Broglie wavelengths  to fit around the  cyclotron orbit. For a given Landau orbit the radius of the circle is termed the {\em cyclotron radius}  and the center of the circle  is termed the {\em guiding center} [see \fig{landauDiracDispersion+cyclotronCoord}(b)]. Landau levels are highly degenerate  in strong magnetic fields because there are many  possible locations for the center of the circle of some radius defining a cyclotron orbit.

\subsection{Degeneracy and level densities}

\noindent
The degeneracy for a Landau level labeled by $n$ is equal to the number of $k$ values associated with that $n$.  If electronic interactions are neglected, the  level density $g(\epsilon)$ is a series of $\delta$-functions at  energies  corresponding to discrete values of $n$, weighted by a degeneracy factor $N$ equal to the number of electrons that can occupy the Landau level,
\begin{equation}
    g(\epsilon) = N \sum _n \delta \left[ \epsilon -\hbar\omega\tsub c
    \left( n + \frac12 \right) \right] .
    \mwgtag{ll1.10}
\end{equation}
 For a 2D sample of  monolayer graphene having width $w$, length $\ell$, and magnetic length $\Lambda$,
\begin{equation}
     N = \frac{B\ell w}{hc/e} = \frac{\ell w}{2\pi \Lambda^2}.
    \mwgtag{ll1.11}
\end{equation}
Equation (\ref{eq:ll1.11}) has a simple physical interpretation.  The magnetic flux through a 2D sample of area $A = \ell w$  is $B\ell w$, so $N$ is  the magnetic flux in units of the quantum of magnetic flux, $hc/e$, where $h$ is Planck's constant, $e$ is the electronic charge, and $c$ is the speed of light. From \eq{ll1.11}, the number of states per unit area in each Landau level is
\begin{equation}
    n_B = \frac{B}{hc/e} = \frac{1}{2\pi \Lambda^2} ,
    \mwgtag{ll1.12}
\end{equation}
which does not depend on the Landau level but scales linearly with the  strength of the magnetic field.  Thus, \textit{large magnetic fields imply large degeneracies}.  
If the actual number density of electrons  is $n_e$,  a {\em filling factor} $\nu$ may be defined by by
\begin{equation}
    \nu \equiv \frac {n_e}{n_B} = \frac{n_e hc}{eB}.
    \mwgtag{ll1.13}
\end{equation}
If $\nu$ is an integer, the lowest $\nu$ Landau levels will  be filled completely and all other Landau levels will be empty, while if $\nu$ is not  integer a  Landau level will be partially filled. For $\nu$ equal to an  integer $m$ there will be $m$ completely-filled Landau levels and there will be  an energy gap between the $n=m$ and $n=m+1$ Landau levels.  Thus, at $T=0$  the flow of charge is suppressed because electronic excitation is inhibited by the  gap.

\section{\label{IQHEchapter} The Integer Quantum Hall Effect}

\noindent
What happens to the classical Hall effect  for a dilute 2D  electron gas at very low temperature and very strong magnetic field? If the filling factor $\nu$ given by \eq{ll1.13} is an integer  the Hall  conductance  $\sigma\tsub H = 1/R\tsub H$ becomes quantized, since 
\begin{equation}
    \sigma\tsub H = \frac{1}{R\tsub H} = -\frac{ecn_e}{B}
    = -\frac{ec\nu n_B}{B} = \nu \frac{-e^2}{h}
    \mwgtag{ll1.14}
\end{equation}
As electrons are added beyond $\nu$ completely filled Landau levels, they will  go into the next level.  One might expect that the  longitudinal conductance $\sigma\tsub L$ would at first increase with added electrons until the  level becomes half full, and then decrease with addition of further electrons until the level  becomes completely full, as illustrated by the dashed curves in  \fig{conductivitiesLH}(a).%
 \singlefig
          {conductivitiesLH} 
          {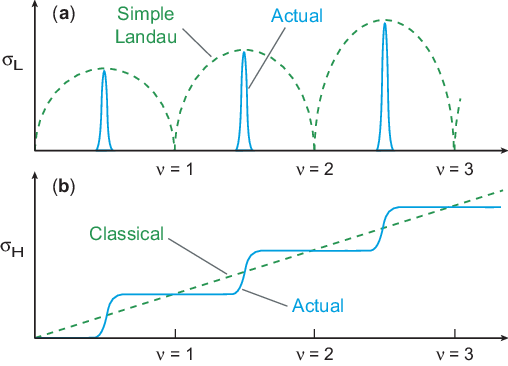}
          {0pt} 
          {0pt}
          {1.0}
          {Conductance as a function of Landau-level filling factors $\nu$.  (\textbf{a}) Longitudinal conductance $\sigma\tsub L$.  (\textbf{b}) Hall conductance  $\sigma\tsub  H$.  Dashed green curves indicate the expectation from simple considerations.   Solid blue curves suggest what is actually observed (compare experimental data in \fig{IQHEsteps}).}
However, that is not what is found. Instead, 
\begin{enumerate}
\item
the  longitudinal conductance is finite only in narrow ranges [solid blue peaks in  \fig{conductivitiesLH}(a)], and  

\item
the Hall conductance $\sigma\tsub H$ has very flat plateaus in the regions around integer filling  factor  $\nu$ that coincide with regions of very small longitudinal conductance, as  illustrated in \fig{conductivitiesLH}(b).
\end{enumerate}
The next section will  describe these quantum Hall experiments and attempt to understand these properties.

\subsection{\label{integalQuantumHallEffect}Discovery of the integer quantum 
Hall effect}

\noindent
The \iqhe\ (IQHE) was discovered in 1980 by von Klitzing,  Dorda, and Pepper  \cite{vonk1980}. Experimental data exhibiting the IQHE are  shown in \fig{IQHEsteps}%
 \doublefig
          {IQHEsteps}  
          {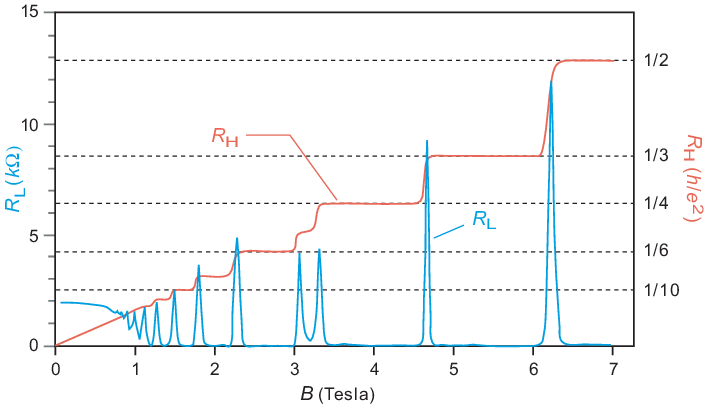}
          {0pt} 
          {0pt}
          {1.0}
          {Data showing the integer quantum Hall effect \cite{vonk1980}.  The Hall resistance $R\tsub H$ varies  stepwise between plateaus with changes in magnetic field B (upper curve). Step height is given by  the physical constant $h/e^2$  divided by an integer $i$. The figure shows steps  for $i =2,3,4,5,6,8$ and 10.  The lower curve with multiple peaks represents the  ohmic (longitudinal) resistance $R\tsub L$, which is finite at transitions between plateaus, but drops to almost zero over each plateau. }         
and a schematic experimental apparatus  is illustrated in \fig{hallBar}.%
\singlefigbottom
          {hallBar}  
          {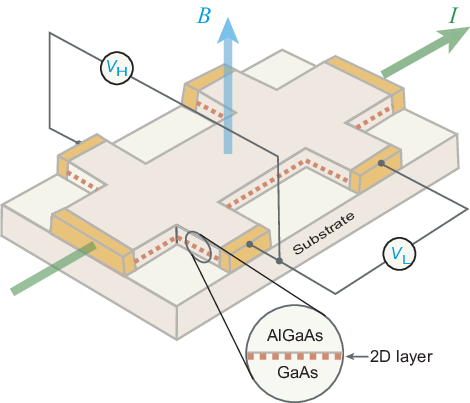}
          {0pt} 
          {0pt}
          {1.0}
          {Schematic diagram of a  Hall bar experimental apparatus  employing a gallium arsenide heterostructure. The dotted red line indicates the  2D electron gas at the interface between the gallium arsenide (GaAs) and  aluminum gallium arsenide (AlGaAs) layers. The magnetic field is $B$, the  electrical current is $I$, the Hall voltage is $V\tsub H$, and the longitudinal  voltage is $V\tsub L$.  Adapted from Ref.\  \cite{eise1990}.}
In the classical Hall effect the Hall resistance $R\tsub H$ should increase  linearly with field strength, according to  \eq{clasicalHall1.8}, while the longitudinal resistance $R\tsub L$ should be  independent of the magnetic field, according to \eq{classicalHall1.9}. This is approximately true for magnetic fields with strength less than about 1 tesla in \fig{IQHEsteps}, but for stronger fields the Hall resistance $R\tsub H$ develops plateaus  where it remains constant with increasing magnetic field at values 
\begin{equation}
    R\tsub H = \rho_{yx} = \frac{V_y}{I_x} = \frac{h}{ne^2} 
    \mwgtag{iqhe1.1}
\end{equation}
(see Appendix  \ref{ResistanceAndStuff} for the definition of the resistivity tensor $\rho_{ij}$), where $n$ is an integer, and in the center of regions where $R\tsub H$ is  constant the longitudinal resistance $R\tsub L$ behaves as $R\tsub L \simeq \exp(-\Delta/ 2k\tsub B T)$, which tends to zero as $T \rightarrow 0$, indicating the onset of charge  transport without dissipation. This is suggestive of an energy gap of magnitude  $\Delta$ suppressing the scattering at low temperature. The plateaus in $R\tsub  H$ are associated with quantization of the Hall conductance, with the precision  of the quantization being remarkably high (of order one part in a billion). This behavior constitutes the \textit{\iqhe} (IQHE).  The IQHE diverges from classical behavior at larger magnetic fields; let's see if we can understand this using the quantum Landau-level theory of Section \ref{landauLevels}.

\subsection{\label{iqheUnderstand}Understanding the integer quantum Hall effect}

\noindent
The \iqhe\ does not require electron--electron or electron--phonon correlations for its explanation.  It related solely to the filling of Landau  levels with non-interacting electrons and our discussion of quantized Landau levels in Section \ref{landauLevels}, supplemented by impurity scattering effects and topological constraints to be discussed below, provides an understanding of its essential features. As implied by \eq{ll1.11}, the number of electrons that can be accommodated by each Landau level  is governed by the magnetic field strength $B$. At certain special values of the field strength $B$ a  match between the number of electrons and the capacity of the Landau levels causes an integer number of Landau levels to become exactly filled,  producing (in outline) the \iqhe.  At these special values of $B$  the ratio of the number of electrons per  unit area to the number of units of flux $h/e^2$, which is the filling factor $\nu$  defined in \eq{ll1.13}, takes integer values.
However, this simple Landau-level picture is not sufficient to understand all features of \fig{IQHEsteps}.  Level filling and the associated ``shell closures'' suggest the broad outlines of the  IQHE, but there are two essential details of  \fig{IQHEsteps} that remain unexplained.

\begin{enumerate}

\item
The longitudinal resistance is very near zero over the entire filling-factor 
range of a plateau in the Hall resistance.

\item
The individual plateaus of \fig{IQHEsteps} have heights quantized with a precision as high as one part in $10^9$ for broad ranges of filling factors.

\end{enumerate}

\noindent
The first suggests that only some states in a  Landau level contribute to  longitudinal  conductance. The second hints at  a fundamental principle responsible for  quantizing the Hall resistance that renders it insensitive to details.    As will now be shown, the first is explained by the effect of impurities on the transport properties of the electron gas, and  the second by a conserved topological quantum number that protects quantization of the Hall resistance.

 \doublefig
          {landauStateDensity} 
          {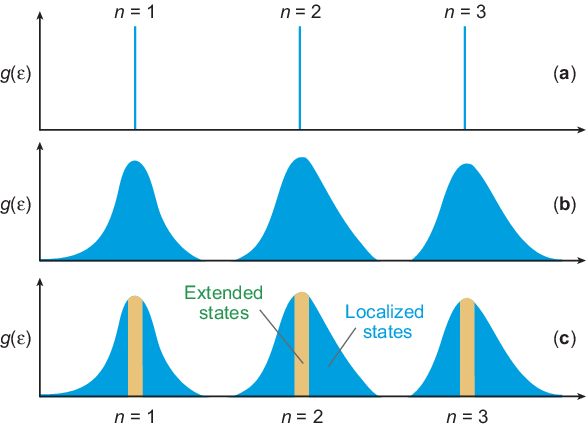}
          {0pt} 
          {0pt}
          {1.0}
          {Level densities $g(\epsilon)$  for a 2D electron gas in a strong  magnetic field. (\textbf{a}) In the clean limit,   Landau level densities correspond to $\delta$-functions, as in \eq{ll1.10}.   (\textbf{b}) Interactions with impurities lead to a broadening of the $\delta$-function distributions.   (\textbf{c}) The broadened peaks consist of   extended states near the centroids and localized states in the wings. Adapted from a figure in Ref.\ \cite{guid2022}.
}
 \singlefig
          {disorderedDensityStates}  
          {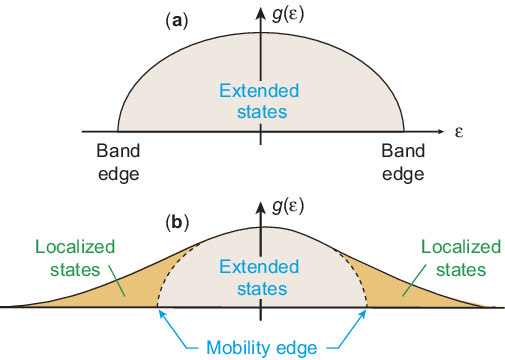}
          {0pt} 
          {0pt}
          {1.0}
          {Density of states (\textbf{a}) for  a band  in the clean  limit and (\textbf{b}) for a band in a disordered system. The mobility edges mark the boundary between localized and extended states.}

\subsubsection{\label{mobilityGaps}Impurity scattering and mobility gaps}

\noindent
The \iqhe\ is observed for impure samples  and the  effect increases up to a point with increased impurity concentration.   Let us try to understand that.  Impurities break translational invariance and increase  resistance by scattering charge carriers, so we must consider the influence of  disorder and impurity scattering on the results of \fig{IQHEsteps}.  First note that  the electronic states in a sample may be separated broadly into two  categories.

\begin{enumerate}

\item 
Some states are {\em extended (metallic) states}, with wavefunctions that fall off slowly with distance; such states facilitate charge transport and  {\em increase the conductance}. 

\item
Some states are {\em localized (insulating) states,} with wavefunctions that are finite only in a small region; such states suppress charge transport and  {\em increase the resistance}.

\end{enumerate}

\noindent
 The effect of  impurities on charge transport is often framed in terms of {\em Anderson localization}, which is described briefly in Appendix \ref{andersonLoc}. The Anderson model of localization is difficult to solve exactly but qualitative  insight from it will suffice for our discussion.
 
 Figure \ref{fg:landauStateDensity}(a) 
illustrates the density of  states implied by \eq{ll1.10} for a clean system, which corresponds to $\delta$-functions for each Landau level $n$, weighted by a degeneracy factor $N$ depending on the magnetic field strength.  For disordered samples the $\delta$-functions are broadened into peaks of finite width by impurity scattering, as illustrated in \fig{landauStateDensity}(b).  Intuitively, impurity scattering tends to localize a state and impede charge transport, with stronger impurity scattering implying greater deviation of the energy from the clean limit. Thus,  as suggested  in \fig{landauStateDensity}(c), only states near  the centers of the broadened peaks in \fig{landauStateDensity}(b) are extended spatially and can carry a current, with the states  in the wings of the peaks localized by impurity scattering so that they are insulating.  

The general idea is illustrated in \fig{landauStateDensity}(c) and \fig{disorderedDensityStates}.
In \fig{disorderedDensityStates}(a) the density of states expected as a  function  of energy for a band in a clean system is illustrated.  Figure \ref{fg:disorderedDensityStates}(b) indicates that the effect of impurity scattering  is to spread those states in energy, with states near the center of the resulting  distribution being spatially extended and those on the wings  being spatially localized, with a {\em mobility edge} characterizing the boundary between the two regions.   This leads to {\em mobility gaps,} corresponding to ranges  of energies having localized states that do not support charge transport.

The influence of the magnetic field is crucial for the present argument. Normally in a 2D system all current-carrying states are destroyed by even a tiny amount of impurity  scattering.  However, {\em this conclusion can be invalidated by the presence of a magnetic field,} which  breaks time-reversal symmetry and interferes with scattering processes that promote localization.  Thus, {\em 2D systems subject to a magnetic field can carry  current, even in the presence of impurities}.  More detailed studies than our  simple discussion here confirm that in 2D systems current-carrying states near the energy of the unperturbed Landau level survive in the presence of a magnetic field, as  illustrated schematically in \fig{landauStateDensity}(c).

Figure \ref{fg:landauStateDensity}(c) explains the Hall plateaus of \fig{IQHEsteps}. Only states near the unperturbed Landau-level energy are delocalized and   carry current, which causes $R\tsub H$ to jump discontinuously  as the chemical  potential is tuned through those states, but $R\tsub H$ remains constant as the chemical potential is  tuned through  localized states in the wings of the peak, as illustrated in \fig{hallPlateausReason}.%
 \singlefig
          {hallPlateausReason}  
          {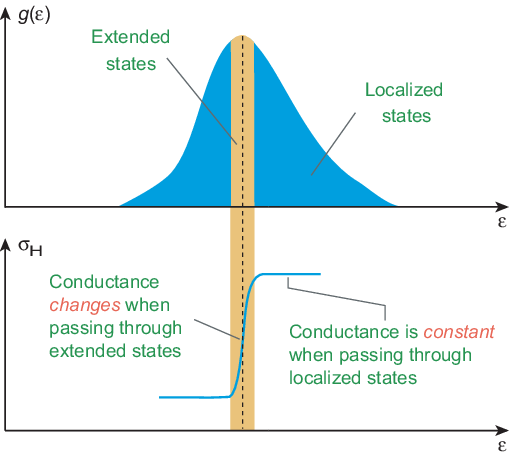}
          {0pt} 
          {0pt}
          {1.0}
          {Origin of Hall plateaus. Extended states (near the centroid of the peak) can carry current; localized states (in the wings of the peak) cannot. Thus conductance $\sigma\tsub H$ changes only when the energy $\epsilon$ is near the centroid of a Landau peak.}
Thus the Hall plateaus of \fig{IQHEsteps} correspond to ranges of magnetic field strength where the  population of extended states is constant because the chemical potential is  not near the center of the broadened Landau level. In effect, when the Fermi  level of a disordered sample  lies in a mobility gap between Landau  levels $n$ and $n+1$, the transport characteristics mimic those  of a pure sample with $\nu = n$.  The plateaus in \fig{IQHEsteps} are  broad so the ranges of extended states must be narrow.   Note that thermal excitation invalidates the preceding argument by scattering electrons between localized and  extended states. Thus, quantum Hall  experiments require very low temperatures, typically less than several  kelvin.

\subsubsection{\label{edgeStatesConduction}Edge states and conduction}

\noindent
For an electron gas confined to two dimensions with a magnetic field 
orthogonal to the 2D plane, the classical Lorentz force causes electrons to move in circular (cyclotron) orbits.  But as suggested by \fig{edgeStates}(a),
 \doublefig
          {edgeStates}   
          {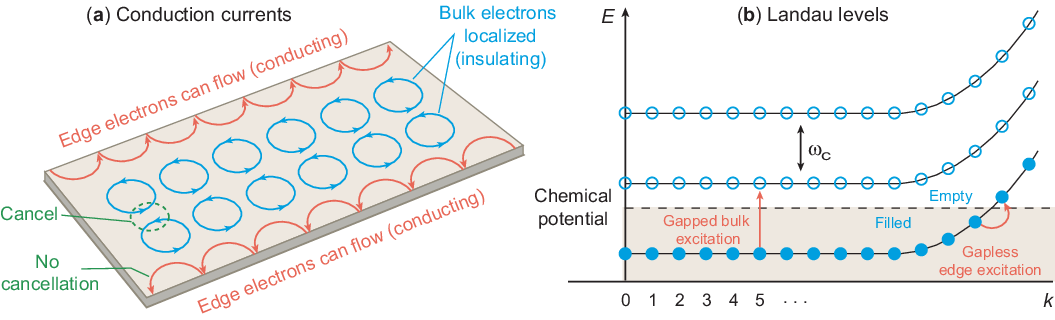}
          {0pt} 
          {0pt}
          {1.0}
          {(\textbf{a}) Conduction by edge states for a 2D electron gas in a  perpendicular magnetic field.  (\textbf{b}) First three Landau levels with a smooth confining potential on the  right  \cite{wen2004}. States in each level are labeled by $k$ and indicated by circles.   Filled states are shaded, empty states are unshaded, and the chemical potential  is indicated by the dashed horizontal line. Bulk excitations are inhibited by a gap $\Delta  E = \hbar \omega\tsub c$ (where $\omega\tsub c$ is the cyclotron frequency), but the gap for edge excitations tends to zero in the  thermodynamic limit, implying gapless edge excitations.
}
no net current flows in the central region because the motion  in one cyclotron orbit is offset by the opposite motion in the adjacent  cyclotron orbit.  Thus, for the 2D electron gas in a magnetic field the {\em  current vanishes in the bulk.}  However, because of the confinement provided by the edge of  the sample, states near the edge  do not suffer this cancellation.  Thus boundary electrons can ``skip'' along the edges and in confined 2D electron gases subject to a transverse magnetic field the {\em current is carried entirely by edge states.} This edge current is {\em  chiral}:  it is right-going along the upper edge and left-going along the lower edge in \fig{edgeStates}(a).  

The preceding argument is  basically correct, even though it is classical and \fig{edgeStates} is just a cartoon.   Solution of the Schr\"odinger equation with  an  edge confining potential  leads to a delocalization of the wavefunction parallel to the edge and  currents that are chiral because they flow in a single direction along each edge.  Energy levels in the first few  Landau levels are illustrated in \fig{edgeStates}(b) for a confining potential along one edge   \cite{wen2004}. States are \textit{gapped by $\Delta E = \hbar\omega\tsub c$ in the bulk}, but are \textit{gapless near the edge},  where  energy between states tends to zero in the thermodynamic limit of many particles.

The {\em chiral nature of the  edge states} protects their  extended character, even if there is impurity scattering (Figs.\  \ref{fg:edgeStates} and \ref{fg:landauStateDensity}). Normally, scattering between time-reversed states of  opposite momentum  in low-dimensional  systems  leads to Anderson localization  and insulating character (Appendix \ref{andersonLoc}).   But if all  states are of a single chirality, scattering of a particle of definite chirality into a time-reversed state would change the chirality, which is forbidden if no states of opposite chirality are accessible.   Thus, if  tunneling across the device is negligible,  impurity scattering is ineffective in localizing  edge states.

\subsubsection {\label{sh:brilltop} Topology of the 2D Brillouin Zone}

\noindent
Let us now understand why quantization of conductance in  integer multiples of $e^2/h$ is so precise across varying experimental conditions, with samples having  different geometries, electron densities, and impurity  concentrations.  
The independent variables are momenta $\bm k$ in the Brillouin zone (BZ), but these are turned into   angular variables by the constraint $k(0) = k(2\pi)$ arising from  the periodicity of Bloch waves on the lattice. Taking the  Brillouin zone for  a 1D crystal to range from $k=-\pi/a$ to $+\pi/a$ for   lattice spacing $a$, the momentum dependence of a band energy can be  displayed as in \fig{brillouinComposite2D}(a).
\doublefig
    {brillouinComposite2D}
    {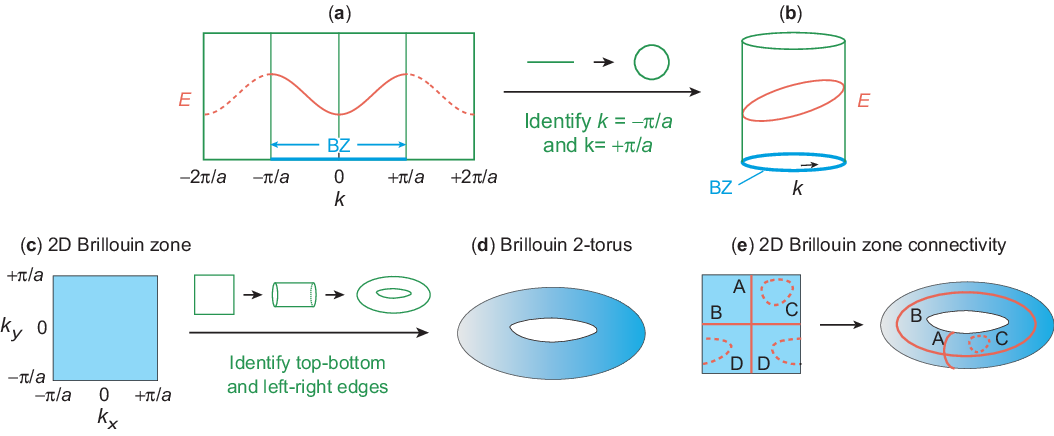}
    {0pt}
    {0pt}
    {1.0}
    {Topology of the Brillouin zone.  (\textbf{a}) Momentum dependence of a single band in a 1D crystal. (\textbf{b}) The periodicity of the crystal requires that $k=-\pi/a$ and $k=+\pi/a$ be identified, which  converts the Brillouin zone to a circle and the band-energy plot to a cylinder. (\textbf{c}) The 2D Brillouin zone.  (\textbf{d}) Because of crystal periodicity in the $x$ and  $y$ directions, the top and bottom edges, as well as the left  and right edges, of the 2D BZ must be identified, giving the geometry of a (closed) 2-torus for the Brillouin zone. (\textbf{e}) Different  closed paths in the toroidal 2D Brillouin zone, implying non-trivial topology. Adapted from Refs.\ \cite{guid2022,van2018}.}
But  this doesn't indicate clearly the {\em equivalence} of the right and left sides of the BZ at $k=\pm \pi /a$. 
We can improve things by identifying $k=+\pi /a$ and $k=- \pi /a$, which maps the 1D BZ from an interval on the real number line to a circle and the band-energy plot to a cylindrical, as  in  \fig{brillouinComposite2D}(b). 

The Brillouin zone for a 2D crystal is displayed in  \fig{brillouinComposite2D}(c). Periodicity in $k_y$ identifies top and bottom edges, converting the square to a cylinder, and periodicity in $k_x$  identifies left and right edges, joining the ends of the cylinder to form the 2-torus of \fig{brillouinComposite2D}(d). The toroidal geometry  of the 2D Brillouin zone implies different  classes of closed paths on the manifold, as illustrated in \fig{brillouinComposite2D}(e). Loops C and D can be shrunk continuously to a point (D is closed  because of  the periodic boundary condition); loops A and B cannot. Thus  the loops A, B, and C lie in distinct topological  sectors that cannot be deformed continuously into one another.  The non-trivial connectivity of the 2D Brillouin zone in \fig{brillouinComposite2D} implies  states characterized by topological quantum numbers that cannot be deformed smoothly into each other.  Thus the Hilbert space  for the integer quantum Hall effect is defined on a momentum-space torus and this leads to inequivalent  topological sectors labeled 
by Chern numbers, as we now discuss.

\subsubsection{\label{topoProtectPlateau} Topological Protection of Hall Plateaus}

\noindent
Topological protection  accounts for the remarkable flatness of the plateaus in \fig{IQHEsteps}, because it can be shown that, as a consequence of the non-trivial topology of the Brillouin zone and the \textit{Chern theorem} (Appendix \ref{gaussBonnet}), the Hall conductance $\sigma\tsub{H} = 1/R\tsub{H}$ contributed by a single non-degenerate band is given by \cite{tong2016}
\begin{equation}
    \sigma\tsub H = \frac{e^2}{2\pi h} \int_S \Omega dS = C_1 \frac{e^2}{h},
    \mwgtag{holyGrailEquation}
\end{equation}
where $e$ is electrical charge, $h$ is Planck's constant,  $S$ is a closed surface in the manifold of quantum states, $\Omega$ is local Berry curvature in that manifold, and $C_1$ is a topological Chern number that takes integer values.  Thus the exquisite flatness of the plateaus for $R\tsub H$ in \fig{IQHEsteps}  follows from topological protection within a particular quantum topological phase ensured by \eq{holyGrailEquation}, and the vertical jumps between plateaus correspond to quantum phase transitions between topological states characterized by different Chern numbers (see Appendix \ref{gaussBonnet}).

\subsubsection{\label{bulk-boundary}Bulk--Boundary Correspondence}

\noindent
In the quantum Hall effect, and more generally in topological matter (material characterized by topological quantum numbers), it is important to distinguish the \textit{bulk} (interior) of a sample and the \textit{boundary} (surface in 3D, edges in 2D, and ends in 1D). In the quantum Hall effect the strong magnetic field localizes  electrons in the bulk, while forcing boundary electrons into delocalized chiral edge states [\fig{edgeStates}(a)]. Thus the bulk states are insulating but the boundary states are conducting (metallic). This relationship between the bulk states and the boundary states for the IQHE is an example of \textit{bulk--boundary correspondence,} which is  a characteristic feature of topological matter  and implies that topological quantum numbers characterizing the bulk also predict boundary properties.

\subsubsection{\label{essenceIQHE} Incompressible States}

\noindent
When $N$ electrons exactly fill an integer number $\nu$ of Landau levels there  is an energy gap $\hbar \omega\tsub c$ between filled and empty states.  Now a decrease in  the area $A$ of the system at constant magnetic field and electron density will decrease the number of flux quanta $N_\phi$ piercing the sample. By \eq{ll1.12}, this reduces the capacity of the Landau levels and requires that electrons be promoted to higher Landau levels.  There  is an energy cost $\hbar \omega\tsub c$ for each promoted electron,  so the system resists compression and is said to be {\em  incompressible}.  Thus,  a quantum Hall system is an  {\em incompressible quantum fluid.}

\subsection{\label{summaryIQHE}Summary: Integer Quantum Hall Effect}

\noindent
In summary, for the \iqhe\ the crucial points are  (1)~Landau-level quantization in the magnetic field,  (2)~a random  scattering potential caused by impurities, and that  (3)~each Landau level contributes a fixed value to the Hall  conductance as a consequence of mobility gaps created by the  impurity scattering. Thus conductance is quantized because effectively it counts the  number of filled Landau levels. The robustness of this quantization derives  from the non-trivial Brillouin zone topology and the Chern theorem, as described in Sections \ref{sh:brilltop} and \ref{topoProtectPlateau}, and  Appendix \ref{gaussBonnet}.

\section{\label{fractionalQuantumHallEffect} The Fractional Quantum Hall Effect}

 \noindent
 Tsui, Stormer,  and Gossard  \cite{tsui1982} discovered the \fqhe\ (FQHE) in 1982 when they found that at higher magnetic  fields and lower temperatures the quantum Hall  effect can occur for the fractional value of $\tfrac13 (e^2/h)$, corresponding  to a  filling fraction $\nu = 1/3$, unlike the \iqhe\ (IQHE), which occurs only at integer multiples of $e^2/h$. Laughlin  \cite{laug1983} proposed that the FQHE corresponds to a new state of matter formed by strong electron--electron correlations  within a single Landau level. It was shown in later work  that the FQHE could occur for whole series of fractional values $\tfrac1m  (e^2/h)$, with $m$ an even or odd integer.

\subsection{\label{propertiesFQHE} Properties of the fractional quantum Hall state}


\noindent
Let us consider a theoretical understanding of this \fqhe. First, note that fractional filling factors in themselves are not unusual.   The conductance is expected to  be $\nu  e^2/h$ if we add non-interacting electrons to the lowest Landau level up to  a fractional occupation $\nu$.  But in the absence of interactions there will be no gap at the Fermi surface and adding electrons will cost almost no energy, destroying the  plateau structure of the Hall resistance in \fig{IQHEsteps}.  If we assume interacting electrons instead, they will scatter into empty Landau  states, leading to a finite longitudinal resistance, which contradicts  experiments.    
Thus, explanation of the fractional quantum Hall effect requires a ground state that 

\begin{enumerate}
\item
has a \textit{fractionally-filled Landau level} and
\item
has an {\em energy gap}  with respect to  excitation at that fractional filling.
\end{enumerate}
As noted above, a state with an energy gap resists compression, so we may also term such a state {\em incompressible.}   Ground states with a gap and small resistance were known when the  \fqhe\ was discovered, but none had the right properties to explain the FQHE.    A fundamentally new kind of state is required for which strong electronic correlations  in a partially-filled Landau level produce a ground state corresponding to  an incompressible  electronic liquid, with a fractional filling factor and an energy gap for all excitations. 

Both the IQHE and the FQHE  result from quantizing electronic motion for 2D electron gases in strong  magnetic fields, but their mechanisms are fundamentally different. The IQHE involves weakly-interacting electrons and depends on impurity  scattering for its plateau structure indicating quantization of conductance, but the FQHE  requires  very pure samples because it is a consequence of strong Coulomb interactions  between the degenerate electrons in Landau levels.   Strongly-correlated electrons in a magnetic field represent an inherently difficult problem; it was solved only by an educated guess for the form of the wavefunction. The difficulty of solving this  problem derives from some features of  the FQHE that distinguish it from most other problems in many-body  physics \cite{jain2007}.

\smallskip
\noindent
(1)~It is generally  assumed  that the FQHE corresponds to partial filling of the lowest Landau level (LLL).  Absent interactions  there are many ways to fill the LLL partially  that give the  same energy.  What principle should guide us to  choose the linear combination of these states that will  define the physical ground state when the interactions are turned on? 

\medskip
\noindent
(2)~In the very high magnetic fields that characterize FQHE experiments the problem reduces approximately to Coulomb interactions between  electrons in the lowest Landau level.  The strength of the Coulomb interaction just sets an  energy scale, so the fractional quantum Hall problem contains no obvious parameters to adjust in order to  gain intuition about the physics responsible for the solution.

\smallskip
\noindent
(3)~In many-body physics it is common to view an emergent state as resulting from an instability of a parent ``normal state'' that would be favored in the absence of 
interactions.   But turning off the interactions for the FQHE state does 
not lead to a normal state that is the obvious parent of the FQHE state. 

\medskip
\noindent
For these reasons the solution of the FQHE problem cannot be obtained by successive small steps or other standard approximations and we are reduced to \textit{making an educated guess} for a wavefunction that leads to the observed FQHE. The best such educated guess to date is the \textit{Laughlin wavefunction}.

\subsection{\label{laughlinWF} The Laughlin wavefunction}

\noindent
Let us consider a  wavefunction to describe the fractional quantum Hall state, guided by the discussion in Phillips \cite{phil2003}.  We  assume the magnetic field to be large and as a first approximation ignore the  Coulomb interactions between electrons, and assume the electron spins to be  completely polarized by the field and drop the corresponding (constant) spin term. This gives again  the Hamiltonian \eqnoeq{ll1.1}, but now let's  work in the symmetric gauge defined in \eq{ll1.2b}. It is convenient to introduce the complex coordinate $z_i = x_i  + iy_i = re^{i\phi}$ for the $i$th particle, in which case the solution of the  Shr\"odinger equation in symmetric gauge  gives the  $N$-body wavefunctions
%
%
\begin{equation}
    \psi_N = \prod^N_{j<k} f(z_j- z_k)
    e^{- \sum_{j=1}^N 
    |z_j|^2 / 4\Lambda^2},
    \mwgtag{lwf1.1}
\end{equation}
 where $f(z_j, z_k)$ is a polynomial in the electron coordinates $z_i$ and $\Lambda$  is the magnetic length defined in \eq{ll1.7}. Analysis of this  wavefunction for two and three electrons led Laughlin to propose  what is now  termed the {\em Laughlin wavefunction,}
 %
%
\begin{equation}
     \psi_{1/m}  = \prod^N_{j<k} (z_j-z_k)^m \, e^{-\sum_{j=1}^N 
    |z_j|^2/ 4\Lambda^2},
     \mwgtag{laughlinWF}
\end{equation}
which specifies the FQHE states at filling factors $\nu = 1/m$, with the integer $m$ required to be odd to ensure antisymmetry of the fermionic wavefunction. Thus, the  first  FQHE state discovered with filling factor $\nu = \tfrac13$ corresponds to the wavefunction \eqnoeq{laughlinWF}  with $m=3$. 

The exponent $m$ was introduced originally as a variational parameter  but  $\psi_{1/m}$ is an eigenstate of the angular momentum  operator with angular momentum $M$  given by 
\begin{equation}
M = \frac12 N(N-1) m 
\label{laughlin_m}
\end{equation} 
for $N$ particles.
Thus $m$ is  related directly to the angular momentum content of  the Laughlin wavefunction \eqnoeq{laughlinWF}, which may be  viewed as a superposition of states from the lowest Landau level having the same angular momentum.

The Laughlin  wavefunction given by \eq{laughlinWF} has several features that make it a plausible candidate for a  FQHE  wavefunction.

\begin{enumerate}

\item
It minimizes the kinetic energy by placing all electrons in the lowest Landau 
level.

\item 
It vanishes if the coordinates of any two electrons are the same  because of the $(z_k - z_\ell)^m$ factor, which is desirable for a  wavefunction in a strongly-correlated system with repulsive interactions  because it reduces the total potential energy.

\item
It is antisymmetric under the exchange of any pair of particles if 
$m$ is odd.

\item
It can be shown to describe a circular droplet with $\nu = \tfrac13$ and an 
approximately uniform density.

\item
Simple models and numerical simulations suggest that it describes an incompressible state that has an energy gap to all  excitations.

\end{enumerate}

\noindent
The wavefunction \eqnoeq{laughlinWF} is now accepted broadly as a  valid description of the first $\nu=\tfrac13$ FQHE state, and by inference of other  fractional quantum Hall states discovered later. 

The Laughlin wavefunction provides only limited fundamental insight into the underlying  microscopic physics responsible for the FQHE state since it is in essence a  highly-educated guess at the many-body wavefunction (similar in spirit to the  ``guess'' by Bardeen, Cooper, and Schrieffer for the form of the BCS  wavefunction that led to comprehensive understanding of the superconducting problem).  However, the influence of the Laughlin wavefunction on understanding the  \fqhe\ has been seminal.

\section{\label{grapheneMag}Graphene in Strong Magnetic Fields}

\noindent
As we have seen, rather remarkable quantum Hall physics  is  observed when a cold 2D electron gas is subject to a strong  magnetic field.  Since a single layer of graphene is the ideal 2D material, it is of interest whether similar quantum Hall effects could be  observed  in graphene. To begin addressing that question we consider in this section the  quantum states of electrons confined in 2D and governed by a massless Dirac  single-particle Hamiltonian, with a strong magnetic field placed perpendicular  to  the 2D plane. Our discussion will have some overlap with the earlier treatment of a similar  system with electrons governed by a Schr\"odinger equation, but there will be  substantial differences associated with the Dirac rather than Schr\"odinger  physics appropriate for graphene. We shall use the review by Goerbig  \cite{goer2011} for guidance, and often use $c=1$ and $\hbar=1$ units.

\subsection{\label{quantizationDiracLL}Quantization of Dirac Landau levels}

%
%
\noindent
By the usual minimal substitution, electromagnetism may be included in a Hamiltonian by replacing the canonical momentum $\bm p$ with the  gauge-invariant kinetic momentum $\bm\Pi$,
\begin{equation}
    \bm\Pi = \bm p + e \bm A(\bm r),
    \mwgtag{diracLL1.1}
\end{equation}
where the magnetic field $\bm B$ is related to the vector potential $\bm A$ by
\begin{equation}
    \bm B = \bm\nabla \times \bm A.
    \mwgtag{diracLL1.2}
\end{equation}
From \eq{tbind1.34}, to 
lowest order in $|\bm q| a$ the Hamiltonian with the minimal substitution is
%
%
\begin{equation}
    H = \xi v\tsub F (\Pi_x \sigma^x + \Pi_y \sigma^y).
    \mwgtag{diracLL1.3}
\end{equation}
The electron energies will also split into two  branches according to the Zeeman  energy in the magnetic field $\Delta_z = g \mu\tsub B B$, where $g \sim 2$ for graphene.  In this initial discussion we ignore $\Delta_z$, but  will return to the role of spin later.

The canonical method may be used to quantize this system.  In the absence of 
the vector potential the coordinates and momenta would obey the commutation 
relations
\begin{equation}
\begin{gathered}
    \comm{x}{p_x} = \comm{y}{p_y}= i\hbar
    \\
    \comm{x}{y} = \comm{p_x}{p_y} = \comm{x}{p_y} = \comm{y}{p_x} = 0.
    \mwgtag{diracLL1.5}
\end{gathered}
\end{equation}
From Eqs.\ \eqnoeq{diracLL1.1} and \eqnoeq{diracLL1.5}, the commutator of $\Pi_x$ and $\Pi_y$ is \cite{goer2011}
\begin{align}
    \comm{\Pi_x}{\Pi_y} 
    =
    -ie\hbar \left( \pardiv{A_y}{x} - \pardiv{A_x}{y} \right).
\mwgtag{diracLL1.6}
\end{align}
But the magnetic field is assumed aligned along the $z$-axis, $\bm B (B_x, B_y, B_z) =  (0,0,B)$ and from \eq{diracLL1.2}, 
$$
    \pardiv{A_y}{x} - \pardiv{A_x}{y} = B_z \equiv B,
$$
and therefore \eq{diracLL1.6} becomes
\begin{equation}
    \comm{\Pi_x}{\Pi_y}
    = -ie\hbar B = -i\hbar^2 \frac{eB}{\hbar}
    = -i \,\frac{\hbar^2}{\Lambda^2} ,
    \mwgtag{diracLL1.8}     
\end{equation}
with the magnetic length $\Lambda$ introduced in \eq{ll1.7}  given  by $\Lambda = \sqrt{\hbar/eB}$ in $c=1$ units.
It will be convenient to introduce the raising and lowering operators
\begin{equation}
    a^\dagger = \frac{\Lambda}{\sqrt 2 \, \hbar}
    \left(\Pi_x + i\, \Pi_y \right)
    \qquad
    a = \frac{\Lambda}{\sqrt 2 \, \hbar}
    \left(\Pi_x - i\, \Pi_y \right),
    \qquad
    \mwgtag{diracLL1.10}
\end{equation}
where the normalization factor is chosen so that
\begin{equation}
    \comm{a}{a^\dagger} = 1.
    \mwgtag{diracLL1.11}
\end{equation}
Adding and subtracting Eqs.\ \eqnoeq{diracLL1.10} gives the  inverse relations
\begin{equation}
    \Pi_x = \frac{\hbar}{\sqrt 2 \Lambda}(a^\dagger + a)
    \qquad
    \Pi_y = \frac{\hbar}{\sqrt 2 \Lambda i}(a^\dagger - a) ,
    \mwgtag{diracLL1.12}
\end{equation}
and inserting this result into the Hamiltonian \eqnoeq{diracLL1.3} gives
\begin{align}
    H &=
    \xi v\tsub F \left(
    \Pi_x \,\sigma^x + \Pi_y  \,\sigma^y
    \right)
    \nonumber
    \\
    &= \xi v\tsub F \left[
    \Pi_x \paulix + \Pi_y \pauliy
    \right]
    \nonumber
    \\
    &=
    \xi v\tsub F 
    \twomatrix
    {0}
    {\Pi_x - i\Pi_y}
    {\Pi_x + i\Pi_y}
    {0}
    \nonumber
    \\
    &=
    \frac{\sqrt 2 \,\hbar}{\Lambda} \, \xi v\tsub F
    \twomatrix{0}{a}{a^\dagger}{0}
    \nonumber
    \\
    &=
    \hbar \,\xi \omega \twomatrix{0}{a}{a^\dagger}{0} ,
    \mwgtag{diracLL1.13}
\end{align}
where Eqs.\ \eqnoeq{diracLL1.10} were used and  $\omega \equiv \sqrt2 v\tsub F/\Lambda$ is the analog of the cyclotron frequency \eqnoeq{classicalHall1.3} for the  present relativistic problem.  Introducing the 2-component spinor
\begin{equation}
     \psi_n =
     \onematrix{u_n}{v_n}, 
     \mwgtag{diracLL1.14b}
\end{equation}
we must then solve the  eigenvalue equation $H \psi_n = \epsilon_n \psi_n$, which takes the form
\begin{equation}
    \hbar \xi \omega \twomatrix{0}{a}{a^\dagger}{0}
     \begin{pmatrix}u_n \\ v_n \end{pmatrix}
     = \epsilon_n  
     \begin{pmatrix}u_n \\ v_n \end{pmatrix} ,
    \mwgtag{diracLL1.15}
\end{equation}
implying the  simultaneous equations
\begin{subequations}
\begin{align}
    \hbar \xi \omega a v_n &= \epsilon_n u_n,
    \mwgtag{diracLL1.16a}
    \\
    \hbar \xi \omega a^\dagger u_n &= \epsilon_n v_n.
    \mwgtag{diracLL1.16b}
\end{align}
\end{subequations}
Solving \eq{diracLL1.16a} for $u_n$ and substituting into \eq{diracLL1.16b}  gives $a^\dagger a v_n = (\epsilon_n/\hbar\omega)^2 v_n$. But because of \eq{diracLL1.11} this is just the usual harmonic oscillator  number equation $a^\dagger a \ket\psi = n \ket\psi$ with $n =  (\epsilon_n/\hbar\omega)^2$, and up to numerical factors the state $\ket{v_n}$  may be identified with a number eigenstate $\ket n$. Solving for $\epsilon_n$  in terms of $n$ gives  $\epsilon_n = \pm \hbar \omega \sqrt n$, which may be written
\begin{align}
    \epsilon_n =  \lambda \hbar \omega \sqrt n
    = \lambda \frac{\hbar v\tsub F}{\Lambda} \sqrt{2n}
    \propto \lambda \sqrt{nB}
    ,
    \mwgtag{diracLL1.18}
\end{align}
where the quantum number $\lambda = \pm 1$ has been introduced to distinguish  the positive and negative square roots; it plays the same role as the band  index in the zero-field case [see \eq{tbind1.9}], corresponding to the  graphene two-fold valley degeneracy for $B=0$.  Thus, the energy  disperses as $\epsilon_n\sim \pm (nB)^{1/2}$ with magnetic field strength, as  illustrated in \fig{landauDiracDispersion+cyclotronCoord}(a).%
%
%
 \doublefig
          {landauDiracDispersion+cyclotronCoord} 
          {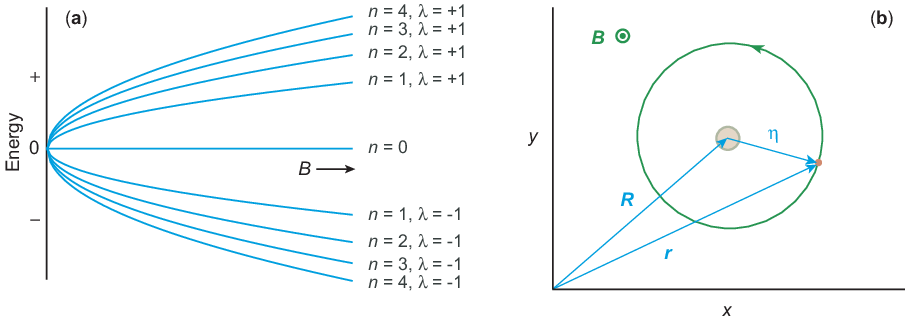}
          {0pt} 
          {0pt}
          {1.0}
          {(\textbf{a}) Relativistic Landau energy levels as a function of magnetic field  strength $ B$. (\textbf{b}) Guiding center coordinates $\bm R$ and cyclotron coordinates $\bm  r$. Classically the guiding center coordinates define the center of the  cyclotron orbit and the cyclotron coordinates describe the motion of the  electron relative to the guiding center. The shaded area in the center of the  circle represents the uncertainty in the position of the guiding center implied  by \eq{LLdegen1.8}. Figure adapted from Ref.\ \cite{goer2011}.}

From \eq{diracLL1.18} and \fig{landauDiracDispersion+cyclotronCoord}(a) the $n=0$ state is not split into positive  and negative branches, as is the case for all other values of $n$.  From \eq{diracLL1.16a}  we have $u_n \propto av_n \sim a\ket n \sim \ket{n-1}$, because of the usual oscillator operator requirements that
\begin{gather*}
a^\dagger \ket n = \sqrt{n+1} \ket{n+1}
\qquad a\ket{ n>0} = \sqrt n \ket{n-1} 
\\
 a\ket{n=0} = 0 .
\end{gather*}
Thus, for $n=0$ the eigenspinor \eqnoeq{diracLL1.14b} has only a single 
component,
\begin{equation}
    \psi_n = \onematrix{u_n}{v_n} =
     \begin{pmatrix}0 \\ \ket{n=0} \end{pmatrix},
    \mwgtag{diracLL1.120}
\end{equation}
with the non-zero component corresponding to the B sublattice in the $K$ valley 
($\xi = +1$) and the A sublattice in the $K^\prime$ valley ($\xi = -1$). This 
implies that the valley pseudospin and the sublattice pseudospin coincide for 
$n=0$.  For all other values of $n$, there are two spinor components and 
positive and negative energy solutions.

\subsection{\label{LLdegeneracy}Degeneracy of Landau levels}

\noindent
Just as for the non-relativistic case, Landau levels for relativistic  fermions in a strong magnetic field become \textit{highly degenerate}.  In  both  the non-relativistic and relativistic cases the degeneracy $N_B$ is equal to  the  number of flux quanta threading the 2D electron gas,
\begin{equation}
    N_B = \frac{B}{(h/e)}\, S ,
    \mwgtag{LLdegen1.1}
\end{equation}
where $S$ is the area of the 2D sample, $BS$ is the total flux, and $h/e$ is  the  flux quantum (in $c=1$ units); see \eq{ll1.11} and the discussion following it.  Let's investigate this degeneracy in more detail for the relativistic case appropriate for graphene. 

%
%

It is convenient to separate the relativistic cyclotron motion of the  electrons into {\em guiding center coordinates} $\bm R = (X, Y)$ that  commute with the Hamiltonian, and relative coordinates termed {\em cyclotron  variables} $\bm \eta = (\eta_1, \eta_2)$, according to
\begin{equation}
    \bm r = \bm R + \bm \eta,
    \mwgtag{cyclotronCoordinates1.1}
\end{equation}
as illustrated in \fig{landauDiracDispersion+cyclotronCoord}(b). Classically, the guiding center coordinates represent the center of the  cyclotron motion for an electron and the cyclotron coordinates describe the  time-dependent (dynamical) component of motion.  From  \fig{landauDiracDispersion+cyclotronCoord}(b)  the cyclotron variable $\bm\eta$ is orthogonal to the electron velocity and  hence is related to the kinetic momentum $\bm \Pi$ defined in \eq{diracLL1.1} by
\begin{equation}
    \eta_x = \frac{\Pi_y}{eB}
    \qquad
    \eta_y = -\frac{\Pi_x}{eB}.
    \mwgtag{LLdegen1.2}
\end{equation}
From this relation and \eq{diracLL1.8}, the commutator of the cyclotron  variables is
\begin{align}
    \comm{\eta_x}{\eta_y} 
    =
    \frac{\comm{\Pi_x}{\Pi_y}}{(eB)^2}
    =
    -i\Lambda^2.
    \mwgtag{LLdegen1.3}
\end{align}
This can be used to find the commutation relation for the guiding  center coordinates $(X,Y)$.  From \eq{cyclotronCoordinates1.1} with $r = (x,y)$ we have
$
x = X + \eta_x
$
and
$
y = Y + \eta_y ,
$
and since $\comm xy = 0$,
\begin{align*}
    \comm xy 
    &=
    \comm{X+\eta_x}{Y+\eta_y}
    \\
    &=
    \comm{X}{Y} + \comm{X}{\eta_y} + \comm{\eta_x}{Y} + \comm{\eta_x}{\eta_y}
    = 0.
\end{align*}
Assuming the guiding center and cyclotron coordinates to commute, this yields
\begin{equation}
    \comm XY = -\comm{\eta_x}{\eta_y} = i\Lambda^2,
    \mwgtag{LLdegen1.4}
\end{equation}
where \eq{LLdegen1.3} was used.  Thus $X$ and $Y$ are conjugate  variables and  \eq{LLdegen1.4} is an uncertainty relation associated  with simultaneous specification of $X$ and $Y$.  Introduce 
\begin{equation}
    b = \frac{1}{\sqrt 2 \Lambda} ( X + iY)
    \qquad
    b^\dagger = \frac{1}{\sqrt 2 \Lambda} ( X - iY),
    \mwgtag{LLdegen1.5}
\end{equation}
which satisfy the commutation relation $\comm{b}{b^\dagger} = 1$ and commute with the Hamiltonian (since $X$ and $Y$ do). Because of this  commutation relation this is a second set of harmonic-oscillator operators,  so we expect $b^\dagger b$ to be a number operator with $b^\dagger b \ket {m_k} = m_k  \ket {m_k}$, where the integer $m_k \ge 0$ (separate from $n$) counts the states created  by operating repeatedly on the vacuum with $b^\dagger$. Since $b$ and  $b^\dagger$ commute with the Hamiltonian,  $m_k$ is an additional  quantum number that labels electronic states in the magnetic field and we  conclude that the states on a given branch of \fig{landauDiracDispersion+cyclotronCoord}(a) may  be  labeled by  the quantum numbers $n$ and $m_k$.

 As noted above,  \eq{LLdegen1.4} is an uncertainty relation indicating that we cannot  measure the $X$ and $Y$ coordinates of the guiding center for an electron  simultaneously with arbitrary precision. The  location of the guiding center is found to be uncertain according to \cite{goer2011}
\begin{equation}
    \Delta X \Delta Y = 2\pi \Lambda^2,
    \mwgtag{LLdegen1.8}
\end{equation}
which is represented by the shaded area in the center of \fig{landauDiracDispersion+cyclotronCoord}(b). Let the total area of the 2D gas be $S$ and introduce the flux density $n_B$  measured  in units of the flux quantum $h/e$  through
\begin{equation}
    n_B = \frac{N_B}{S} = \frac{B}{(h/e)},
    \mwgtag{LLdegen1.9}
\end{equation}
where \eq{LLdegen1.1} was used. The number of quantum electronic states found  in the  area $S$ is 
\begin{equation}
    N_B = \frac{S}{\Delta X \Delta Y} = \frac{S}{2\pi \Lambda^2} = n_B S,
    \mwgtag{LLdegen1.10}
\end{equation}
and so
\begin{equation}
    n_B = \frac{B}{(h/e)} = \frac{1}{2\pi \Lambda^2} .
    \mwgtag{LLdegen1.11}
\end{equation}
Thus, the flux density and the density of possible electronic states in a 
magnetic field coincide  for the 2D gas.  The filling factor $\nu$ can then 
be defined by the ratio of the electron number density $n_e$ and the flux 
density $n_B$,
\begin{equation}
    \nu \equiv \frac{n_e}{n_B} = \frac{hn_e}{eB}.
    \mwgtag{LLdegen1.12}
\end{equation}
This  treatment assumes the only degeneracies to be those of the
Landau levels for massless Dirac electrons.  In the following sections we shall 
see how this is modified by additional degeneracies associated with sublattice and 
valley degrees of freedom in graphene.

\section{\label{FQHEgraphene}Quantum Hall Effects in Graphene}

\noindent
The quantum Hall effect occurs for 2D electron gases in strong magnetic fields.  Because of degrees of freedom associated with the honeycomb lattice structure, we may expect that quantum Hall effects in graphene could  be  richer than those in the usual 2D electron gas. From the preceding discussion we may expect that analogs of the integer quantum Hall states might occur for weakly-interacting electrons in  graphene, but the fractional quantum Hall state can be produced only by  strongly-correlated electrons. Most of the low-energy properties of graphene are described well assuming weakly interacting massless Dirac fermions in two dimensions, but we may ask  whether there are particular situations in which monolayer graphene might exhibit strong  electron--electron correlations. The relative  effectiveness of correlations are expected to depend on (1)~the intrinsic strength of the  electronic interactions, and (2) the density of states at the Fermi  surface.

\subsection{\label{strengthOfCorrelations}Strength of  correlations in  graphene}

%
%
\noindent
For the Coulomb gas of electrons characterizing a layer of graphene, the average interaction energy at a characteristic length scale $k_{\rm F}^{-1}$ is $E\tsub{int} \sim e^2 k\tsub F/\epsilon$, where $e$ is the electronic charge, $\epsilon$ is the effective dielectric constant of the environment, and $k\tsub{F}$ is the Fermi momentum. The average kinetic energy at the same length scale is $E\tsub{kin} \sim \hbar \epsilon v\tsub F$, where $v\tsub F$ is the Fermi velocity.
The  ratio of interaction energy and kinetic energy at this characteristic length scale defines a coupling constant for graphene\cite{goer2011,goer2012},
\begin{equation}
     \alpha\tsub G \equiv \frac{E\tsub{int}}{E\tsub {kin}}
     = \frac{e^2 k\tsub F/\epsilon}{\hbar v\tsub F k\tsub F} 
     = \frac{e^2}{\hbar \epsilon v\tsub F}.
     \mwgtag{effectiveCouplingConstantEq}
\end{equation}
This has the same form as the fine-structure constant  $\alpha$  of quantum electrodynamics (QED) if  the  speed of light $c$ is exchanged for the Fermi velocity $v\tsub  F$.  The ``fine structure constant'' $\alpha\tsub G$  of graphene is  $c/v\tsub F \sim 300$ times that for QED ($\alpha \sim 1/137$)  at  comparable energies, implying a relatively large Coulomb interaction. However, the low-energy excitations in graphene obey an approximately linear  dispersion relationship and in the undoped compound the density of states  vanishes as the Fermi surface [at the apex of the Dirac cones in \fig{dispersionGraphene}(b)] is approached.  This low density of states near the Fermi surface suppresses  electron--electron correlations and much of low-energy  physics for monolayer graphene is dominated by  weakly-correlated particles.  If we wish to search for strongly-correlated  physics in graphene, it is necessary to find circumstances that correspond to increased  level density near the Fermi surface  \cite{goer2012}. 

One way to attain higher local level density is to place a strong magnetic field $B$ perpendicular to the sample.  Then the electronic levels become quantized into 
Landau levels that are highly degenerate, with a strongly-peaked density of 
states
\begin{equation}
     \rho(E) = gn_B \sum_n f(E-E_n),
     \mwgtag{densityStatesLL}
\end{equation}
where $g$ is a degeneracy factor for internal degrees of freedom (in graphene,  $g=4$ because of the four-fold spin--valley degeneracy), $n_B = eB/h$ is the  flux density in units of the flux quantum $h/e$, and $ f(E-E_n)$  is  strongly peaked, tending to a $\delta$-function  in the clean limit. In this limit, each LL may then be approximated as an  infinitely flat (dispersionless, because kinetic energy is negligible) energy  band in which the state density grows linearly with  the magnetic field strength \cite{goer2012}.  The effect of Fermi  surface position relative to the Landau levels may be studied by fixing the  magnetic field and sweeping a gate voltage applied to the sample, or by fixing the  gate voltage and sweeping the magnetic field.

\subsection{\label{IQHEgraphene}The graphene integer quantum Hall effect}

\noindent
Electron spin has largely been ignored in the discussion of the quantum Hall  effect to this point, but if it is not neglected IQHE conditions are satisfied when both spin  branches (split by the Zeeman term) of the last Landau level (LLL) are  completely filled.  Thus $\nu\tsub{IQHE} = 2n$, where $n$ is an integer  and filling factors are even integers in the  usual IQHE.
The graphene IQHE is expected to resemble the usual IQHE, except for two key points.

\begin{enumerate}

\item 
In addition to the 2-fold spin degeneracy (if Zeeman splitting is neglected), 
there is a 2-fold $K$ and $K^\prime$ valley degeneracy   for graphene.  Thus the filling 
factor changes in steps of four between plateaus in the Hall resistance for 
graphene.

\item
For graphene the filling factor
\begin{equation}
    \nu = \frac{n_e}{n_B} = \frac{hn_e}{eB}
    \mwgtag{iqhegraph1.1}
\end{equation}
vanishes at the Dirac point for particle--hole symmetric half filling of the  graphene lattice, because the electron density $n_e$ tends to zero at the Dirac point. Therefore, absent a Zeeman effect or electron correlations, one expects no \iqhe\  effect in graphene for $\nu = 0$.

\end{enumerate}

\noindent
Thus the signature of a graphene IQHE is Hall resistance 
quantization for filling factors 
%
%
\begin{equation}
    \nu  = \pm 4\left(n+\frac12\right) = \pm (4n + 2)
    = \pm2, \pm6,\pm10, \ldots
    \mwgtag{iqhegraph1.2}
\end{equation}
where $n$ is an integer, the factor of 4 is because of the 4-fold valley and  spin degeneracy, and  the added $\frac12$ (not present in non-relativistic 2D systems) is because of  the special status of the $n=0$ state for massless Dirac fermions in graphene.  

Evidence for an \iqhe\ in graphene was reported in Refs.\ \cite{novo2005,zhang2005}, and
is illustrated in \fig{iqhe+fqheComposite}(a).%
\doublefig
    {iqhe+fqheComposite} 
    {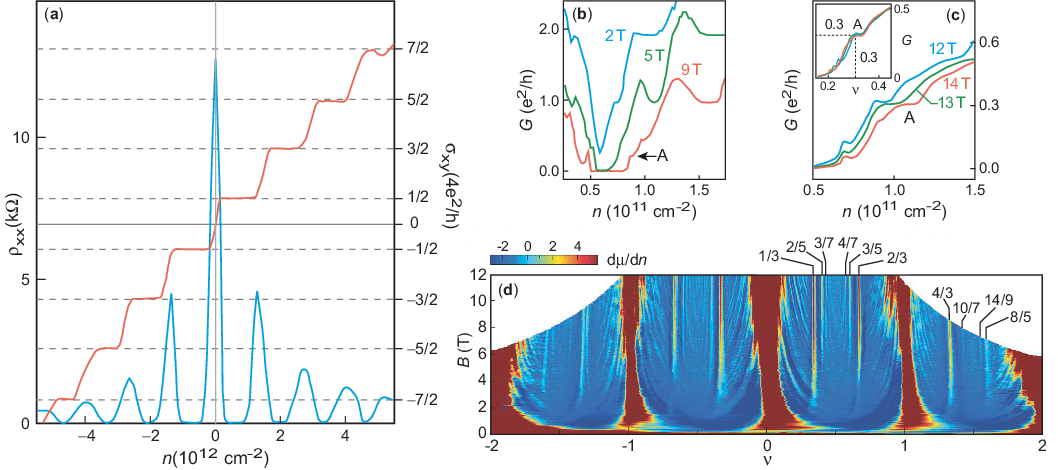}
    {0pt} 
    {0pt}
    {1.0}
    {Evidence for integer and fractional quantum Hall effects in monolayer graphene \cite{novo2005,bolo2009,feld2012,zhang2005}.  (\textbf{a})~Integer quantum Hall effect \cite{novo2005}.  The filling factors are anomalous, as described in the text.  (\textbf{b})~Fractional quantum Hall effect; lower fields \cite{bolo2009}. (\textbf{c})~Fractional quantum Hall effect; higher fields \cite{bolo2009}.  The plateau  labeled A has  features expected for a $\nu=1/3$ fractional quantum  Hall state. (\textbf{d})~Incompressible states with fractional filling numbers (marked),  as a function of magnetic field $B$ and carrier density $n$ \cite{feld2012}; contours of $d\mu/dn$ in units of $10^{-10} \units{meV\,cm}^2$ are displayed.} 
These experiments confirmed the predicted anomalous spacing of \eq{iqhegraph1.2} for a graphene \iqhe.  Later experiments observed fragile IQHE states at  filling factors $0,\,\pm 1, \pm 4$, that are thought to be caused by Zeeman  splitting and electron interactions breaking degeneracies of the $n=0$ Landau level.

\subsection{\label{effectStrongCorrelations}The effect of strong correlations 
in graphene}

\noindent
Graphene is a unique environment because of two aspects not found in non-relativistic quantum Hall experiments:  (1)~its charge carriers are effectively massless chiral fermions, and (2)~the symmetries associated with the four-fold  spin and valley degrees of freedom introduce new features for states and correlations. As we have seen in Section \ref{strengthOfCorrelations}, low-energy excitations in normal monolayer graphene occur in regions of reduced electron density, which disfavors electron  correlations.  But by applying a strong  magnetic field  the resulting Landau quantization leads to a bunching of levels into regions of high degeneracy,  giving conditions more amenable to the development of strong electronic correlations if the Fermi surface lies in one of those regions.

Landau levels (LL) become strongly correlated when inter-LL excitations  may be neglected and  the low-energy  excitations involve only the same level.  Then the kinetic energy  is  a  constant that can be ignored. This limit of strong electronic correlations  leads  to two important physical effects  \cite{goer2012}.

\begin{enumerate}

 \item 
 The 4-fold approximate spin--valley degeneracy of the graphene Landau levels 
leads to \textit{quantum Hall ferromagnetic states}.

\item
The strong correlations can lead to a \textit{fractional quantum Hall effects (FQHE)}.

\end{enumerate}

\noindent
Because of the approximate 4-fold degeneracy, these effects can be discussed  within the framework of an SU(4) symmetry that we shall elaborate  in Section \ref{symmetriesFQHEgraphene}. Let us now consider the \fqhe\ in graphene, which is a clear signal of a strongly-correlated state.

\subsection{\label{fqheGraphene} The graphene fractional quantum Hall effect}

\noindent
The \fqhe\ in graphene can be observed only in samples  extremely free of impurities. Strong electron--electron correlations in  general,  and the FQHE in particular, were initially very difficult to observe in  graphene  because of  samples that were not sufficiently clean to suppress impurity scattering. The  \fqhe\ in graphene was discovered in 2009   \cite{bolo2009,du2009}, when it became possible to make measurements on  suspended  graphene sheets, thus avoiding impurities and scattering associated with  substrate contact. Evidence for a \fqhe\ in graphene from those experiments  is presented in \fig{iqhe+fqheComposite}(b)-(c). A plateau labeled A at a density dependent on the magnetic field emerges for $B  > 11$ T. The inset to \fig{iqhe+fqheComposite}(c) plots this feature as a function of  filling factor.  The different traces for different magnetic field strength  collapse into a single universal feature, suggesting that A be identified with  a  $\nu=\tfrac13$ fractional quantum Hall state. 

In \fig{iqhe+fqheComposite}(d), further evidence for incompressible states at fractional filling in graphene is  shown \cite{feld2012}. The inverse compressibility is defined by $\kappa^{-1} =  d\mu/dn$, where $n$ is the local carrier density and $\mu$ is the chemical potential. In  \fig{iqhe+fqheComposite}(d) a scanning single-electron transistor was used  to measure the local incompressibility of electrons in a suspended graphene  sample and the derivative $d\mu/d n$ was plotted versus  density and magnetic field strength. High incompressibility is indicated by vertical red bands.  Localized states are broad and curve as the magnetic field is varied but the fractional  quantum Hall states are narrow and vertical,  and are labeled by fractional  filling numbers. The states that are observed  follow the standard sequence for filling factors $\nu =  0-1$, but only even-numerator fractions are seen for $\nu = 1-2$. It is thought  that these sequences and the corresponding energy gaps for the incompressible  states reflect the interplay of strong electron correlations and the  characteristic symmetries of graphene that will be discussed further in Section \ref{symmetriesFQHEgraphene} \cite{feld2012}.

\section{\label{symmetriesFQHEgraphene}SU(4) Quantum Hall Ferromagnetism}

\noindent
Graphene FQHE states can exhibit various symmetries because of possible degeneracies associated with the pseudospin and valley  degrees of freedom. In the normal two-dimensional electron gas (2DEG) produced in semiconductor  devices the Landau levels (LL) can contain $eB/h$ states, where $e$ is the  electronic charge, $B$ is the magnetic field strength, and $h$ is Planck's constant. In graphene, there is an  additional 4-fold degeneracy associated with the spin and valley degrees of  freedom. It is common to unite these four degrees of freedom (often termed  {\em flavors}) in terms of an  \su4 symmetry called \textit{quantum Hall ferromagnetism (QHF)}.

\subsection {Degeneracies and filling of Landau levels}

\noindent
Single-particle states within a   Landau  level may be  labeled by quantum numbers  $(n,m_k)$, with  $n$ indicating the Landau level and $m_k$ labeling degenerate states  associated with that level.  These Landau states  $(n,m_k)$  hold a maximum of  $2\Omega_k  = BS/(h/e)$ electrons if spin and valley degrees of freedom are neglected, where $B$ is the strength of the magnetic field, $S$ is the area of the  two-dimensional sample, and
$
h/e  = 4.136 \times 10^{-15} \units {Wb}
$
is the magnetic flux quantum. But  the 4-fold degenerate internal $\ket{\rm spin}\otimes\ket{\rm isospin}$ space for graphene implies that there are four copies of each Landau level $(n,m_k)$ and  the total electron degeneracy  $2\Omega$ for graphene is given by
\begin{equation}
    2\Omega = 4(2\Omega_k) = \frac{4BS}{(h/e)}.
    \mwgtag{20degen1.2}
\end{equation}
For a single Landau level the {\em fractional occupation} $f$ of the single Landau level is
\begin{equation}
    f \equiv \frac{n}{2\Omega} = \frac{N}{\Omega},
    \mwgtag{20fillingFraction}
\end{equation}
where $n$ is the electron number and $N=\tfrac12 n$ is the   electron pair number. Note that the fractional occupation $f$ and the  quantum Hall filling factor $\nu$ defined in \eq{ll1.13} are related by \reference{wu2017}
\begin{equation}
\nu = 4\left(f- \frac12\right).
\mwgtag{fillFacNurelation}
\end{equation}
 For the ground state of undoped graphene the $n=0$ Landau level  located at the Fermi  surface  is half filled and the electron number $n\tsub{gs}$ is  then
\begin{equation}
    n\tsub{gs}=\Omega = \frac{2BS}{(h/e)}.
    \mwgtag{20degen1.3}
\end{equation}
These degeneracies and occupation numbers are standard results for relativistic Landau electrons in a 2D electron gas, but now modified by the intrinsic graphene  degrees of freedom.

\subsection{\label{su4FQHE}SU(4) quantum Hall states in graphene}

\noindent
The graphene honeycomb lattice is bipartite, corresponding to the interlocking 
A and B sublattices shown in \fig{A-B_sublattices}.%
\doublefig
    {A-B_sublattices} 
    {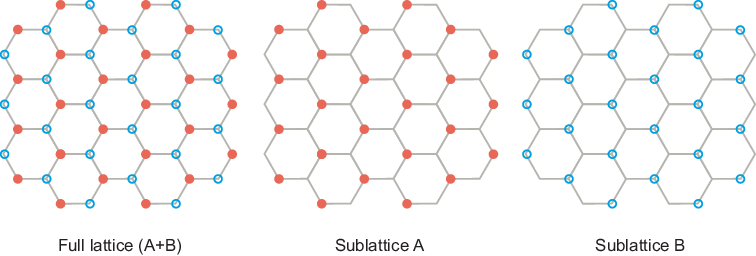}
    {0pt} 
    {0pt}
    {1.0}
    {The full bipartite lattice for graphene and the A (solid red circles) and B (open blue circles) sublattices.  }
The A and B sublattices are related by inversion, as illustrated in  \fig{Kequivalence}(a). The $n=0$ LL is located exactly at the Dirac point corresponding to $\epsilon =  0$. For low-energy excitations in each  $K$ or $K'$ valley the  inter-valley tunneling may be ignored and the wavefunctions in the valley  reside entirely on either the A or B sublattice, as  in  \fig{valleySublattices_n0}.%
\singlefigbottom
    {valleySublattices_n0}
    {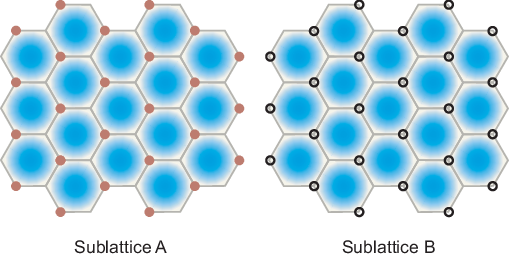}
    {0pt} 
    {0pt}
    {1.0}
    {Localization of valley wavefunctions in the $n=0$ LL.  In each valley $K$ or $K'$, the wavefunction resides on only one sublattice, either A (solid red circles on left) or B (open black circles on right).  }
For the $n=0$ LL, the valley isospin (labeling whether the electron is in a $K$ 
or $K'$ valley) is identical to the sublattice pseudospin (labeling whether the 
electron is on the A or B sublattice). Thus, this is analogous to a N\'eel antiferromagnetic state with 
spins on two different sublattices, with a N\'eel order defined by the 
difference in spins on the A and B sublattices.

\subsubsection{\label{effectiveHam}Effective Hamiltonian}

\noindent
The two largest energy scales for monolayer graphene in a strong magnetic field are 

\begin{enumerate}
\item
the Landau-level separation,  $H_{\Delta} = \sqrt 2 \hbar v\tsub F/\Lambda$, and 
\item
the Coulomb energy $H\tsub C$.
\end{enumerate}
 The separation of Landau levels is  typically several times larger than the Coulomb energy, which is in turn  considerably larger than any other terms in the interaction.  Therefore,  we adopt a strategy of ignoring excitations between Landau levels and  projecting onto the $n=0$ LL.   Such an  approximation gives the correct qualitative physics, which will be sufficient  for our discussion.  Within this single LL we assume that the Hamiltonian is dominated by an \su4-symmetric, long-range Coulomb interaction,  with  shorter-range spin and valley isospin interactions originating in  electron--electron interactions and  electron--phonon interactions that break \su4 symmetry. We shall use an approximate graphene Hamiltonian for the $n=0$  Landau level that was proposed in Ref.\ \cite{khar2012} and employed in Ref.\  \cite{wufe2014},
\begin{equation}
     H = 
     H\tsub C + H\tsub v + H\tsub Z ,
     \mwgtag{so5_1.1a}
\end{equation}
with the valley-independent Coulomb interaction $H\tsub C$, the Zeeman energy $H\tsub  Z$, and the short-range, valley-dependent interaction $H\tsub v$ given respectively by
\begin{equation}
\begin{gathered}
     H\tsub C =
     \frac12 \, \sum_{i \ne j} \, \frac{e^2}{\epsilon|\bm r_i - \bm r_j|} 
     \qquad
      H\tsub Z = -\mu\tsub B B \sum_i \sigma_z^i,
    \\
      H\tsub v =
     \frac12 \sum_{i\ne j}\left[
     g_z\tau_z^i \tau_z^j + g_{\perp} (\tau_x^i \tau_x^j + \tau_y^i \tau_y^j)
     \right]
     \delta(\bm r_i - \bm r_j) ,
 \end{gathered}
 \mwgtag{so5_1.1c}
\end{equation}
where $g_z$ and $g_\perp$ are coupling constants,  $\mu\tsub B$ is the Bohr magneton, $B$ is magnetic field strength,  the Pauli matrices $\tau_\alpha$ and $\sigma_\alpha$ operate on valley isospin and  spin, respectively, and
the $z$ direction for the spin space is assumed to be aligned with the magnetic field.

\subsubsection{\label{symmetryBreakingSU4} Symmetry and explicit symmetry breaking}

\noindent
Labeling states in a single Landau level by $m_k$ and defining  composite indices $\alpha \equiv\{x, y, z\}$ and $\beta\equiv\{x,y\}$,  we may introduce the 15 operators
\begin{equation}
\begin{aligned}
     \spin_\alpha &=
     \sum_{m_k}
    \sum_{ \tau \sigma \sigma'} \mel{\sigma'}{\sigma_\alpha}{\sigma}
    c^\dagger_{\tau\sigma'm_k} c_{\tau\sigma m_k} ,
    \\
    T_\alpha &=
    \sum_{m_k}\sum_{\sigma\tau\tau'} \mel{\tau'}{\tau_\alpha}{\tau}
    c^\dagger_{\tau'\sigma m_k} c_{\tau\sigma m_k} ,
    \\
    N_\alpha &=
    \frac12 \sum_{m_k}\sum_{\sigma \sigma' \tau }
    \mel{\tau}{\tau_z}{\tau}
    \mel{\sigma'}{\sigma_\alpha}{\sigma} 
    c^\dagger_{\tau\sigma' m_k}
    c_{\tau\sigma m_k} ,
    \\
    \Piop\alpha\beta &= 
    \frac12 \sum_{m_k}\sum_{ \sigma \sigma' \tau \tau'}
    \mel{\tau'}{\tau_\beta}{\tau}
    \mel{\sigma'}{\sigma_\alpha}{\sigma}
    c^\dagger_{\tau' \sigma' m_k}
    c_{\tau\sigma m_k} .
\end{aligned}
\mwgtag{20algebra1.4}%
\end{equation}
 The operator $\spin_\alpha$ represents the total spin and the operator  $T_\alpha$ represents the total valley pseudospin. In the $n=0$ Landau level for  graphene there is an equivalence between valley and sublattice degrees of  freedom, so $N_\alpha$ can be identified as a N\'eel vector in the $n=0$ Landau  level. The  operators $\Pi_{\alpha\beta}$ couple the  spin and valley isospin  and will be discussed further below. 
 
 Under commutation the operators of \eq{20algebra1.4} satisfy an \su4 Lie algebra that commutes with the Coulomb interaction  $H\tsub  C$ \cite{wufe2014}. Thus, if $H\tsub v$ and $H\tsub Z$  are much smaller than  $H\tsub C$, the Hamiltonian \eqnoeq{so5_1.1a} has an approximate \su4 symmetry.  If Zeeman splitting is ignored,  \su4 symmetry is broken explicitly by $H\tsub v$ in \eq{so5_1.1c}, with the degree of symmetry breaking depending on  the coupling parameters $g_z$ and $g_\perp$. For characteristic graphene lattice spacings, terms involving $g_z$ and $g_\perp$ are much  smaller than the \su4-symmetric Coulomb term and  explicit breaking of \su4 should be  small in realistic systems. Four basic  symmetry-breaking patterns have been studied,  as summarized in  \fig{subgroupChains_QHF_color} \cite{khar2012,wufe2014}.%
\doublefig
     {subgroupChains_QHF_color}
     {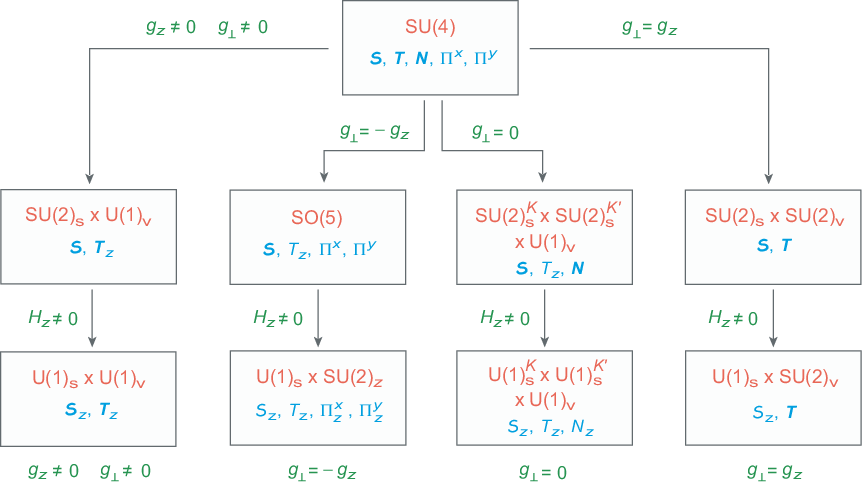}
     {0pt}
     {0pt}
     {1.0}
     {Symmetry-breaking pattern for \su4 quantum Hall ferromagnetism  generated by the operators in \eq{20algebra1.4} \cite{khar2012,wufe2014}. The \su4 symmetry is broken by explicit terms $H\tsub v$ and $H\tsub C$ in the Hamiltonian \eqnoeq{so5_1.1a} that depend on the parameters $g_z$, $g_\perp$, and $H_z$.
}

\medskip
\noindent
1. For  $g_z\ne 0$ and $g_\perp\ne 0$, the symmetry is broken 
to 
\begin{equation}
    {\rm SU(4)} \supset {\rm SU(2)}\tsub s \times 
    {\rm U(1)}\tsub v \supset {\rm U(1)}\tsub s \times {\rm U(1)}\tsub v,
    \mwgtag{so5_1.13}
\end{equation}
where \su2$\tsub s$ is associated with spin conservation, \unitary1$\tsub s$ with conservation of its $z$ component, and \unitary1$\tsub v$ with conservation of $T_z$. If Zeeman splitting is ignored, spin is conserved but  only the $z$ component of valley isospin is conserved.  The full  Hamiltonian \eqnoeq{so5_1.1a} with the Zeeman term  conserves only the $z$ components of spin and valley isospin.

\smallskip
\noindent
2. If $g_\perp=0$ but $g_z \ne 0$, the symmetry is broken in the manner
\begin{align}
    {\rm SU(4)} &\supset {\rm SU(2)}_{\rm\scriptstyle s}^{K} \times 
    {\rm SU(2)}_{\rm\scriptstyle s}^{K'} \times {\rm U}(1)\tsub v
    \nonumber
    \\
    &\supset {\rm U(1)}_{\rm\scriptstyle s}^{K} \times
    {\rm U(1)}_{\rm\scriptstyle s}^{K'} \times {\rm U(1)}\tsub v.
     \mwgtag{so5_1.14}
\end{align}
If Zeeman splitting is neglected this implies conservation of spins  independently in the $K$ and $K'$ valleys, but that valley isospin is broken to  U(1)$\tsub v$. If the Zeeman term is included, the full  Hamiltonian \eqnoeq{so5_1.1a} conserves only the $z$ components of spin in each valley, and  the  $z$ component of valley isospin.

\smallskip
\noindent
3. If $g\tsub z = g_\perp \ne 0$, the symmetry-breaking pattern is
\begin{equation}
    {\rm SU(4)} \supset
     {\rm SU(2)}\tsub s \times {\rm SU(2)}\tsub v
     \supset {\rm U(1)}\tsub s \times {\rm SU(2)}\tsub v.
     \mwgtag{so5_1.15}
\end{equation}
In the absence of  Zeeman splitting, this corresponds to full rotational symmetry in the spin and valley isospin spaces. If the Zeeman term is included in the  full Hamiltonian \eqnoeq{so5_1.1a}, the  \su2 isospin symmetry is conserved but only the $z$ component of  the spin is conserved.

\smallskip
\noindent
4. If $g_\perp = -g_z \ne 0$, the 10 generators $\{ \Pi_\alpha^\beta, \vec S, T_z \}$ commute with the Hamiltonian, forming the Lie group \SO5.  Thus 
\begin{equation}
    {\rm SU(4)} \supset {\rm SO(5)} \supset {\rm U(1)}\tsub s \times {\rm SU(2)_z},
     \mwgtag{so5_1.16}
\end{equation}
where   $\{T_z, \Pi_z^x, \Pi^y_z\}$ generates the \su2$_z$ symmetry \cite{wufe2014}.
If Zeeman splitting is neglected, the system exhibits an \SO5 symmetry  generated by spin and valley isospin. The full Hamiltonian  \eqnoeq{so5_1.1a} with the Zeeman term included conserves  \su2$_z$  but only the $z$ component of spin. 
At a mean-field level, the 10 generators of the \SO5 symmetry in \eq{so5_1.16} produce rotations on a vector space defined by order  parameters associated with the remaining five generators of SU(4): $\{ T_x,  T_y,  N_x, N_y, N_z \} $ \cite{wufe2014}. The vector space and the rotations  generated  by \SO5 are illustrated schematically in \fig{so5Rotations+bloch}(b), which generalizes the Bloch sphere of \fig{so5Rotations+bloch}(a). 
\singlefigbottom
    {so5Rotations+bloch}
    {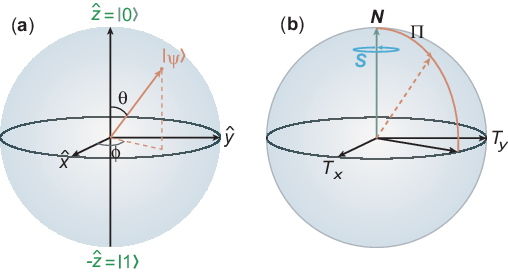}
    {0pt} 
    {0pt}
    {1.0}
    {(\textbf{a}) Bloch sphere representing a system with two basis states $\ket 0$ and $\ket 1$.  (\textbf{b}) The 5D space operated on by the SU(4) $\supset$ SO(5) subgroup and  the actions of the generators on this space. This represents a  generalization of the Bloch sphere in (a).  Adapted from  Ref.\ \cite{wufe2014}.}
The spin  operators  $\bm S$ generate rotations in the N\'eel vector  space $\bm  N$, the operator $T_z$ generates rotations in the valley vector space defined  by $T_x$ and $T_y$, and the $\Pi$ operators rotate between the N\'eel and  valley spaces. Thus \SO5 is a symmetry interpolating between   N\'eel states associated with $\bm N_\alpha$ and states associated  with valley isospin degrees of freedom $T_x$ and $T_y$.

\section{\label{possibleGroundStates} Emergent States in Magnetic Fields}

\noindent
Little evidence exists for  emergent states in  neutral monolayer graphene, as expected because of the low density of states near the  Fermi surface. However, in a strong magnetic field this situation changes  since electrons in the resulting highly degenerate Landau levels can undergo strong electron--electron and electron--phonon interactions that can \textit{break symmetries spontaneously rather than explicitly}.  The spontaneous breaking of symmetry can produce highly collective states that differ qualitatively from low-lying states of the weakly interacting system with explicit symmetry breaking, since the emergent and weakly-interacting states are typically separated by quantum phase transitions. 

As we have discussed in Ref.\ \cite{wu2016}, there is  evidence that the ground state of monolayer graphene in a strong magnetic field is such a (spontaneously) broken-symmetry state. In particular the ground state for monolayer graphene in a magnetic field is known to be very  strongly insulating, exhibiting a rapid divergence of the longitudinal  resistance $R\tsub L$ at a critical magnetic field $B\tsub c$, with $B\tsub c$  smaller for cleaner samples.  This behavior suggests that the resistance  in this state is a consequence of  emergent intrinsic  properties of the state itself and not of impurity scattering. 

With this motivation, let us turn to a discussion of emergent states in graphene  created through strong electron--electron and electron--phonon correlations that break symmetries spontaneously rather than explicitly. We give first a more conventional view.  Then we shall reformulate the discussion of graphene emergent states in terms of powerful new approaches  based on fermion dynamical symmetries.  In all of this discussion ``graphene'' will be shorthand for monolayer graphene in a strong magnetic field.

\subsection{\label{roleEE} The role of  correlations}

\noindent
Some general basis states for collective states that could be realized by placing two  electron spins in $n=0$ graphene valleys are illustrated in  \fig{collectiveValleyModes}.%
\doublefig
    {collectiveValleyModes} 
    {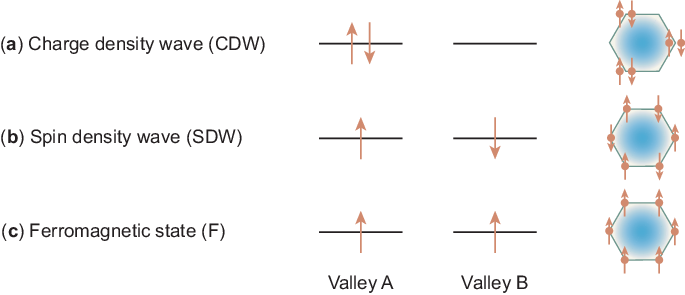}
    {0pt} 
    {0pt}
    {1.0}
    {Some possible  basis states for collective states in the graphene $n=0$ Landau level.  Electron  occupation for adjacent valleys A and B are shown, with up and down arrows indicating  spin-up electrons and spin-down electrons, respectively.  In the  absence of Landau level mixing valley indices and the sublattice indices A and  B coincide for $n=0$. The corresponding lattice occupation is illustrated on the right.  (\textbf{a}) A charge  density wave (CDW). (\textbf{b}) A spin density wave  (SDW). (\textbf{c}) A ferromagnetic state (F). Figure  adapted  from Ref.\ \cite{jung2009}.
}

(1)~In \fig{collectiveValleyModes}(a), both electrons  are  in the same  valley  (with opposite spins) and the adjacent valley is unoccupied, so valleys  alternate in charge $+2e,\, 0,\, +2e,\, 0 \ldots,$ around the carbon ring.  The right side of  \fig{collectiveValleyModes}(a) illustrates the general lattice occupation, with  adjacent sites alternating between being occupied by two spin-singlet electrons  and by no electrons. This is  a {\em charge density  wave} (CDW). 

(2)~In \fig{collectiveValleyModes}(b), the electrons are  in  adjacent valleys with opposite spin projections.  The right side of  \fig{collectiveValleyModes}(b) illustrates the lattice occupation, with  adjacent  sites alternating between spin-up  and spin-down  electrons.  This is  an \textit{antiferromagnetic} (AF) \textit{spin-density wave} (SDW). 

(3)~In  \fig{collectiveValleyModes}(c), the electrons are in adjacent valleys  with the same spin projections.  The right side of  \fig{collectiveValleyModes}(c) illustrates the lattice occupation, with a  spin-up electron at each site. This is a  \textit{ferromagnetically polarized state} (F). 

(4)~A fourth possible collective mode on the lattice is
{\em Kekul\'e distortion} (KD), which induces alternating  shorter and longer bonds around the ring in the graphene structure  (a  ``bond density wave'' \cite{cham2000}).   The terminology references August  Kekul\'e, who proposed the alternating single and double bond structure for the  benzene ring in \fig{sp2-orbitals}(b). A Kekul\'e distortion state will be illustrated  in \fig{grapheneCDW-ferro-sdw-kekule}(d) below.

 The collective modes described above are illustrated schematically in  \fig{grapheneCDW-ferro-sdw-kekule}.
\doublefig
          {grapheneCDW-ferro-sdw-kekule}
          {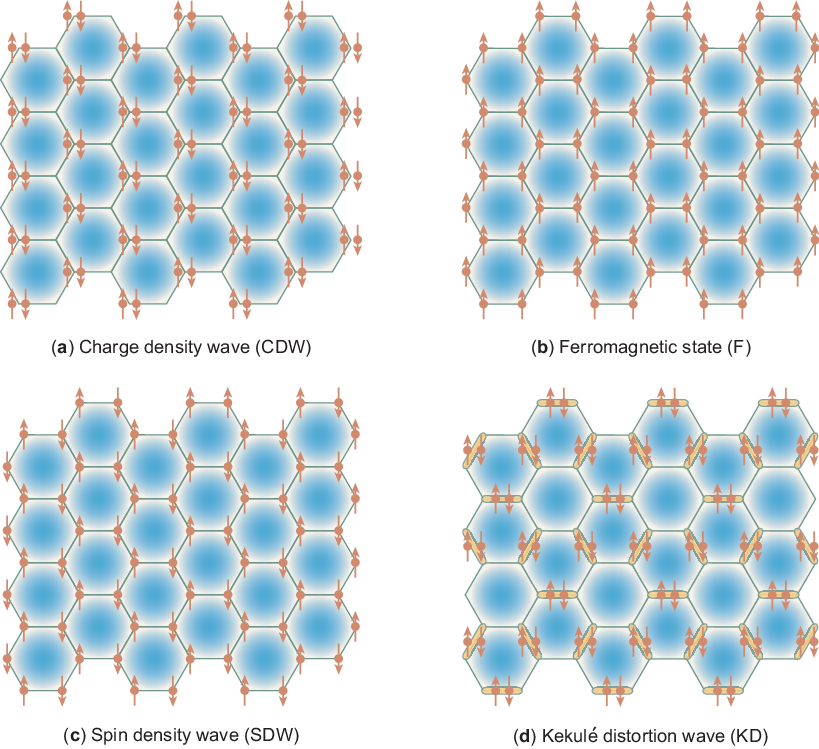}
          {0pt} 
          {0pt}
          {1.0}
          {Some collective states in the monolayer graphene, adapted from Ref.\ \cite{khar2012}.  (\textbf{a}) Charge density wave (CDW).   (\textbf{b}) Ferromagnetic state (F). (\textbf{c}) Spin density wave (SDW), corresponding to an antiferromagnetic N\'eel state. (\textbf{d}) Kekul\'e distortion (KD) mode (``bond density wave'').  }
Assuming the effective Hamiltonian of \eq{so5_1.1c} to capture the essential physics, we  expect in the simplest picture that the system could be in one of four possible phases  exhibiting different forms of long-range order \cite{khar2012}.

\begin{enumerate}

\item
A spin-singlet charge density wave phase CDW   illustrated schematically in \fig{grapheneCDW-ferro-sdw-kekule}(a) that is not  affected by the Zeeman term $H\tsub Z$ in the Hamiltonian.

\item 
A ferromagnetic phase F,  illustrated schematically in  \fig{grapheneCDW-ferro-sdw-kekule}(b). The symmetry is broken to \unitary1 by $H\tsub Z$, corresponding to conservation of spin projection. 

\item
An antiferromagnetic or N\'eel phase AF, illustrated schematically in  \fig{grapheneCDW-ferro-sdw-kekule}(c). The symmetry is broken to U(1)  by $H\tsub Z$, corresponding to  conservation of spin projection.

\item
A spin-singlet Kekul\'e distortion phase KD  illustrated schematically in \fig{grapheneCDW-ferro-sdw-kekule}(d) that is not affected by the  Zeeman term in the Hamiltonian.

\end{enumerate}

\noindent
These competing emergent modes are a result of correlations.  Which, if any, of these modes are realized depends on which correlations dominate, which will now be discussed.

\subsection{\label{roleEP}Symmetry-breaking interactions}

\noindent
Symmetries may be broken spontaneously by short-range electron--electron (e--e) interactions, and  short-range electron--phonon (e--ph) interactions between the lattice and the  electrons, with  the  electron--electron interactions generally being repulsive while  the  electron--phonon interactions are attractive. The most important e--ph interaction is thought to be Kekul\'e distortion, which perturbs the valley locations  and contributes to the $g_\perp$  coefficient  in \eq{so5_1.1c}. Hence the valley-isospin $T_x$ and $T_y$ degrees of  freedom should couple to Kekul\'e distortion modes.
 A  systematic analysis by Kharitonov \cite{khar2012} suggests that the e--e interactions could  favor any of the states described above:  F, AF, CDW, or KD, depending on  details of the symmetry-breaking interactions, but the leading terms in the  attractive e--ph interactions always favor the KD mode. 

 Thus, we deal with a system having multiple possible emergent ground  states, 
with the actual ground state determined by a delicate balance among the  different e--e and e--ph interactions.  This is the hallmark of {\em complexity:}  a choice between competing emergent ground  states that is sensitive to even weak external perturbations.  Complexity is more common in soft condensed matter systems, but can occur in  hard condensed matter systems when  there are  multiple  collective ground-state candidates that are nearly degenerate, as we expect to be the case here.

\section{\label{dynamical_su4} Low-Energy Collective Modes}

\noindent
In the preceding section we have seen that graphene in a strong magnetic field  behaves as an \su4 ferromagnet if only the long-range Coulomb interaction is  retained in the Hamiltonian.  In the presence of shorter-range valley and spin interaction  terms this symmetry is broken  to an approximate \su4 symmetry.  We discussed the classification of states with these terms in the  Hamiltonian, and the effect of adding to the Hamiltonian the Zeeman term.   Those considerations led to  \su4 symmetry and  symmetry-breaking classification schemes and serve as a basis for implementing  numerical simulations or low-energy field theory approximations.  In this section those symmetries and general principles  will be used to elucidate wavefunctions for possible collective modes that could develop in  graphene  because of non-perturbative electron--electron and  electron--phonon interactions.

\subsection{\label{basisCollective} Basis states for collective modes}

\noindent
Let us address some plausible basis states corresponding to the modes  discussed in  the preceding section, guided by the presentation in Kharitonov \cite{khar2012}. If attention is restricted to a single Landau level labeled  by $n$, the degeneracy of states for a 2D electron gas in a magnetic field   corresponds to the possible values of the quantum number $m_k$ distinguishing the states for that $n$.   However, for graphene  the degeneracy of the Landau states $(n,m_k)$ will be modified by the  spin, sublattice pseudospin, and valley isospin degrees of freedom.  In the  simplest case where valley mixing is neglected we may equate the valley isospin  and sublattice pseudospin labels, leaving potentially four additional degrees of  freedom (valley isospin $\otimes$ real spin)  for each Landau state $(n,m_k)$.  If these  four ``internal'' degrees of freedom are taken to be degenerate, this leads to  a  Hamiltonian with \su4 symmetry for  quantum Hall states.  Physically, this corresponds to four copies of a Laughlin-like incompressible  state as described in Section \ref{laughlinWF}.

We begin by defining a Dirac 4-component field in the valley and  sublattice degrees of freedom with fixed spin polarization $\sigma$ \cite{khar2012},
%
%
\begin{equation}
    \psi_\sigma(\bm r) = 
    \begin{pmatrix}
     \psi_{KA\sigma}(\bm r) \\[2pt]
     \psi_{KB\sigma}(\bm r) \\[2pt]
     \psi_{K'B\sigma}(\bm r) \\[2pt]
     -\psi_{K'A\sigma}(\bm r)
    \end{pmatrix}
     _{KK' \otimes AB} ,
     \mwgtag{dirac1.1}
\end{equation}
where the ordering and signs are chosen to give the most symmetric  representation of the Dirac Hamiltonian and $KK' \otimes AB$ denotes the  direct product of valley ($KK'$) and sublattice ($AB$) degrees of freedom. Letting the spin polarization index $\sigma$ take the two  values $\uparrow$ and $\downarrow$, the 8-component operator
\begin{equation}
    \psi(\vec r) = 
    \begin{pmatrix}
     \psi_\uparrow(\bm r) \\[2pt]
     \psi_\downarrow(\bm r)
    \end{pmatrix}
    _s
     \mwgtag{dirac1.2}
\end{equation}
defines the most general spinor in the direct product of valley  ($KK'$), sublattice ($AB$), and spin ($s$) spaces.

The $n=0$ Landau level is located exactly at the Dirac point, and  in each valley  labeled by $K$ or $K'$ the wavefunctions reside entirely on one actual  sublattice, $K \leftrightarrow A$ or $K' \leftrightarrow B$. Therefore, for  the $n=0$ level the field operator has only two non-vanishing components for a  given spin polarization $\sigma$.
Accordingly, in the $n=0$ LL the non-vanishing components of \eq{dirac1.2} can 
be collected into a 4-component vector
%
%
\begin{equation}
    \psi^{(0)}(\vec r) =
    \begin{pmatrix}
     \psi_{KA\uparrow}^{(0)} (\bm r) \\[3pt]
     \psi_{KA\downarrow}^{(0)}(\bm r) \\[3pt]
     \psi_{K'B\uparrow}^{(0)}(\bm r) \\[3pt]
     \psi_{K'B\downarrow}^{(0)}(\bm r)
    \end{pmatrix}
    _{KK'\otimes s}
     \mwgtag{dirac1.4}
\end{equation}
in the valley isospin and spin space, where the superscript $(0)$ indicates 
that this is valid specifically for the $n=0$ LL.

\subsection{\label{generalManyBodyState} Most general many-body wavefunction}

\noindent
Let the operator $c^\dagger_{nm_k\lambda\sigma}$ create an electron in the  $(n,m_k)$ Landau level with $\lambda$ the sublattice or valley index and  $\sigma $ the spin polarization index.  For the $\nu = 0$ state in  graphene, corresponding to two electrons per 4-fold degenerate $n=0$ Landau  level, a general many-body wavefunction can be written \cite{khar2012}
%
%
\begin{equation}
    \Psi = 
    \prod_{m_k} \bigg(
    \sum_{\lambda \sigma, \lambda' \sigma'}
    \Phi^*_{\lambda\sigma, \lambda'\sigma'}
    c^\dagger_{0m_k\lambda\sigma}  c^\dagger_{0m_k\lambda'\sigma'}
    \bigg)
    \ket 0 ,
    \mwgtag{basis1.0}
\end{equation}
where the range of $m_k$ is given by \eq{LLdegen1.10}, and where the vacuum state $\ket 0$ corresponds to completely filled Landau levels  for $n<0$ and completely empty Landau levels for $n\ge 0$.  Each  factor in the product $\prod_{m_k}$ creates a pair of electrons in the state $\Phi = \{\Phi_{\lambda\sigma, \lambda'\sigma'} \}$ at orbital $m_k$ of the  $n=0$ Landau level,

\begin{enumerate}
\item
with $\lambda$ and $\lambda'$ equal to sublattice labels A or B, 

\item
with $\sigma$ and $\sigma'$  equal to spin-up ($\uparrow$) or spin-down ($\downarrow$), 
\item
and with the valley isospin and sublattice pseudospin labels identified: $K \leftrightarrow A$ and $K' \leftrightarrow B$.
\end{enumerate}
  Letting $V$ denote the valley isospin symmetry and $S$ the electron spin  symmetry, the wavefunction $\Phi$ in \eq{basis1.0} describes how the 4-fold degenerate $V  \otimes  S$ spin--isospin subspace of each orbital is occupied by two electrons.

\subsection{\label{explicit-spontaneous} Explicit vs.\ spontaneous symmetry breaking}

States like the  ground state in graphene are thought to be the  result of {\em spontaneous symmetry breaking}, where the symmetry is broken by an unsymmetric vacuum rather than by explicit terms in the Hamiltonian. Then there are likely multiple solutions of the form \eqnoeq{basis1.0} that have similar energy but represent emergent states having very different character, with the one of lowest energy  determined by details of the symmetry-breaking  correlations. Kharitonov \cite{khar2012} has given an overview examining states based on the $n=0$ Landau level in the spirit of this section, with a general discussion of which are expected to be favored energetically. We refer readers to that discussion for more details of that approach.  

We shall instead take a fundamental point of view that emergent states in graphene (or any many-body fermionic system) are separated from the weakly interacting states by a phase transition; thus they have no direct connection with states generated by explicit breaking of some symmetry.  
One needs exact solutions, or reasonable approximations, for states with symmetry broken spontaneously to investigate such emergent states.  Within the framework discussed to this point, these typically require numerical solutions.
In Section \ref{fds} we shall  introduce a different approach to determining the nature of low-lying emergent states based on dynamical symmetries  for monolayer graphene in a magnetic field. The corresponding results  will have two important implications.  
\begin{enumerate}
\item
We will be able to obtain \textit{analytical solutions that} correspond to spontaneously broken symmetry, and thus to matrix elements of observables for possible emergent graphene states.
\item
Our analysis will suggest that the most useful symmetry to break spontaneously for emergent states in  monolayer graphene  isn't \su4, as assumed in most discussions in the literature, but rather is an \SO8 parent symmetry of \su4.
\end{enumerate}
  One important consequence is the prediction of more, and more varied, emergent states in graphene from breaking \SO8 spontaneously than from breaking \su4 spontaneously.

\section{\label{fds} Fermion Dynamical Symmetry}

\noindent
The preceding material has given a general introduction to monolayer graphene in a strong magnetic field from a traditional point of view.  The remainder of this review describes an alternative way to view the emergent states of this system, by solving for relevant matrix elements in a Hilbert space that has been truncated to a manageable collective subspace using the properties of Lie groups and  Lie algebras.  Reviews of these methods applied to various problems in nuclear structure physics and high-temperature superconductors may be found in Refs.\ \cite{wu1986,wu1987,wu1994,guid2020} and references cited there,  with a textbook overview given in Ref.\ \cite{guid2022}. Specific applications for graphene are given in Refs.\ \cite{wu2016,wu2017,guid2017} and in Ch. 20 of Ref.\ \cite{guid2022}.

\subsection{The microscopic dynamical symmetry method}

\noindent
The dynamical symmetry method uses generators of Lie groups and Lie algebras to construct a microscopic theory imposing a truncation of  the full Hilbert space to a tractable subspace;  \fig{dynamicalSymmMicroscopic}(a) illustrates.
\doublefig
    {dynamicalSymmMicroscopic}
    {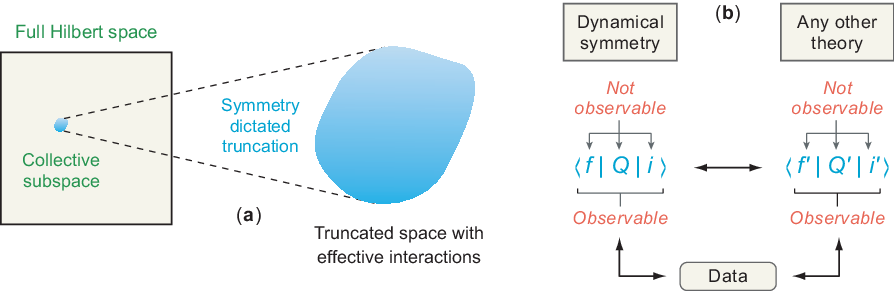}
    {0pt}
    {0pt}
    {1.0}
    {(\textbf{a}) Emergent-symmetry truncation of the full fermionic Hilbert space to a collective  subspace using principles of dynamical symmetry. (\textbf{b})~Comparison of  matrix elements among different theories and data. Wavefunctions and operators  are {\em not observables}.  Only {\em matrix elements} are related directly to  experimental data and serve as valid comparison criteria.  }
Such a drastic truncation is justified if it leads to correct matrix elements  for physical observables, as illustrated in \fig{dynamicalSymmMicroscopic}(b).  Two  quantum theories may use  different methodologies, but each must produce matrix elements of observables  as  physical output; valid comparisons between theories, and of theories to data, must be through {\em matrix elements corresponding to observables}. Wavefunctions and operators considered individually are {\em not observables} and thus are not a reliable basis of comparison.  

A formal statement of the  microscopic dynamical symmetry method begins with the following hypothesis \cite{wu1994}:
\graybox{{\bf\em Dynamical Symmetry Hypothesis:}  Strongly-correlated emergent states  imply a  dynamical symmetry described by a Lie algebra that is generated by commutation of operators representing the emergent mode. }
%
Dynamical symmetries may be defined for fermions or bosons, but our interest here will be in fermionic systems.
The \textit{Dynamical Symmetry Hypothesis} is supported by a large amount of   evidence from various fields of many-body physics \cite{bi94,guid01,guid2020,guid2022,iac87,guid2017,iac95,iac99,wu1994,wu2016,wu2017}.

\subsection{\label{solutionAlgorithm} Dynamical symmetry solution algorithm}

\noindent
Assuming the \textit{Dynamical Symmetry Hypothesis}, microscopic solutions for  emergent states in quantum many-body systems may be found by the following  procedure \reference{guid2020,wu1994,guid2022}.

\smallskip\noindent
1.~Use phenomenology and theory to identify emergent degrees of freedom that are
relevant physically for the problem.  

\smallskip\noindent
2.~\textit{Close a commutation algebra}  on a set of second-quantized operators creating,  annihilating, and counting these modes. This Lie algebra is specified by the  microscopic generators for  emergent physical states and is termed the {\em  highest symmetry}. 

\smallskip\noindent
3.~Truncate the full Hilbert space to a {\em collective subspace} by requiring  that matrix elements of the operators found in the preceding step don't cause  transitions out of the collective subspace.  This dramatic reduction of the  space is termed {\em symmetry-dictated truncation}  (see \fig{dynamicalSymmMicroscopic}). Collective  states in this  subspace have low energy but their wavefunction components  are {\em selected  by symmetry, not energy.} This means that they may contain both low-energy  and  high-energy components of a basis appropriate for the weakly interacting system.

\smallskip\noindent
4.~Identify subalgebra chains of the highest symmetry ending in algebras imposing expected conservation laws, like those for charge and spin. Associated with  these Lie algebras will be corresponding {\em Lie groups.} 

\smallskip\noindent
5.~Each  subalgebra chain and corresponding subgroup chain specifies a {\em dynamical  symmetry} associated with the highest symmetry, and each dynamical symmetry defines a \textit{distinct quantum phase}. Thus a given highest symmetry can birth multiple quantum phases, each distinct physically but  \textit{all related through a common highest symmetry}.

\smallskip\noindent
6.~Construct {\em  dynamical symmetry Hamiltonians} that are polynomials in the Casimir invariants for subgroup chains. Each  chain defines a wavefunction basis labeled  by eigenvalues of chain invariants (Casimirs and  elements of the Cartan subalgebras), and a Hamiltonian that is diagonal in this basis, because it is constructed  from invariants. 

\smallskip\noindent
7.~Thus, the dynamical symmetries allow the \textit{the Sch\"odinger equation to be solved analytically} for each subgroup chain.

\smallskip\noindent
8.~Examine the physical content of each  dynamical symmetry by \textit{calculating   matrix elements of observables}. This is possible because of the  eigenvalues and eigenvectors obtained in step 6, and because  consistency requires that any transition operators be related to group generators;  otherwise transitions would mix irreducible multiplets and break the symmetry.

\smallskip\noindent
9.~Construct the most general Hamiltonian in the model space as  a linear combination of terms in the Hamiltonians for each symmetry group chain. Invariant operators of different subgroup chains don't generally commute, so an invariant for one  chain may be a source of symmetry breaking for another.  

\smallskip\noindent
10.~Thus the competition and quantum phase transitions between different dynamical symmetry chains may be studied.

\smallskip\noindent
11.~More ambitious formulations that may include terms breaking the dynamical symmetries can be solved by 

\begin{itemize}

\item[a)]
coherent state or other approximations of the  dynamical symmetry described  in step (7), or

\item[b)]
perturbation theory around  the symmetric solutions (which are  non-perturbative vacuum states, so these ``perturbative'' solutions are actually non-perturbative with respect to the non-interacting ground state), or

\item[c)]
numerical solutions incorporating symmetry-breaking  terms.

\end{itemize}

\noindent
Reviews and applications of this  methodology to both fermionic and bosonic systems in many fields may be found in   Refs.~\cite{bi94,guid01,guid2020,guid2022,iac87,guid2017,iac95,iac99,wu1994,wu2016,wu2017}, and references cited there.

\subsection{Validity of the dynamical symmetry approach}
 
\noindent
The {\em only  approximation} in the microscopic dynamical symmetry approach is the symmetry-dictated truncation described in point (3) above and indicated schematically in \fig{dynamicalSymmMicroscopic}; if all  degrees of freedom are accounted for the   method is microscopic and exact.  Practically, only select degrees of  freedom can be accommodated and the effect of the excluded Hilbert space must be  incorporated through renormalized (effective) interactions operating in the  truncated space. Thus the utility of the approach depends on

\begin{enumerate}

 \item 
making a wise initial choice for the emergent degrees of freedom and their  symmetries,   and 

\item
the availability of sufficient information from  theoretical understanding and phenomenology to specify  the  effective interactions in the truncated space.

\end{enumerate}

\noindent
As for any microscopic theory, the validity of this approach then stands on 
whether calculated matrix elements are consistent with experimental
observables.

\section{Dynamical Symmetry in Graphene}

\noindent
In most theoretical approaches, studying the emergent modes for monolayer graphene in a strong magnetic field requires numerical calculations, since the corresponding states are likely to be highly non-perturbative. Let us now address these states that were described in Sections \ref{possibleGroundStates} and \ref{dynamical_su4}, using the fermion dynamical symmetry  method  outlined in Section \ref{fds}. This will allow determining  properties of various emergent states {\em analytically} rather than by computer calculations.

\subsection{Basis states}

\noindent
 Spin and valley isospin quantum number assignments for a useful basis are  illustrated in \fig{grapheneBasis_withTableBW} [compare with the wavefunction in \eq{dirac1.4}].%
\doublefig
     {grapheneBasis_withTableBW}
     {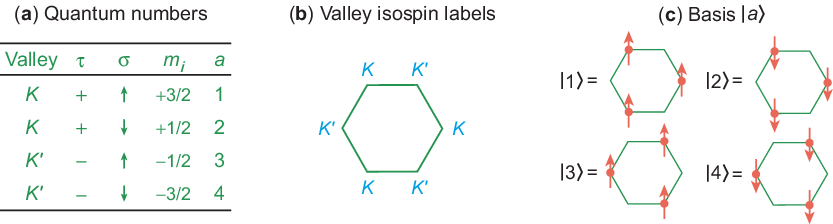}
     {0pt}
     {0pt}
     {1.0}
     {(\textbf{a}) Valley isospin ($\tau$) and spin ($\sigma$) quantum numbers. 
     Each row of the table labeled by $a$ corresponds to a basis state $\ket a$ displayed in (\textbf{c}).
     (\textbf{b}) The  valley isospin labels $K$ ($\tau=+$) and $K'$ $(\tau = -$). (\textbf{c}) Four  basis  states corresponding to the rows of table (\textbf{a}).
     }
For example, consider the state labeled $\ket 2$ in  \fig{grapheneBasis_withTableBW}(c). The electron occupies valleys labeled by  valley isospin $K$ ($\tau = +$) [see \fig{grapheneBasis_withTableBW}(a-b)] with  spin down ($\sigma=\,\downarrow$), so  this basis state corresponds to quantum numbers in the second row of the table in  \fig{grapheneBasis_withTableBW}(a). The table of \fig{grapheneBasis_withTableBW}(a)  also displays a mapping  of the four basis states labeled by $a$ to a label $m_i$ that takes values of the four possible projections $m_i=\{ \pm \tfrac12, \pm \tfrac32 \}$ of a fictitious  angular momentum $i=3/2$; this mapping will allow adaptation of some existing results in the literature to the graphene problem.

\subsection{SO(8) generators and Lie algebra}

\noindent
Let $A^\dagger_{ab}$ create an electron pair with one electron in the $a=(\tau_1,\sigma_1)$ level and one in the $b=(\tau_2,\sigma_2)$  level,  with $m_k$  labeling $n=0$ Landau states coupled to  zero term by term,
\begin{equation}
    A^\dagger_{ab}  = \sum_{m_k} c^\dagger_{a m_k} c^\dagger_{b 
    -m_k} .
    \mwgtag{20algebra1.1}
\end{equation}
Then the hermitian conjugate $A_{ab} = (A_{ab}^\dagger)^\dagger$  annihilates a pair.  [Don't confuse $m_k$ in \eqnoeq{20algebra1.1}  with the label $m_i$ in \fig{grapheneBasis_withTableBW}(a).] There are  $4\times 4=16$ possible combinations $ab$ in \eq{20algebra1.1}, but  antisymmetry of fermionic wavefunctions reduces this to six independent operators  $A^\dagger$ and  six independent hermitian conjugates $A$. Introduce also  the  16 particle--hole  operators,
\begin{equation}
    B_{ab} = \sum_{m_k} c^\dagger_{am_k} c_{bm_k}\phd - \frac14 
    \delta_{ab} \Omega ,
    \mwgtag{20algebra1.2}
\end{equation}
where $\delta_{ab}$ is the Kronecker delta and $\Omega$ is the total degeneracy  of the Landau level.  Then the commutator algebra for  the 28 operators $A$, $A^\dagger$, and $B$  is \reference{chen86}
\begin{subequations}
    \begin{align}
    \comm{A_{ab}\phd}{A^\dagger_{cd}} &=
    -B_{db} \delta_{ac} - B_{ca}\delta_{bd}
    + B_{cb} \delta_{ad} + B_{da} \delta_{bc},
    \mwgtag{20algebraTotal-a}
    \\
    \comm{B_{ab}}{B_{cd}} &=
    \delta_{bc} B_{ad} - \delta_{ad}B_{cb},
    \mwgtag{20algebraTotal-b}
    \\
    \comm{B_{ab}\phd}{A^\dagger_{cd}} &=
    \delta_{bc}\phd A^\dagger_{ad} + \delta_{bd}\phd A^\dagger_{ca},
    \mwgtag{20algebraTotal-c}
    \\
    \comm{B_{ab}}{A_{cd}} &=
    -\delta_{ac} A_{bd} - \delta_{ad} A _{cb} ,
    \mwgtag{20algebraTotal-d}
    \end{align}
    \mwgtag{20algebraTotal}%
\end{subequations}
which is isomorphic to the Lie algebra \SO8.

\subsection{A more physical generator basis}

\noindent
Generators of a Lie algebra span a linear vector space, so any independent linear combination generates the same algebra in a different basis.
To facilitate interpretation for graphene, it is useful to  express  the  generators of \eq{20algebraTotal} in a new basis. First note that the  operators  in \eq{20algebra1.2} can be replaced by the operators in \eq{20algebra1.4} through comparing definitions (Section 4 of the Supplement \cite{supplementMaterials}).   By such methods the spin operators of \eq{20algebra1.4} are given in terms of the $B_{ab}$ generators as 
\begin{subequations}
\begin{align}
    \spin_x &=
    B_{12} + B_{21} + B_{34} + B_{43},
    \mwgtag{algebra1.8a}
    \\
    \spin_y &=
    -i\left(B_{12} - B_{21} + B_{34} - B_{43} \right),
    \mwgtag{algebra1.8b}
    \\
    \spin_z &=
    B_{11} - B_{22} + B_{33} - B_{44} ,
    \mwgtag{algebra1.8c}
\end{align}
\mwgtag{algebra1.8}
\end{subequations}
the valley isospin operators of \eq{20algebra1.4} as
\begin{subequations}
\begin{align}
    T_x &=
    B_{13} + B_{31} + B_{24} + B_{42},
    \mwgtag{algebra1.9a}
    \\
    T_y &=
    -i\left(B_{13} - B_{31} + B_{24} - B_{42} \right),
    \mwgtag{algebra1.9b}
    \\
    T_z &=
    B_{11} + B_{22} - B_{33} - B_{44} ,
    \mwgtag{algebra1.9c}
\end{align}
\mwgtag{algebra1.9}
\end{subequations}
the N\'eel vector of \eq{20algebra1.4} as
\begin{subequations}
\begin{align}
    N_x &=
    \tfrac12 \left( B_{12} + B_{21} - B_{34} - B_{43} \right),
    \mwgtag{algebra1.10a}
    \\
    N_y &=
    -\tfrac i2 \left(B_{12} - B_{21} - B_{34} + B_{43} \right),
    \mwgtag{algebra1.10b}
    \\
    N_z &=
    \tfrac12 \left( B_{11} - B_{22} - B_{33} + B_{44} \right),
    \mwgtag{algebra1.10c}
\end{align}
\mwgtag{algebra1.10}
\end{subequations}
and the interpolating operators $\Piop\alpha\beta$ of \eq{20algebra1.4} as
\begin{subequations}
\begin{align}
    \Piop xx &=
    \tfrac12 \left( B_{14} + B_{41} + B_{23} + B_{32} \right),
    \mwgtag{algebra1.11a}
    \\
    \Piop yx &=
    -\tfrac i2 \left(B_{23} - B_{32} + B_{41} - B_{14} \right),
    \mwgtag{algebra1.11b}
    \\
    \Piop zx &=
    \tfrac12 \left( B_{13} + B_{31} - B_{24} - B_{42} \right),
    \mwgtag{algebra1.11c}
    \\
    \Piop xy &=
    -\tfrac i2 \left( B_{32} - B_{23} + B_{41} - B_{14} \right),
    \mwgtag{algebra1.11d}
    \\
    \Piop yy &=
    -\tfrac 12 \left(B_{41} + B_{14} - B_{23} - B_{32} \right),
    \mwgtag{algebra1.11e}
    \\
    \Piop zy &=
    -\tfrac i2 \left( B_{31} - B_{13} - B_{42} + B_{24} \right) .
    \mwgtag{algebra1.11f}
\end{align}
\mwgtag{algebra1.11}%
\end{subequations}
Inverse transformations expressing the $B_{ab}$ in terms of the  $\{\spin_\alpha,\,  T_\alpha,\, N_\alpha, \,\Piop \alpha x, \,\Piop  \alpha y\}$ are  given in Section 4 of the Supplement  \cite{supplementMaterials}. From Appendix \ref{so8u4su4}, the \su4 algebra generated by the operators in \eq{20algebra1.4} is a  subalgebra of the \SO8 algebra, with its generators corresponding to particular  linear combinations of the subset of \SO8 generators defined by the  particle--hole operators $B_{ab}$ in \eq{20algebra1.2}.

\subsection{Coupled representations for pair operators} 

\noindent
Next, observe that valley  isospin is approximately conserved for low-lying states and spin is conserved if Zeeman splitting can be ignored.  Hence it is useful to  employ  pairing operators coupled to states of good total spin and good total valley isospin.  An operator creating an electron pair  coupled to good spin and  isospin may be written
\begin{align}
    \Adagcoupled{M_S}{M_T}{S}{T} &\equiv
    \sum _{m_1m_k} \sum _{n_1n_2}
    \clebsch{\tfrac12}{m_1}{\tfrac12}{m_2}{S}{M_S } 
    \nonumber
    \\
    &\quad\times \clebsch{\tfrac12}{n_1}{\tfrac12}{n_2}{T}{M_T } c^\dagger_{m_1 n_1 m_k}
    c^\dagger_{m_2 n_2 -m_k},
    \mwgtag{20coupled1.0}
\end{align}
where in this equation

\begin{itemize}
\item
$S$ is the total spin, with projection $M_S$, 

\item
$T$ is the  total valley isospin, with projection $M_T$, 

\item
the  \su2 Clebsch--Gordan  coefficient   $\clebsch{{\scriptstyle\frac12}}{m_1}{{\scriptstyle\frac12}}{m_2}{S}{M_S  }$ couples the spins to good total spin $\ket{S M_S}$, and 

\item
the \su2  Clebsch--Gordan  coefficient $\clebsch{{\scriptstyle\frac12}}{n_1}{{\scriptstyle\frac12}}{n_2}{T}{M_T }$  couples  the isospins to good total isospin $\ket{T M_T}$. 
\end{itemize}
The fermion pair wavefunctions are restricted to total spin and  isospin combinations $(S=1,\, T=0)$ or  $(S=0,\, T=1)$ by  the antisymmetry requirement.  The  six coupled pairing  operators constructed from \eq{20coupled1.0} that satisfy this condition are 
\begin{equation}
\begin{gathered}
     S^\dagger = 
    \frac{1}{\sqrt2} \left( A_{14}^\dagger - A_{23}^\dagger \right)
    \qquad
    D_0^\dagger 
    = \frac{1}{\sqrt2} \left( A_{14}^\dagger + A_{23}^\dagger \right),
    \\
    D_1^\dagger 
    = A_{13}^\dagger
    \qquad
    D_{-1}^\dagger 
    = A_{24}^\dagger,
    \\
    D_2^\dagger 
    = A^\dagger_{12}
    \qquad
    D_{-2}^\dagger 
    = A_{34}^\dagger,
    \end{gathered} 
 \mwgtag{20coupled1.4}
\end{equation}
while $S=(S^\dagger)^\dagger$ and  $D_\mu = (D_\mu^\dagger)^\dagger$ represent six corresponding hermitian-conjugate annihilation operators. The six equations \eqnoeq{20coupled1.4}, their six hermitian conjugates, the 15 equations \eqnoeq{20algebra1.4}, and the number operator $S_0$ defined in  \eq{20numberOp1.2}  (see also Appendix \ref{so8u4su4})
represent independent linear combinations of the 28 generators of \SO8 defined in Eqs.\ \eqnoeq{20algebra1.1} and  \eqnoeq{20algebra1.2}. Therefore, since \eq{20algebraTotal} defines an \SO8 Lie algebra, the 28 operators
\begin{equation}
    G_{\scriptscriptstyle{\rm SO(8)}}=
    \{ \spin_\alpha\phd,\,  T_\alpha\phd,\,  N_\alpha\phd,\, \Piop \alpha x\phd, 
    \,\Piop \alpha y\phd, 
    \,
    S_0\phd, \,
    S, \,S^\dagger, \,D\phantomdagger_{\mu},\, D^\dagger_\mu \}
    \mwgtag{20grapheneBasis}
\end{equation}
constitute an equally valid set of  generators for \SO8, expressed in a new generator basis.

\subsection{SO(8) pair states in graphene}

\noindent
Configurations resulting from applying the pair creation operators in \eq{20grapheneBasis} to the  vacuum $\ket0$ are shown in \fig{S_D_pairs_brillouin_withQ}, as are configurations generated by the linear combinations
\begin{align}
     \ket{Q_\pm} = Q_\pm^\dagger\ket0 \equiv 
     \frac12 \big(S^\dagger \pm D_0^\dagger \big) \ket0
    =\frac12 \big(\ket S \pm |D_0\phd\rangle \big) ,
     \mwgtag{20Qdef}
\end{align}
which will prove to be  useful later. 
\singlefig
     {S_D_pairs_brillouin_withQ}
     {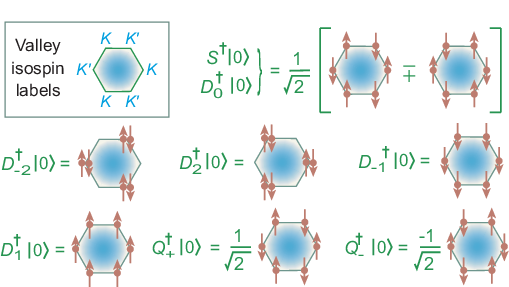}
     {0pt}
     {0pt}
     {1.0}
         {Configurations resulting from the pair creation operators of \eq{20grapheneBasis} operating on  the vacuum $\ket0$.    Location of the dots ($K$ or $K'$ site) indicates valley  isospin; arrows indicate  spin polarization. }
The physical meaning of the states in  \fig{S_D_pairs_brillouin_withQ} may be clarified by constructing the  corresponding electronic configurations.  For example,  consider $D_2^\dagger$ applied to $ \ket0$ for a single Landau level (so we omit a LL index).  From Eqs.\  \eqnoeq{20coupled1.4} and \eqnoeq{20algebra1.1},
\begin{equation*}
    D_2^\dagger = \frac{1}{\sqrt2}\, \Adagcoupled1010 = A^\dagger_{12}
    =
    \sum_{m_k} c^\dagger_{1 m_k} c^\dagger_{2 -m_k}
    = \sum_{m_k} c^\dagger_{K\uparrow m_k} c^\dagger_{K \downarrow -m_k} \ ,
\end{equation*}
where the correspondence between  the labels $a=1,2,3,4$ and the valley  and spin labels in    \fig{grapheneBasis_withTableBW}(a) was used.  We conclude that $D^\dagger_2 \ket 0$ creates  a state with one spin-up and one spin-down electron on each equivalent site $K$.   For the graphene hexagon the spin is zero on each site for this state but the charge  alternates from $0$ (no electrons) to $-2$ (two electrons) between adjacent sites; thus $D^\dagger_2 \ket 0$ generates a {\em charge  density wave} on the $K$ sites. By similar considerations, application of the generator $D_{-2}^\dagger$  to the pair vacuum $ \ket0$ generates a charge density wave  on  the $K'$ sites. 

It is  convenient to introduce \textit{order parameters} to keep track of emergent order on the lattice,
\begin{equation}
\begin{gathered}
    \ev {\spin_z} \equiv 
    \ev {\hat n_1} - \ev{\hat n_2} + \ev{\hat n_3} - \ev{\hat n_4}
    \\
    \ev{T_z} \equiv
    \ev {\hat n_1} + \ev{\hat n_2} - \ev{\hat n_3} - \ev{\hat n_4},
    \\
    \ev{N_z} \equiv
    \ev {\hat n_1} - \ev{\hat n_2} - \ev{\hat n_3} + \ev{\hat n_4} ,
\end{gathered}
 \mwgtag{20orderp1.1}%
\end{equation}
where $\hat n_i $ counts electrons in basis state  $\ket i$. These  have the following interpretation.
 
 \begin{enumerate}
 
\item 
$\ev{\spin_z}$ measures net spin  ({\em ferromagnetic order).}
 
\item
$\ev{T_z}$ measures the difference in charge between the $K$ and $K'$ sites ({\em charge order}).

\item
$\ev{N_z}$ measures the difference in spins between the $K$ and $K'$ sites ({\em antiferromagnetic order}, also known as \textit{N\'eel} or \textit{spin density  wave order}).

\end{enumerate}

\noindent
For example,  $D^\dagger_{2} \ket 0$ in  \fig{S_D_pairs_brillouin_withQ} was interpreted above as a component of a charge density wave.  Evaluating \eq{20orderp1.1} for this state gives $\ev{T_z} = 2$ and  $\ev{\spin_z}= \ev{N_z}=0$, which characterizes a state  having charge density wave order but no ferromagnetic or antiferromagnetic  order, consistent with our earlier interpretation. Evaluation of the order parameters using \eq{20orderp1.1} for the other pair states  in \fig{S_D_pairs_brillouin_withQ}  suggests  the following physical interpretations for these operators applied to the pair vacuum.

\begin{enumerate}

 \item 
$D_{\pm2}^\dagger$ generates {\em charge density waves,} with charge concentrated on alternate sites.

\item
$D_{\pm 1}^\dagger$ generates {\em ferromagnetic states}  in which electrons are distributed equally on all sites, with all spins aligned. 

\item
$S^\dagger$, $D_0^\dagger$, and   $Q_{\pm}^\dagger$ generate {\em antiferromagnetic spin waves} with a single spin  on each site, but spin direction alternating between adjacent sites. 

\end{enumerate}

\noindent
These operators can create emergent states  corresponding to highly collective pairing condensates exhibiting a rich variety of  structure, as will now be demonstrated.

\subsection{The SO(8) collective subspace}

\noindent
Let us work in a single $n=0$ Landau level, so we suppress the LL index and assume
 $2k+1$ degenerate states in the Landau level labeled by the integer quantum number $m_k$, with
 \begin{equation}
 m_k = \{-k, -k+1, \ldots, k-1, \,k\}.
 \mwgtag{m_kvalues}
 \end{equation}
A $2N$-fermion state   is produced by  acting   $N$ times on the vacuum with \SO8 pair creation operators of \eq{20coupled1.4} \cite{guid2020,wu1994,wu2016,wu2017},
\begin{equation}
    \ket{\SO8} = (S^\dagger)^{N_S} (D^\dagger)^{N_D}
    \ket 0 ,
    \mwgtag{so8PairState}
\end{equation}
where in this expression
\begin{itemize}
\item
$N_S$ is the number  of $S$ pairs (created by $S^\dagger$ operators), 
\item
$N_D$ is the number of $D$  pairs (created by $D^\dagger$ operators),  
\item
$N= N_S +N_D$ is the total number of pairs, and
\item
the pair creation operators $S^\dagger$ and $D^\dagger$ are defined in \eq{20coupled1.4}.
\end{itemize}
 A more general wavefunction could have $u$ broken pairs,  but states with no broken pairs are expected to dominate at low energy and we consider only  $u=0$ here.  As will be shown in Section \ref{pauliPairs}, the wavefunction \eqnoeq{so8PairState} written as a product of collective fermion pairs is in fact {\em equivalent} to the wavefunction \eqnoeq{basis1.0} used often in the literature for the \su4 subspace---despite their superficially different structure---as a consequence of the Pauli exclusion principle.

\subsection{\label{pauliPairs}Structure of SO(8) pairs}

\noindent
In this section we demonstrate the equivalence of the \SO8 pair state defined by \eq{so8PairState} and the   state defined by \eq{basis1.0}. From Eqs.\ \eqnoeq{20coupled1.4} and \eqnoeq{so8PairState}, an $N$-pair state in the $u=0$ subspace (no broken pairs) is given by
\begin{align}
\ket{\Psi} =
    \AdagPower12 \AdagPower13 \AdagPower14 \AdagPower23 
    \nonumber
    \\
    \times \AdagPower24 \AdagPower34 
    \, \ket0  ,
    \mwgtag{u0pairs1.0}
\end{align}
where the total pair number $N$ is
\begin{equation}
    N = \frac12 n = N_{12} + N_{13} + N_{14} + N_{23} + N_{24} + N_{34},
    \mwgtag{u0pairs1.1}
\end{equation}
with $n$ the total number of electrons and $N_{ab}$  the number  of electron pairs created by $\Adag ab$ operating on the vacuum state.  From \eq{20algebraTotal-b}, the $B_{ab}$  form an \su4 subalgebra of the \SO8 algebra \eqnoeq{20algebraTotal} under commutation (see Appendix \ref{so8u4su4}).  Let us investigate the \SO8 pair structure by considering the irreducible representations (irreps)  associated with the  $\SO8 \supset \unitary4 \supset \su4$ subgroup chain.

\subsubsection{\label{ss:hwsState} Pauli restrictions on collective pairs}

\noindent
If no pairs are broken ($u=0$), at half filling the number of pairs is $ N=\tfrac12 \Omega = 2k+1 $. The highest-weight \unitary4 representation is  given by $(\tfrac12 \Omega,  \tfrac12 \Omega, 0, 0)$, where we use \SO6 quantum numbers to label the \su4 states, since \su4 and \SO6 are homomorphic.  Define a highest-weight (HW) state in this  space to be the pair state with maximal value of $m_i$  from the table in \fig{grapheneBasis_withTableBW}(a).  This state has one electron in  the $a=1$ level and one electron in the $a=2$ level. Thus, for $N=2k+1$ pairs  the highest-weight state is
\begin{align}
    \ket{\rm HW} &=
    \frac{1}{(2k+1)!} \left( \Adag12 \right)^{2k+1} \ket0
    \nonumber
    \\
    &= \frac{1}{(2k+1)!} \left( \sum_{m_k}c^\dagger_{1 m_k}
    c^\dagger_{2,-m_k}
    \right)^{2k+1} \ket0
    ,
    \mwgtag{hwstate}
\end{align}
with the sum running over the $2k+1$ states in the Landau level labeled by the  quantum number $m_k $ of \eq{m_kvalues}. The other states of the irreducible representation may then be created by successive judicious application of lowering and raising  operators, beginning with the highest weight state (the standard \textit{Cartan--Dynkin algorithm}) \cite{guid2022}. The highest weight state \eqnoeq{hwstate} seems quite complicated, involving a sum  with many terms raised to  a high power.  
However, \textit{because of the Pauli principle} the actual structure of this state is much simpler  than \eq{hwstate} might suggest.

We may illustrate by a simple example, constructing explicitly the  highest-weight state for $k=1$, which corresponds to $2k+1 = 3$ pairs in a  single Landau level.  Writing out the sum over $m_k = \{-1,0,+1\}$ in \eq{hwstate}  gives
\begin{align*}
    \ket{\rm HW} &=  \frac{1}{3!} \left( 
    c_{1,-1}^\dagger
    c_{21}^\dagger
    + 
    c_{10}^\dagger
    c_{20}^\dagger
    + 
    c_{11}^\dagger
    c_{2,-1}^\dagger
    \right)^3 \ket0
    \\
    &=
    c_{10}^\dagger \, c_{20}^\dagger \, c_{11}^\dagger \,
    c_{21}^\dagger \, c_{1,-1}^\dagger \, c_{2,-1}^\dagger \,
    \ket 0
    \nonumber
    \\
&=
    \prod_{m_k=-k}^{m_k =+k}
    c_{1m_k}^\dagger c_{2m_k}^\dagger \ket 0,
\end{align*}
where the \textit{Pauli  principle} (antisymmetry  of the fermion creation operators) causes any products having two  or more creation operators with the same index to vanish in raising the sum   inside the parentheses to the $2k+1$ power.  Applying similar considerations for arbitrary $k$, the most general highest-weight state is 
\begin{align}
    \ket{\rm HW} &= \frac{1}{N!} 
    (\Adag12)^{N} \ket 0
    =
    \frac{1}{N!}
    \bigg(\sum_{m_k} c^\dagger_{1m_k} c^\dagger_{2 -m_k}\bigg)^{N}
    \ket0 
    \nonumber
    \\
    &=
     \prod_{m_k = -k}^{m_k=+k} c^\dagger_{1m_k} c^\dagger_{2m_k} \ket 0
    ,
    \mwgtag{u0pairs1.4}
\end{align}
where the simplification in the last step again follows from the Pauli  principle.  Thus the highest-weight state is a  product state of pairs, one  for each of the $N=2k+1$ states labeled by  $m_k$ in the Landau level.
Other states can be  constructed by applying the Cartan--Dynkin algorithm.  These will  be functions   of the generators $B_{ab}$, so for an arbitrary state $\ket\psi$ in the  weight space
\begin{equation}
\ket\psi = F(B_{ab}) \ket{\rm HW}  ,
\mwgtag{arbstate1.0}
\end{equation}
where the function $F(B_{ab})$ is specified by the Cartan--Dynkin procedure. For example, from Eqs.\ \eqnoeq{algebra1.9} and  \eqnoeq{20algebra1.2} the lowering operator $T_-$ is given by 
$$
    T_- \equiv \frac12 (T_x - iT_y)
    =
    \sum_{m_k} \big(c_{{\scriptscriptstyle K'\uparrow} m_k}^\dagger 
    c_{{\scriptscriptstyle K\uparrow} m_k}^{\vphantom{\dagger}} + 
    c_{{\scriptscriptstyle K'\downarrow} m_k}^\dagger c_{{\scriptscriptstyle 
    K'\downarrow} m_k}^{\vphantom{\dagger}} \big) , 
$$
and the state produced by acting with $T_-$ on the highest-weight state is
\begin{align}
    \ket{\psi} &= T_- \hws
    \nonumber
    \\
    &=
    \prod_{m_k}
    \bigg[ 
    \sum_{n_k} \big(c_{{\scriptscriptstyle K'\uparrow} n_k}^\dagger 
    c_{{\scriptscriptstyle K\uparrow} n_k}^{\vphantom{\dagger}} + 
    c_{{\scriptscriptstyle K'\downarrow} n_k}^\dagger c_{{\scriptscriptstyle 
    K'\downarrow} n_k}^{\vphantom{\dagger}} \big)
    \bigg]
    c_{{\scriptscriptstyle K\uparrow} m_k}^\dagger
    c_{{\scriptscriptstyle K\downarrow} m_k}^\dagger
    \, \ket 0
    \nonumber
    \\
    &
    =
    \prod_{m_k} 
    \big(
    c_{3 m_k}^\dagger
    c_{2 m_k}^\dagger
    +
    c_{4 m_k}^\dagger
    c_{1 m_k}^\dagger
    \big)
    \ket0 
    ,
    \mwgtag{u0pairs1.7}
\end{align}
where the  only terms that survive in the final expression are characterized by having an annihilation  operator in a factor inside the square brackets that is  balanced by a  creation operator from the factor outside the square brackets. All  other $u=0$ states  follow from  successive applications of raising and lowering operators  formed from the  generators  in Eqs.\ \eqnoeq{algebra1.8}--\eqnoeq{algebra1.11}.

\subsubsection{\label{pair-product} Pair and product wavefunctions}

From the preceding discussion, the states at half filling corresponding to $N=\tfrac12 \Omega$ take the form
\begin{align}
    \ket\psi &= 
    F(B_{ab}) \hws
    \nonumber
    \\
    &= 
    \prod_{m_k} \bigg(
    \sum_{\tau\sigma\tau'\sigma'} \Phi^*_{\tau\sigma\tau'\sigma'}
    c_{\tau\sigma m_k}^\dagger c_{\tau' \sigma' m_k}^\dagger
    \bigg) \ket0 ,
    \mwgtag{upairs1.8}
\end{align}
where $\tau,\tau'$ are valley isospin projection quantum numbers and  $\sigma,\sigma'$ are spin projection quantum numbers. This is the same form  \eqnoeq{basis1.0} as the most general collective pair state used by Kharitonov in Ref.\ \cite{khar2012} to classify possible broken symmetry states for the $n=0$ Landau  level in graphene.   

\graybox{Thus, the general pairing wavefunction of \eq{so8PairState} that characterizes  the SO(8) $\supset$ SU(4) fermion dynamical symmetry is {\em equivalent to the product form in \eq{basis1.0}} appearing in  standard discussions of quantum Hall ferromagnetism, for which the summations are  over the internal $(\tau,\sigma)$ rather than Landau $(m_k)$ degrees of freedom.}

\noindent
The equivalence of Eqs.\ \eqnoeq{so8PairState} and \eqnoeq{basis1.0} is a consequence of the Pauli principle and the fundamental restriction that it places on allowed pair configurations   within the collective  subspace of fermion pairs.

\subsection{\label{beyondQHF} Beyond quantum Hall ferromagnetism}

\noindent 
The $B_{ab}$ operators  of  \eq{20algebra1.2} are in one-to-one  correspondence with the operators used to formulate the effective low-energy  Hamiltonian \eqnoeq{so5_1.1a} describing explicit breaking of \su4 symmetry in quantum Hall ferromagnetism.  Thus all   physics discussed in  the prior literature using this effective Hamiltonian (see \cite{khar2012,wufe2014} and references cited therein)  is implicit in the present formalism.   Moreover, the discussion of Section \ref{pair-product} shows that the pair  basis \eqnoeq{u0pairs1.0} spanning the collective subspace of the \SO8 fermion dynamical symmetry is \textit{equivalent} to the most general wavefunction  \eqnoeq{basis1.0} that has been proposed \cite{khar2012} for collective states  breaking the \su4 symmetry spontaneously, despite an apparent difference in form.  

Thus the  collectively-paired SO(8)  subspace truncation of the full Hilbert space recovers the  understanding in the existing  literature of the classes of states to be expected from spontaneous breaking of  the \su4 symmetry.   However, the existing discussions of these collective states  have been largely qualitative, and have turned to numerical  simulations or simplified models to discuss the actual structure  of these emergent states.  We shall now show that the present Fermion dynamical symmetry formalism is capable  of addressing the quantitative nature of those collective  states {\em analytically.}

\subsection{Graphene SO(8) dynamical symmetries}

\noindent
The subset of the full Hilbert space spanned by the states  specified by \eq{so8PairState} is assumed to correspond to the {\em collective subspace} of  \fig{dynamicalSymmMicroscopic}(a).   Following the procedure outlined in  Section \ref{solutionAlgorithm}, the \SO8 symmetry can be used  to  construct  effective Hamiltonians that are diagonal in this subspace, the \SO8  generators don't couple the collective subspace to the rest of the Hilbert space, and effective interactions operate in the collective subspace that represent the average effect of the excluded Hilbert space.  The pair basis displayed in \fig{S_D_pairs_brillouin_withQ} can exhibit  charge density wave, antiferromagnetic, and ferromagnetic order.  Therefore,  the pair condensate defined by \eq{so8PairState} can produce a rich variety of  strongly-correlated quantum phases that correspond in the mean-field limit to spontaneously  broken symmetries.  Let's examine some of those solutions.

The graphene \SO8 subgroup chains that end in  the group  $\su2_\sigma\times\unitary1\tsub c$ imposing spin and charge conservation are  shown in \fig{dynamicalChains_graphene_so8},%
\doublefig
     {dynamicalChains_graphene_so8}
     {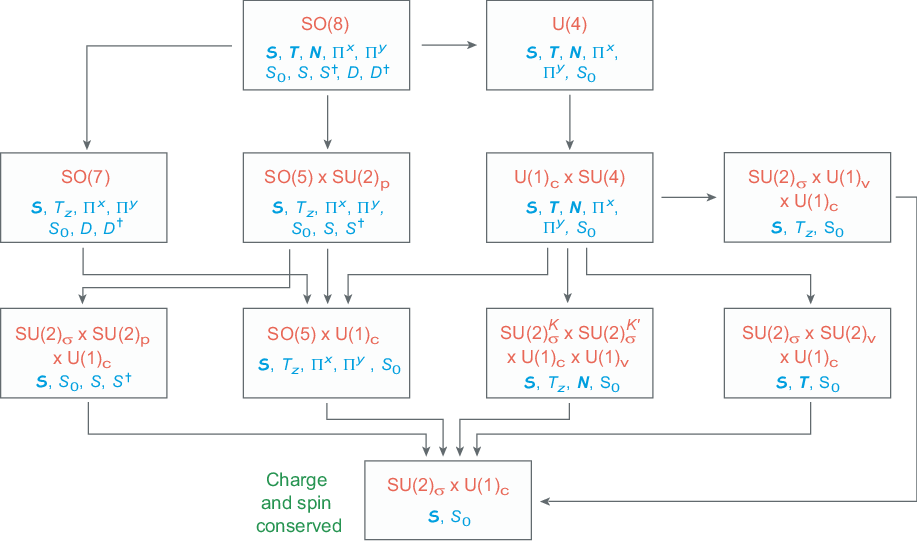}
     {0pt}
     {0pt}
     {1.0}
     {\SO8 fermion dynamical symmetry subgroup chains for monolayer graphene in a  magnetic field \cite{guid2017,wu2016,wu2017}. Each box is labeled by a Lie group in red and the group generators in blue.  Seven subgroup chains are indicated by directed paths beginning at the highest symmetry \so8 and ending with the group $\su2_\sigma \times \unitary1\tsub c$ corresponding to conservation of charge and spin. These  subgroup chains define  seven distinct fermion  dynamical symmetries that correspond to seven different emergent states or quantum phases, that are compatible with the \SO8 highest symmetry and that conserve charge and spin (If the Zeeman term is ignored in the Hamiltonian). If the Zeeman term is included it affects only the spin sector and breaks $\su2_\sigma$ down to the $\unitary1_\sigma$ group generated by $\spin_z$.
}
with Zeeman splitting ignored. If Zeeman splitting were included it would  influence directly only the spin sector and  break  $\su2_\sigma$ down to the $\unitary1_\sigma$ group generated by $\spin_z$. Seven subgroup chains are  displayed, representing seven distinct subgroup paths (indicated by directed line segments) connecting the highest symmetry \SO8 to the final symmetry $\su2_\sigma\times\unitary1\tsub c$ imposing spin and charge conservation.  
Each  defines an emergent mode representing a distinct quantum phase for  which matrix elements such as those for energies, order  parameters, and transition rates may be calculated analytically within the dynamical symmetry formalism. The following  examples describe several of these subgroup chains.

1.~The  set $\{S, \,S^\dagger, \,S_0\}$ generates an $\su2\tsub p$ Lie group  and   the set $\{\spin_\alpha,\, \Piop \alpha x, \, \Piop \alpha y,\, T_z\}$  generates an \SO5 Lie group that commutes with the $\su2\tsub p$ group.  Furthermore, the   components of the total spin $\spin_\alpha$ generate an  $\su2_\sigma$ subgroup of \SO5 and $S_0$ generates a $\unitary1\tsub c$  subgroup  of $\su2\tsub p$.  Thus one emergent state with highest symmetry \SO8 that conserves spin and charge corresponds to the dynamical  symmetry subgroup chain
\begin{align}
    \SO8 &\supset 
    \SO5 
    \times
    \su2\tsub {p}
    \nonumber
    \\
    &\supset
    \su2_\sigma\times\su2\tsub p
    \supset
    \su2_\sigma
    \times\unitary1\tsub c 
    \mwgtag{chain1}
\end{align}
displayed in \fig{dynamicalChains_graphene_so8}.
Alternatively, \SO5 may be broken according to the dynamical symmetry pattern
\begin{align}
    \SO8 &\supset\SO5\times
    \su2\tsub p
    \nonumber
    \\
    &\supset
    \SO5 \times
    \unitary1\tsub c
    \supset
    \su2_\sigma\times 
    \unitary1\tsub c ,
\end{align}
which also has highest symmetry \SO8 with conserved spin and charge, but represents  a different  quantum phase than that described by the chain of \eq{chain1}.

2.~Removal of the 12 pairing operators 
 $\{S, \,S^\dagger, \,D\phantomdagger_{\mu},\, D^\dagger_\mu \}$
from the \SO8 generators in
\eq{20grapheneBasis} yields the 16-generators of a  $\SO8\supset\unitary4 \supset \unitary1\tsub c \times \su4$ subgroup, with the $\unitary1\tsub c$ subgroup generated by the 
particle number operator $S_0$ and the \su4 subgroup generated by the 15 
generators of \eq{20algebra1.4}. There are several options for 
 $\SO8 \supset \unitary4 \supset \unitary1\tsub c \times \su4$ subgroup chains.

\begin{itemize}

 \item [(a)]
The subset  $\{ S_\alpha, \,\Piop \alpha x, \,\Piop \alpha y,\,  T_z\}$ generates an  \SO5 subgroup of   \su4;  hence
\begin{align}
    \SO8 &\supset
     \unitary4
     \supset
    \unitary1\tsub c 
    \times 
    \su4
    \nonumber
    \\
    &\supset
    \SO5
    \times
    \unitary1\tsub c
    \supset
    \su2_\sigma \times\unitary1\tsub c 
    \mwgtag{chains1.3}
\end{align}
is one possibility.

\item[(b)]
A $\su2_\sigma^K  \times  \su2_\sigma^{K'}$ symmetry results if inter-valley  scattering is ignored and spin is conserved  within each  valley. Thus a second subgroup chain  is
\begin{align}
    \SO8 &\supset
     \unitary4
     \supset
    \unitary1\tsub c 
    \times 
    \su4
    \nonumber
    \\
    &\supset \su2_\sigma^K \times 
    \su2_\sigma^{K'} \times \unitary1\tsub c \times \unitary1\tsub v
    \nonumber
    \\
    &\supset \su2_\sigma \times \unitary1\tsub c ,
\end{align}
where  $T_z$ generates the $\unitary1\tsub v$ subgroup.

\item[(c)]
A third  subgroup chain is
\begin{align}
     \SO8 &\supset
    \unitary4
     \supset
    \unitary1\tsub{c}
    \times 
    \su4
    \nonumber
    \\
    &\supset
    \su2_\sigma \times \su2 \tsub v \times \unitary1\tsub c
    \nonumber
    \\
    &\supset \su2_\sigma \times \unitary1\tsub c .
    \mwgtag{chainy1.2}
\end{align}
\end{itemize}

\vspace{-5pt}

\noindent
The three chains \eqnoeq{chains1.3}-\eqnoeq{chainy1.2} represent physically distinct emergent states that conserve charge and spin.  They are related by having the same \SO8 highest symmetry, but correspond to different collective excitations that are selected by different effective interaction parameter sets.

3.~The 21 operators 
$$
G_{\scriptscriptstyle{\rm SO(7)}}= \{ \spin_\alpha\phd, \, \Piop \alpha x\phd, \,\Piop \alpha  y\phd, \,  T_z\phd,\, S_0\phd, \, D^\dagger_\mu, D_\mu\phd \}
$$
generate  an \SO7   subgroup of \SO8 and the 11 operators 
$$
G_{\scriptscriptstyle{\rm SO(5)}}= \{  \spin_\alpha, \, \Piop \alpha x, \, \Piop \alpha y,\, T_z, \, S_0\phd \}
$$
generate an  $\SO5 \times \unitary1\tsub c$ subgroup of \SO7, implying a  subgroup chain
\begin{align}
    \SO8 \supset
    \SO7
    \supset
    \SO5
    \times
    \unitary1\tsub c
    \supset
    \su2_\sigma \times \unitary1\tsub c .
    \mwgtag{so7Chain}
\end{align}
This subgroup chain defines a {\em  critical dynamical symmetry,} where an entire  quantum phase  can exhibit critical behavior; this will be discussed further in  Section \ref{criticalDynamicalSymmetries} below.

The states represented by dynamical symmetry chains in   \fig{dynamicalChains_graphene_so8} are \textit{non-perturbative} (\textit{emergent}) and represent \textit{different quantum phases}.  These are strongly-correlated, exact many-body solutions but, using mean-field language,    the  symmetry has been {\em broken  spontaneously} for these states.    The quantum phases of \fig{dynamicalChains_graphene_so8} are separated by quantum phase transitions and cannot be related perturbatively to each other, nor can they be related perturbatively to the states corresponding to small fluctuations around \su4 symmetry of the  quantum Hall ferromagnetism model in \fig{subgroupChains_QHF_color}. We may gain a deeper understanding of  these  broken-symmetry states  by  using   \textit{generalized coherent states}  to approximate the exact dynamical symmetry solutions  corresponding to the subgroup chains of \fig{dynamicalChains_graphene_so8}.

\subsection{\label{20.genCoherentStatesSO8} Generalized coherent  states for graphene }

\noindent
The dynamical symmetry limits represented by subgroup chains in  \fig{dynamicalChains_graphene_so8}  correspond to particular choices of effective interaction parameters and  have \textit{exact analytical solutions} for physical matrix elements.  For arbitrary values of effective interaction parameters,  solutions are  superpositions of the different symmetry-limit solutions that are not easily expressed in concise analytical form. In this more general case there  is a powerful alternative:  the {\em generalized coherent state  approximation}, which permits analytical solutions for values of interaction parameters that need not correspond to symmetry limits. An overview of this method may be found in Ch.\ 21 of Ref.\ \cite{guid2022} and a more comprehensive technical review has been given in Ref.\ \cite{zha90}.

One way to visualize the nature of the collective states that are implied by the  subgroup chains of \fig{dynamicalChains_graphene_so8} is to use coherent states to construct the corresponding  ground-state total energy surfaces. In  generalized coherent state approximation for the present case the energy is determined by the  variational condition $\delta \mel{\eta}{H}{\eta} = 0$, with $H$ the \SO8 Hamiltonian and $\ket \eta$ the coherent state. Let us illustrate the method for the three subgroup  chains of \fig{dynamicalChains_graphene_so8} that contain the  $\SO5\times\unitary1\tsub c$ subgroup. Thus our coherent state  solutions will represent an approximate superposition of symmetry-limit solutions corresponding to the
\begin{gather}
    \SO8\supset\SO5\times\su2\tsub p\supset\SO5\times\unitary1\tsub c ,
    \nonumber
    \\
    \SO8\supset\unitary4\supset\unitary1\tsub{c}\times\su4\supset\SO5\times\unitary1\tsub c ,
    \mwgtag{20dynChainsCoherentSO5}
    \\
    \SO8\supset\SO7\supset\SO5\times\unitary1\tsub c ,
    \nonumber
\end{gather}
dynamical symmetry chains illustrated in \fig{dynamicalChains_graphene_so8}, which for brevity will be  termed the $\SO5\times\su2$, \su4, and \SO7 symmetries, respectively.

\subsubsection{Another useful generator basis}

\noindent
To leverage some results already worked out in the existing literature for coherent states, it is useful to introduce a new basis  $P_\mu^r$ for the particle--hole operators $B_{ab}$ in \eq{20algebra1.2} that is given by
\begin{equation}
    P_\mu^r = \sum_{m_j m_l} 
    (-1)^{{\ttfrac32} + m_l}
    \clebsch{\tfrac32}{m_j}{\tfrac32}{m_l}{
    \,r } {\,\mu }
    B_{m_j -m_l},
    \mwgtag{20multipole1.1}
\end{equation}
with the definition
\begin{equation}
    B_{m_j -m_l} \equiv \sum_{m_k} c^\dagger_{m_j m_k} 
    c^{\vphantom{\dagger}}_{-m_l 
    m_k}
    -\frac14 \delta_{m_j -m_l} \Omega ,
    \mwgtag{20multipole1.2}
\end{equation}
where  the table in \fig{grapheneBasis_withTableBW}(a) provides a one-to-one mapping between $m_i m_j$ values  and $ab$  values for the index on $B$, 
$$
    m_i=\{\tfrac32, \tfrac12, -\tfrac12, -\tfrac32 \}
    \ \ \longleftrightarrow\ \
    a=\left\{ 1,2,3,4\right\} .
$$
As an example, from   \fig{grapheneBasis_withTableBW}(a) 
$
B_{ab} = B_{12}$ and $ B_{m_j m_l} = 
B_{\scriptscriptstyle 3/2,1/2}
$
label the same quantity, which is defined through either \eq{20algebra1.2} or  \eq{20multipole1.2}. 

The index $r$ in \eq{20multipole1.1} can take  values $r = 0, 1, 2, 3$, with  $2r+1$ projections $\mu$ for each possibility, giving a total of 16 operators  $P^r_\mu$. By inserting  explicit values of the Clebsch--Gordan coefficients  in \eq{20multipole1.1}, the 16 independent $P_\mu^r$ may be evaluated in terms  of  the 16 independent $B_{ab}$. These are worked out in  Ref.\ \reference{wu2017} and in Section 3 of the Supplement \cite{supplementMaterials}.  A number operator $n_i$ may be introduced through
\begin{equation}
     n_i \equiv B_{ii} = \sum_{m_k} c^\dagger_{im_k} c_{i m_k}\phd 
     -\frac\Omega4 ,
     \mwgtag{20numberOp1.1}
\end{equation}
where $2\Omega$ is the degeneracy of the space for particles contributing to  \SO8 symmetry.  It is also useful to define
\begin{equation}
    S_0 \equiv \frac{n-\Omega}{2} = P_0^0,
     \mwgtag{20numberOp1.2}
\end{equation}
where a total particle number operator 
\begin{equation}
n\equiv \sum_a n_a = n_1 + n_2 + n_3 + n_4, 
\mwgtag{20numberOp1.3}
\end{equation}
[$a$ labels the 4 basis states in \fig{grapheneBasis_withTableBW}(a)] has been introduced. Physically, $S_0$ may be interpreted as half the particle number measured from half filling  ($n=\Omega$). Summarizing, our coherent state discussion will be formulated in the \SO8 generator basis
(called the nuclear \SO8 basis in the Supplement \cite{supplementMaterials})
\begin{equation}
    G^{\,\prime}_{\scriptscriptstyle{\rm SO(8)}}=
    \{ P^1,\, P^2,\, P^3, 
    \,
    S_0, \,
    S, \,S^\dagger, \,D\phantomdagger_{\mu},\, D^\dagger_\mu \}.
    \mwgtag{20nuclearBasis}
\end{equation}
where $S_0$ is defined in \eq{20numberOp1.2}.
The  \SO8  commutation algebra for the generators  \eqnoeq{20nuclearBasis} is given in Ref.\ \reference{wu2017} and in Section 5 of the Supplement \cite{supplementMaterials}.

\subsubsection{SO(8) coherent state energy surfaces}

\noindent
In terms of the generators  in \eq{20nuclearBasis}, the Hamiltonian may  be written
\begin{equation}
    H = H_0 + G_0 S^\dagger S + G_2 D^\dagger \cdot D +\sum_{r=1,2,3} b_r P^r 
    \tightdot P^r,
     \mwgtag{20coherent1.12}
\end{equation}
where we define
\begin{equation}
D^\dagger  \tightdot D  \equiv  \sum_\mu D^\dagger_\mu D_\mu
\qquad
P^r\tightdot P^r \equiv \sum_\mu P^r_\mu  P^r_\mu,
\mwgtag{20coherent1.12b}
\end{equation}
the  $G_0$, $G_2$, and $b_r$ are effective  interaction strengths in the collective subspace,  and we shall approximate $H_0$ as constant. The  coherent state solution isn't restricted to the  dynamical symmetry limits but it is instructive physically to evaluate the coherent  state energy surface for each dynamical symmetry limit. For the three \SO8 dynamical symmetry chains of  \eq{20dynChainsCoherentSO5} the coherent state total energy surface can be parameterized as   \reference{zha88}
\begin{align}
    E\tsub g(n,\beta) &= \ev{H} 
    \nonumber
    \\
    &= N\tsub g \left[ A\tsub g\beta^4 + B\tsub 
    g(n)\beta^2 + 
    C\tsub g(n) + D\tsub g(n,\beta) \right] ,
    \mwgtag{20generalEnergyFormula}
\end{align}
where  $n$ is  particle number and $\beta$ is the single order parameter  characterizing these states ($\beta$ measures AF order and   indicates the mixture of $S$ and $D_\mu$ pairs  in the ground state;  the alternative AF order parameter $\ev{N_z}$ is  maximal at the values of $\beta$ that correspond to minimum total energy). The parameters appearing in \eq{20generalEnergyFormula} depend on the group $g$ and are  tabulated in Ref.\  \reference{wu2017}. The ground states  at fixed $n/2\Omega$ will be given by those values of $\beta  \equiv \beta_0$ that correspond to minima of the energy surface $E(n,\beta)$. Evaluation of these constraints for  \eq{20generalEnergyFormula} indicates that  minima correspond to \reference{zha88}
\begin{equation}
\begin{gathered}
    \beta^{\small\su2\times\SO5}_0 = 0
    \qquad
    \beta^{\small\SO7}_0 = 0,
    \\
    \beta^{\small\su4}_0 = \pm \sqrt{n/4\Omega} ,
    \mwgtag{20betaMinima}
\end{gathered}
\end{equation}
for the three dynamical symmetries defined in \eq{20dynChainsCoherentSO5} that are labeled $\SO5\times\su2$, \su4, and \SO7, respectively.
Coherent state total energy surfaces for these  symmetries are  shown in  \fig{coherentEnergySurfaces_withWF}(a-c).%
\doublefig
     {coherentEnergySurfaces_withWF}
     {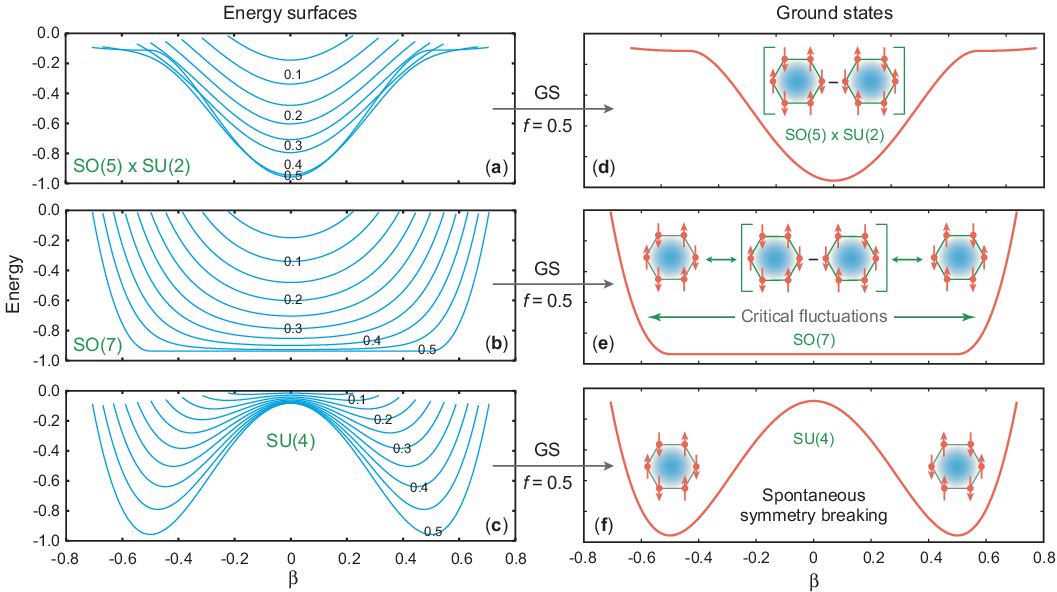}
     {0pt}
     {0pt}
     {1.0}
     {Coherent state energy surfaces for a single layer of undoped graphene in a magnetic field  \reference{wu2016}.  These figures are slices along a particular axis; full diagrams are multidimensional. (\textbf{a})-(\textbf{c}) Energy surfaces vs.\  AF order parameter  $\beta$ for the dynamical symmetry limits of  \fig{dynamicalChains_graphene_so8} that are displayed in \eq{20dynChainsCoherentSO5}.  Curves are labeled by fractional occupation $f=n/2\Omega$ defined in  \eq{20fillingFraction}.  (\textbf{d})-(\textbf{f})  Ground-state ($f=0.5$)  energy surfaces  corresponding to (a)-(c), respectively.   The inset diagrams in (d)-(f) represent  schematically  the corresponding ground-state wavefunctions in terms of the configurations in  \fig{S_D_pairs_brillouin_withQ}. }

\subsubsection{\label{20.physInterpCS}Physical interpretation of energy surfaces}

\noindent
The nature of the energy surfaces displayed in \fig{coherentEnergySurfaces_withWF}(a-c) is interpreted physically in  \fig{coherentEnergySurfaces_withWF}(d-f). The coherent state wavefunction for $N= \tfrac12 n $ pairs  is a  superposition of terms having different pair numbers $p$,  since it is an approximation that conserves only average particle number \cite{wu2017,zha88},
\begin{align}
    \ket{\SO5\times\su2} &= 
    \sum _{p} C_p \left( S^\dagger \right)^p
    \ket0
    \simeq 
     \left( S^\dagger \right)^N
    \ket0 ,
    \nonumber
    \\
    \ket{\su4} &=
    \sum _{p} C_p \left( S^\dagger \pm D_0^\dagger \right)^p \ket0
    \mwgtag{20coherentFN1.4}
    \\
     &=2\sum _{p} C_p \left( Q_\pm^\dagger \right)^p \ket0 
    \simeq
     \left( Q_\pm^\dagger \right)^N  \ket0,
     \nonumber
\end{align}
where the  $C_p$ are given in Ref.\ \cite{zha88},  $S^\dagger$ and $Q_\pm^\dagger$ are defined in Eqs.\  \eqnoeq{20coupled1.4} and \eqnoeq{20Qdef} (and \fig{S_D_pairs_brillouin_withQ}), and the final approximations are justified  because the fluctuation in particle number is small  for extended physical systems and  sums in \eqnoeq{20coherentFN1.4} are dominated by terms with $p \sim N$ \cite{wu2017}. From \eq{20coherentFN1.4},

\begin{enumerate}

\item
The $\SO5\times\su2$ ground state in \fig{coherentEnergySurfaces_withWF}(d) is a  coherent superposition of $S$  pairs, with  order  parameters $\ev{\spin_z}$,  $\ev{T_z}$, and $\ev{N_z}$ or $\ev\beta$ all equal to zero.

 \item 
The \su4 ground state in \fig{coherentEnergySurfaces_withWF}(f) is a coherent  superposition of $Q_-$ or $Q_+$ pairs with finite AF order parameters $\ev{N_z}$ or $\ev\beta$, but vanishing  ferromagnetic  $\ev{\spin_z}$ and charge density wave  $\ev{T_z}$ order.  The symmetry is broken spontaneously  if one of the degenerate energy  minima at finite $\beta$  is chosen as  the physical ground state.

\item
The \SO7 ground state in \fig{coherentEnergySurfaces_withWF}(e) is a superposition of  $S$ and $Q_\pm$ pairs that resembles an $\SO5\times\su2$ state if $\beta\sim 0$, and resembles an \su4 AF state if $\beta \sim \pm \beta_0$:
\begin{subequations}
\begin{alignat}{2}
&\ket{\beta=0} \propto (S^\dagger)^N \ket0 &\quad &[\sim \,\SO5\times\su2],
\\
&\ket{\beta=\beta_0} \propto (Q_+)^N \ket0 &\quad &[\sim\,\su4 \units{AF}],
\\
&\ket{\beta=-\beta_0} \propto (Q_-)^N \ket0 &\quad & [\sim\,\su4 \units{AF}],
\end{alignat}
\mwgtag{20.so7Interpoplate}%
\end{subequations}
where the degenerate energy  minima $\beta_0$ in the \su4 limit are given by
\begin{equation}
\beta=\beta_0 = \pm\left(\frac{n}{4\Omega}\right)^{1/2}.
\mwgtag{gsMinimumBeta}
\end{equation}  
Undoped graphene has $n = \Omega$ in the ground state so that
$
\beta_0 = \pm \tfrac12.
$
It follows that the fluctuations in $\beta$ suggested  by the flat regions of  \fig{coherentEnergySurfaces_withWF}(e) are  {\em maximal:} they represent excursions over the full range  of $\beta$ consistent with $\SO8 \supset \su4$ symmetry for a Landau level with $n$  electrons. The flat energy surface indicates that  in the \SO7 ground state many configurations having large differences in $\beta$  are {\em nearly degenerate in energy}.

\end{enumerate}

\noindent
Thus the $\SO5\times\su2$ and \su4 symmetries may be distinguished by  the order parameters $\ev{N_z}$ or $\ev\beta$, which are finite in the \su4 state but vanish in the  $\SO5\times\su2$ state. The \SO7 dynamical  symmetry is characterized by {\em maximal fluctuations in} $\ev \beta$.  It is  an example of a {\em critical dynamical symmetry,} which will be discussed  further in  Section \ref{criticalDynamicalSymmetries} .

\subsection{\label{20.quantumPhaseSO8} Quantum phase transitions}

\noindent
Transitions between different dynamical symmetry chains  correspond to quantum phase transitions, which can be  studied by varying control parameters in coherent state approximation.   In Ref.\ \reference{wu2017} the approximate \SO8 coherent state Hamiltonian
\begin{equation}
    H = 
    G_0 S^\dagger S + b_2 P^2\tightdot P^2 ,
     \mwgtag{20modelHamiltonian}
\end{equation}
was employed to study transitions among the quantum phases associated with the dynamical  symmetries in \eq{20dynChainsCoherentSO5}. Approximate Casimir expectation  values associated with dominant symmetries of the subgroup chains are  \cite{wu2017}
\begin{gather*}
    \ev{\casimir{{\rm SO(5)} \times {\rm SU(2)}}}  
    \sim \ev{S^\dagger S}   \qquad
    \ev{\casimir{\rm SU(4)}}  \sim \ev{P^2 \tightdot P^2}
    \\
    \ev{\casimir{\rm SO(7)}} \sim \ev{S^\dagger S} + \ev{ P^2\tightdot P^2} .
\end{gather*}
The Hamiltonian \eqnoeq{20modelHamiltonian} can be rewritten,
\begin{equation}
    H = G_0(S^\dagger S + q P^2\tightdot P^2) ,
     \mwgtag{20modelHamiltonian2}
\end{equation}
where a control parameter $q  = b_2/G_0$ has been defined that tunes the Hamiltonian  between $\su2\times\SO5$ and \su4 phases, via an intermediate quantum critical \SO7  phase.

\doublefig
     {quantumPhaseTransitions}
     {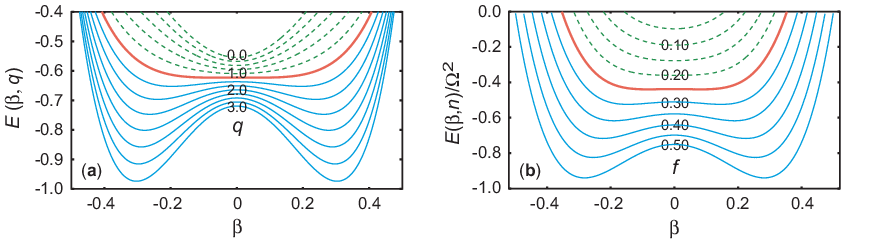}
     {0pt}
     {0pt}
     {1.0}
     {Total energy surfaces illustrating \SO8 quantum phase transitions for  graphene in a magnetic field. Dashed  green curves  correspond to  $\sim  \SO5\times\su2$ symmetry, solid blue curves to  $\sim$ \su4 symmetry, and  the heavier solid red curve corresponds to \SO7 critical dynamical symmetry occurring near the transition from $\SO5\times\su2$ to \su4. (\textbf{a}) Phase transition with  particle number fixed and coupling strength ratio  $q$ as the control parameter. The figure shows coherent state energy surfaces as a  function of  $q \equiv G_0/b_2$ for a fractional occupation $f = n/2\Omega =  0.5$.  (\textbf{b}) Phase transition with coupling strength fixed and particle number as the control parameter. The figure shows coherent state energy surfaces labeled by filling fractions $f=n/2\Omega$ at fixed $q=2.5$. 
}

\begin{enumerate}

\item 
If $q \ll 1$ the ground-state energy surface is approximated by  \fig{coherentEnergySurfaces_withWF}(d), with  a minimum  at $\beta = 0$, no  AF order, and $\su2\times\SO5$ dynamical symmetry.

\item
If $q \gg  1$ the ground-state energy surface is approximated by  \fig{coherentEnergySurfaces_withWF}(f), with a set of degenerate energy minima at $|\beta| \ne  0$, spontaneously broken  symmetry corresponding to   \su4 dynamical symmetry, and antiferromagnetic order.
 
\item
If $q \sim 1$, the ground-state energy is approximated by  \fig{coherentEnergySurfaces_withWF}(e), with  large fluctuations in AF order    parameter $\beta$ and \SO7  critical  dynamical symmetry.

\end{enumerate}

\doublefig
     {nuc_SC_graphene_chains}
     {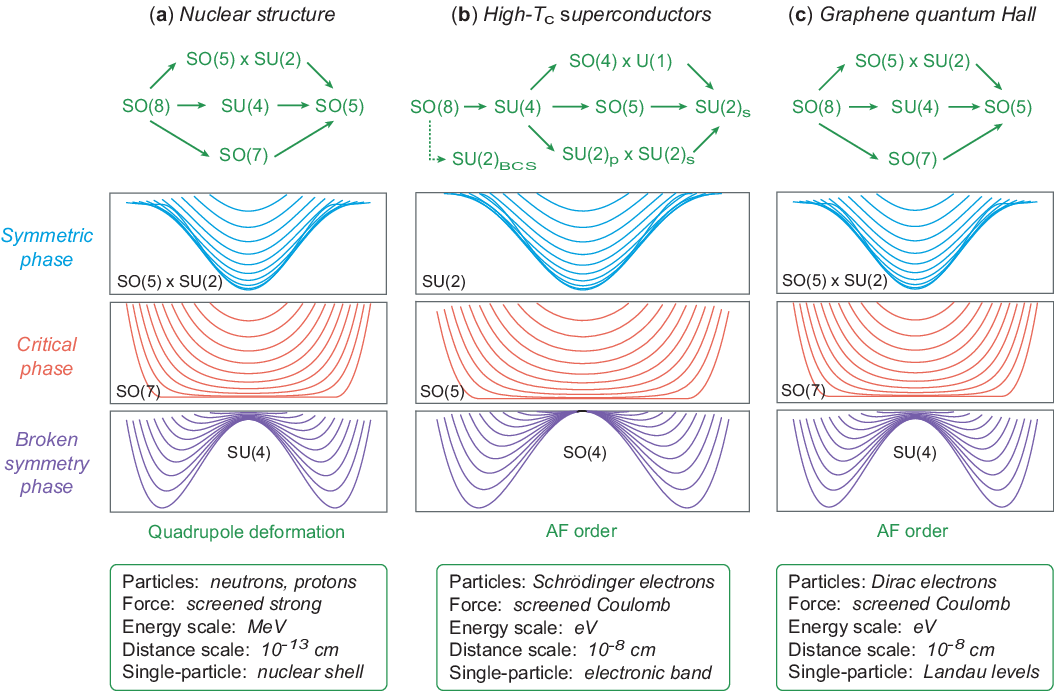}
     {0pt}
     {0pt}
     {1.0}
     {Dynamical symmetry chains and ground coherent-state  energy surfaces for fermion dynamical symmetry in (\textbf{a}) nuclear structure  \reference{wu1994,guid2022}, (\textbf{b}) high-temperature superconductivity \reference{guid01,wu03},  and (\textbf{c}) monolayer graphene in a strong magnetic field \reference{wu2016}. Plots show total energy as a function of an appropriate order parameter,  with each curve corresponding to a different value of a particle number parameter.
     }

\noindent
Figure~\ref{fg:quantumPhaseTransitions}(a)
illustrates \SO8 quantum phase transitions  controlled by varying  $q$ in \eq{20modelHamiltonian2}. The heavy solid red curve near $q \sim 1$ corresponds to  the quantum phase transition
$$
\SO5\times\su2 \ \longleftrightarrow\ \su4,
$$
mediated by the $\sim$ \SO7 critical dynamical symmetry. Quantum phase transitions may also be induced at constant $q$ by using the filling fraction $f$ of \eq{20fillingFraction} to vary particle number, as in \fig{quantumPhaseTransitions}(b). The heavy solid red curve at 
$$
f=\frac{n}{2\Omega} \sim 0.25
$$
then defines the $\sim$\SO7 critical-phase boundary between the $\SO5\times\su2$ and \su4 quantum phases.

\subsection{\label{20.universalityDynSymm} Universality of Emergent States}

\noindent
The fermion dynamical symmetries described for graphene in this review have many similarities with symmetries observed in other many-body systems.
Figure \ref{fg:nuc_SC_graphene_chains} illustrates an impressive  {\em universality of emergent dynamical symmetries}  that is observed over a range of disciplines.   As illustrated at the bottom of each column of figures in \fig{nuc_SC_graphene_chains}, these systems   differ  fundamentally at the microscopic level in properties such as characteristic distance and energy scales, forces responsible for interactions, constituent particles, and the nature of single-particle states. Yet the energy surfaces exhibit a clear similarity  manifested through a similar Lie group structure (and thus a similar symmetry-dictated truncation of the respective Hilbert spaces) for the  symmetries leading to emergent states. Although the symmetries are different for the different physical systems represented in each column, each row of energy surfaces in \fig{nuc_SC_graphene_chains} describes a similar level of symmetry breaking.  In mean field language, the top row corresponds to no breaking of the symmetry measured by the order parameter on the horizontal axis (quadrupole deformation for the nuclear case, and two kinds of antiferromagnetism for the superconductor and  graphene cases, respectively), the bottom row corresponds to the breaking of that symmetry, and the middle row corresponds to a critical dynamical symmetry mediating between symmetric and broken symmetry phases.
These results suggest a new kind of \textit{universality} building on an abstract similarity in Lie algebras for the truncated Hilbert space of emergent modes in physically  very different systems, with the similarities encoded in the Lie algebras and the differences reflected in the effective interaction parameters entering the dynamical symmetry Hamiltonian.

\graybox{{\bf\em Universality of Emergent States Hypothesis:} Fermion dynamical symmetries encode collective macroscopic similarities of emergent  states; effective interactions parameterize smoothly the  differences that follow  from underlying microscopic structure for those macroscopic states that has been averaged out by the symmetry-dictated truncation of the full space. The resulting fermion dynamical symmetries can provide similar descriptions of emergent states in physically very different systems.
}

\noindent
This universality of emergent states through similarity of the Lie algebras controlling their collective emergent properties extends well beyond qualitative descriptions: dynamical symmetries  provide microscopic descriptions of  phenomena within given subfields.  For example, \fig{nuc_SC_graphene_chains} provides a unified picture of emergent  states in  atomic nuclei,  high-temperature superconductors, and  graphene in terms of their total energy surfaces, but at the same  time, these methods provide a {\em quantitative description}  through a systematic methodology for 
calculating matrix elements of the interesting observables in each physical system \cite{wu1994,guid2020,wu2017}.

\subsection{\label{criticalDynamicalSymmetries} Critical Dynamical Symmetries}

\noindent
The  $\SO8 \supset \SO7$ dynamical  symmetry chain shown in \eq{so7Chain}  is an   example of a {\em critical dynamical  symmetry (CDS)} \reference{zha87}, which is observed with strikingly common features  in nuclear structure physics \reference{wu1994,zha87,zha88},  high-temperature superconductors \reference{guid01,guid2020,wu03},  and in the present discussion of  monolayer graphene in magnetic fields \reference{guid2017,wu2016,wu2017}.  
Figure~\ref{fg:nuc_SC_graphene_chains}
illustrates the remarkable universality of critical dynamical symmetries across these varied fields. 

\begin{itemize}

 \item 
The $\SO8\supset\SO7$ symmetry in  \fig{nuc_SC_graphene_chains}(a), which is observed in nuclear structure physics,

\item
the $\su4 \supset \SO5$ symmetry in  \fig{nuc_SC_graphene_chains}(b), which is observed in  high-temperature superconductors, and

\item
the $\SO8 \supset \SO7$  symmetry in  \fig{nuc_SC_graphene_chains}(c), which has been proposed here for monolayer graphene in a magnetic field,
\end{itemize}
 
 \noindent
all exhibit a common set of properties  characteristic of critical dynamical 
symmetry.

\begin{enumerate}

 \item 
 The energy surface is extremely flat as a function of one or more order 
parameters for a range of control parameters.

\item
The flat energy surfaces associated with critical dynamical symmetries imply an infinity of nearly degenerate ground states 
having very different wavefunctions and values of the order parameter(s).

\item
Critical dynamical symmetries can lead to \textit{interpolating phases,}  in which  order-parameter fluctuations can establish a doorway   between two different  phases; see \fig{coherentEnergySurfaces_withWF}(e). Thus they may play a role in facilitating quantum phase transitions.

\item
In all cases observed thus far,
a CDS doorway connects a phase exhibiting spontaneously broken symmetry to one where the symmetry has been restored;  Figs.\ \ref{fg:coherentEnergySurfaces_withWF} and \ref{fg:quantumPhaseTransitions} give examples.

\item
The flat energy surfaces characteristic of critical dynamical symmetry are conducive to fluctuations produced by perturbations and corresponding {\em complexity}  (extreme sensitivity of observables to initial conditions)  \reference{guid2011,guid2020}.

\item
A critical dynamical symmetry may be viewed in some sense as a  generalization of a quantum  critical point that extends critical fluctuations to an entire {\em quantum  critical phase.}

\end{enumerate}

\noindent
Because of these properties, 
we have proposed that critical  dynamical symmetry may be a fundamental organizing principle for quantum  criticality and for quantum phase transitions in complex systems that exhibit  multiple emergent modes \reference{guid2020}.

Note finally that the flat energy surfaces for critical dynamical symmetries imply the existence  of many nearly degenerate classical ground states, distinguished by different  values of order parameters and  by different wavefunction  components.  This suggests that in a quantum variational sense a  better ground state with lower energy could be found  by using the {\em  generator coordinate method} (see Ch.\ 8 of Ref.\ \cite{guid2022} for an  introduction) to obtain a new  ground state that is a superposition of the nearly degenerate quantum critical  states. 
We have suggested that an emergent state obtained by applying the generator coordinate method to the critical \SO5 phase in \fig{nuc_SC_graphene_chains}(b) may be largely responsible for pseudogap features observed in high temperature superconductors \cite{guid2020}. The critical \so7 graphene energy surface in \fig{nuc_SC_graphene_chains}(c) is, up to scaling of axes,
almost identical to the critical \SO5 energy surface in \fig{nuc_SC_graphene_chains}(b).  This raises the question of whether similar generator coordinate states might exist in graphene, and what their properties would be.

\section{Summary and Conclusions}

\noindent
We have given a general overview of monolayer graphene, and of the highly  collective emergent states that appear in the presence of a strong magnetic field. After summarizing the conventional view of these states we have shown that they  also may be described in terms of fermion dynamical  symmetries corresponding to subgroup chains originating in an \SO8 highest  symmetry and that conserve charge and spin. The previously known quantum Hall ferromagnetic \su4 symmetry of  graphene in strong magnetic fields has been extended by adding to the  one-body \su4 operators  a set of  two-body operators that  create or annihilate fermion pairs in either (1)~a total valley isospin triplet, total  spin singlet state, or
(2)~a total valley isospin singlet, total spin triplet state.
This extended set of 28 (16 particle--hole and 12 pairing) operators  closes an \SO8 Lie algebra under commutation. A rich set of low-energy collective modes is found associated with the subgroup structure of \SO8, which has seven dynamical symmetry subgroup chains (implying seven distinct emergent quantum phases) that have \SO8 as a highest symmetry and states that conserve spin and charge.   

It was possible to  decouple a subspace of collective pairs from the full Hilbert space of the problem by exploiting the  established methodology of fermion dynamical symmetries.  This permitted  exact, analytical, many-body solutions to be obtained in the dynamical symmetry limits. In addition to exact solutions in specific  dynamical symmetry limits, a generalized \SO8 coherent state  approximation was introduced that permits a broad range of generic solutions to be  obtained that extend quantitative calculations beyond the dynamical symmetry limits. 

The pairs spanning the collective subspace were shown to be equivalent mathematically  to pairs  that have already been discussed  in the graphene  literature  \cite{khar2012}, and to define possible  ground  states with spontaneously broken symmetry created by strong electron--electron and electron--phonon correlations in the  $n=0$ Landau level.  The present development 
(1)~places these pairs on  a firm, unified  mathematical footing and 
(2)~permits analytical solutions for emergent states that required numerical  simulation for their quantitative description in previous work.

Finally, it has been shown that there are uncanny dynamical symmetry analogies  among emergent states for graphene in a strong magnetic field,  high temperature superconductors, and strongly collective states in atomic  nuclei. This   implies a deep and intriguing mathematical affinity for emergent states among physical  problems that have very different microscopic structures and are not usually viewed as having more than a superficial  connection.

\begin{appendix}

\section{\label{ResistanceAndStuff} Resistivity and conductivity tensors}

\noindent
In simple conductors a current flows in the same direction as an applied electric  field, governed by the basic form of {\em Ohm's Law,} $I = V/R$, with the  resistance $R$, the voltage $V$, and the current $I$ being scalar quantities. This also may be expressed in terms of the electric field $E$ and current density $j$ as $j = \sigma E$ or $E = \rho j$, where $\sigma$ is the conductivity scalar and $\rho = \sigma^{-1}$ is the resistivity scalar. However, in the classical Hall effect  a current flows  perpendicular to an applied electric field  because of an applied magnetic field.   In such more general cases  $\sigma$ and $\rho$ become tensors and Ohm's Law takes the form $\bm j = \sigma \bm E$ or $\bm E = \rho \bm j$, where $\bm j$ is the current density vector and $\bm E$ is the electric field  vector. In 2D the {\em conductivity tensor} $\sigma$ and  {\em resistivity  tensor} $\rho$ can be expressed as $2\times2$ matrices,
\begin{gather}
    \sigma = \twomatrix
    {\sigma_{11}}
    {\sigma_{12}}
    {\sigma_{21}}
    {\sigma_{22}}
    =
    \twomatrix
    {\sigma_{xx}}
    {\sigma_{xy}}
    {\sigma_{yx}}
    {\sigma_{yy}},
    \\
    \rho = \twomatrix
    {\rho_{11}}
    {\rho_{12}}
    {\rho_{21}}
    {\rho_{22}}
    =
    \twomatrix
    {\rho_{xx}}
    {\rho_{xy}}
    {\rho_{yx}}
    {\rho_{yy}} ,
\end{gather}
where the components are $\sigma_{ij} = j_i/E_j$ and $\rho_{ij}=E_i/j_j$. Thus Ohm's Law generalizes to a matrix equation, $j_i= \sigma_{ij} E_j$ or $E_i = \rho_{ij} j_j$ (implied sum on repeated indices), with the interpretation  that a non-zero  $\sigma_{ij}$ means that an  electric field in the $j$ direction produces a current density in the $i$  direction.  Only for the  diagonal elements are the applied field and current in the same direction. The  conductivity and resistivity tensors are now related by matrix inversion:
\begin{align*}
      \rho = \sigma^{-1} &=
    \twomatrix
    {\sigma_{xx}}
    {\sigma_{xy}}
    {\sigma_{yx}}
    {\sigma_{yy}}^{-1}
    =
    \twomatrix
    {\sigma_{xx}}
    {\sigma_{xy}}
    {-\sigma_{xy}}
    {\sigma_{xx}}^{-1}
    \nonumber
    \\
    &= 
    \frac{1}{\sigma_{xx}^2 + \sigma_{xy}^2}
    \twomatrix
    {\sigma_{xx}}
    {-\sigma_{xy}}
    {\sigma_{xy}}
    {\sigma_{xx}} ,
    \mwgtag{rhorelation}
 \end{align*}
where $\sigma_{yx} = -\sigma_{xy}$ and $\sigma_{xx} = \sigma_{yy}$ have been 
assumed, and
\begin{align*}
     \sigma = \rho^{-1} &=
    \twomatrix
    {\rho_{xx}}
    {\rho_{xy}}
    {\rho_{yx}}
    {\rho_{yy}}^{-1} 
    \\
    &=
    \frac{1}{\rho_{xx}^2 + \rho_{xy}^2}
    \twomatrix
    {\rho_{xx}}
    {-\rho_{xy}}
    {\rho_{xy}}
    {\rho_{xx}} .
\end{align*}
For the Hall effect in 2D, the {\em Hall resistance} $R\tsub H$ is related to  the {\em Hall resistivity} by $R\tsub H = -\rho_{xy}$, the {\em Hall  conductivity} $\sigma\tsub H$ is given by $\sigma\tsub H = \sigma_{xy}$, and  the {\em diagonal conductivity} is $\sigma_{xx}$.   The voltages $V\tsub H$  and  $V\tsub L$ in \fig{hallBar} are commonly measured in a Hall experiment.  Thus it is often convenient to  work with resistance  rather than resistivity, with
\begin{gather}
    R\tsub H \equiv \frac{V\tsub H}{I} \ \ \units{(Hall resistance)}
    \\
    R\tsub L \equiv \frac{V\tsub L}{I} \ \ \units{(Longitudinal resistance)}.
\end{gather}
Ordinarily resistance implies energy dissipation through impurity  scattering.  The Hall resistance  $R\tsub H$ has the units of resistance but no energy dissipation is associated  with it in a clean system. It is  {\em defined} by the  ratio of measured quantities $V\tsub H$ and $I$.

\section{\label{andersonLoc} Anderson localization}

\noindent
The effect of impurity scattering on electrical conductivity often is discussed in terms of  the {\em Anderson model} of localization \cite{ande58}.  The  model consists of a system of atomic levels at different lattice sites with a  hopping term allowing electrons to jump from one site to another (in the  simplest approximation, only between nearest-neighbor sites). It is then assumed  that the energy of an electron on each site is not constant, but varies randomly  over some range $w$ to simulate the effect of impurity scattering. The  Hamiltonian for sites labeled by $n$ with one orbital per site is
\begin{equation}
    H =  \sum_n \epsilon_n a^\dagger_n a_n + V \sum_{\{nm\}} a^\dagger_n a_m ,
\end{equation}
where brackets on the summation indices signify a restriction to  nearest-neighbor hopping. If $w=0$, all sites have the same  energy, the resulting eigenstates are Bloch states delocalized over the entire  system, and a band structure results for which electron transport is  approximately ballistic.  On the other  hand, in the limit $V=0$ there is no hopping between sites, the  states are localized, and there is no electron transport. Thus this simple Hamiltonian implements a metal--insulator transition controlled by the ratio $w/V$. If $w/V$ is small the material is conducting but if $w/V$ is  large the states are localized and the material becomes insulating. This  suppression of conductivity by impurity scattering is termed {\em Anderson  localization} and the corresponding transition between conducting and insulating phases is termed the {\em Anderson  transition.} The Anderson model is difficult to solve exactly but concepts  derived from it are common in discussing the  conductor--insulator transition.

It might be expected that this metal--insulator transition would involve a smooth  interpolation between the two regimes near a finite critical value of $w/V$.   This is approximately true in $d=3$ dimensions, but for $d\le 2$ it is found   that {\em an infinitesimal value for $w/V$ is sufficient to destroy conductivity  completely,} except in certain special cases.  Thus, an unperturbed 2D  electron gas is expected to be insulating, except in the clean limit.  However,   this conclusion can be avoided in the presence of  a magnetic field, which breaks  time-reversal symmetry and interferes with  localization.  In 3D, 

\begin{enumerate}
\item
when $w/V > (w/V)\tsub{crit}$, there are no extended states but 

\item
when $w/V  < (w/V)\tsub{crit}$, extended and localized states coexist, separated by  a {\em mobility edge} $E\tsub c$

\end{enumerate}
 Extended (conducting) states  have energy $E > E\tsub c$ and localized (insulating) states have $E < E\tsub  c$, with the conductivity tending smoothly to zero as $w/V$ approaches  $(w/V)\tsub{crit}$ from below.  The 2D electron gas \textit{in a magnetic field} has  similar behavior, with both extended and localized states separated by a  mobility edge; this is central to explaining the \iqhe\ in Section \ref{mobilityGaps}. For further discussion of localization, see Ch.\ 12 of Phillips \cite{phil2003}.

\section{\label{gaussBonnet}The Gauss--Bonnet and Chern theorems}

\noindent
For a 2D surface there is a relationship between its geometry and its  topology  called the {\em Gauss--Bonnet theorem,} which may be expressed through  the  {\em Gauss--Bonnet equation,}
\begin{equation}
    \frac{1}{2\pi}\int_{S} KdA = 2(1-g),
    \mwgtag{gauss--bonnetEq}
\end{equation}
where the integral is over a closed surface $S$, the local curvature  is $K$, and $g$ is the genus of the surface (number of ``holes'',  which characterizes its topology). The Gauss--Bonnet theorem relates the  geometrical properties of the 2-surface (carried by the local curvature $K$ on  the left side) and its topological properties (carried by the genus $g$ on the right  side), with $g$ a topological invariant since the number of holes is unchanged by  smooth deformations of the surface.
For example, consider the 2D manifolds displayed in \fig{torus-sphere}. 
\singlefig
    {torus-sphere}
    {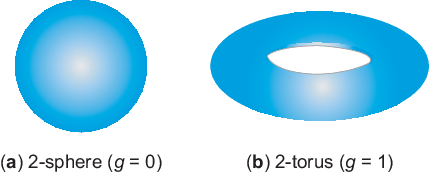}
    {0pt}
    {0pt}
    {1.0}
    {(\textbf{a}) The 2D sphere, which has genus $g=0$.  (\textbf{b}) The 2D torus, which has genus  $g=1$.}
For the  sphere, inserting $g=0$ and the Gaussian curvature $K=R^{-2}$   in  \eq{gauss--bonnetEq} gives the expected result, $A = 4\pi R^2$. For the torus the  integral on the left side of \eq{gauss--bonnetEq} vanishes (because the torus  has regions of positive and negative curvature $K$ that cancel exactly), which  is consistent with \eq{gauss--bonnetEq} only if $g=1$, as expected for the  torus.

One fundamental implication of the linkage of geometry and topology for a 2-surface implied by \eq{gauss--bonnetEq} is illustrated in \fig{genusInvariance}. Because the integer $2(1-g) = 0$ for $g=1$ on the right side of \eq{gauss--bonnetEq} cannot change  continuously into another integer for a small, smooth deformation of the surface, \textit{the  integral on the left side cannot change under this deformation either.} Thus, in \fig{genusInvariance}  the  left side is {\em topologically protected}  against change for smooth deformations that do not  change the genus.

\singlefigbottom
    {genusInvariance}
    {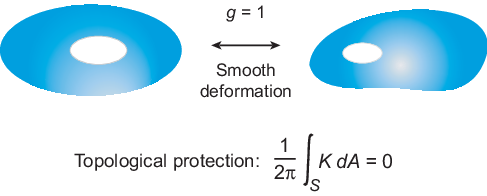}
    {0pt}
    {0pt}
    {1.0}
    {Smooth deformations that do not change the genus $g$ cannot change the 
    Gauss--Bonnet topological invariant defined by the curvature $K$ integrated 
    over the surface in \eq{gauss--bonnetEq}. The 
    curvature may be changed locally by a smooth deformation but $\int_S KdA$ 
    is topologically protected  if the deformation does not alter the 
    genus for the surface. Adapted from Ref.\ \cite{guid2022}.}

The Gauss--Bonnet equation \eqnoeq{gauss--bonnetEq} relates geometry and  topology for spatial manifolds, but in 1944 S.-S. Chern generalized the  Gauss--Bonnet equation to the {\em Gauss--Bonnet--Chern equation} (which for  brevity we will term the {\em Chern equation}),
\begin{equation}
     \frac{1}{2\pi}\int_{S} \Omega \,dA = C_n.
     \mwgtag{28.berryHall1.4}
\end{equation}
This  resembles the Gauss--Bonnet formula \eqnoeq{gauss--bonnetEq},  except that for $n$ a positive integer,

\begin{enumerate}

 \item 
Equation (\ref{eq:28.berryHall1.4})  is valid for any $2n$-dimensional closed Riemannian manifold.

\item
The {\em Chern index} $C_n$ on the right side of  \eqnoeq{28.berryHall1.4} is an integer that labels a {\em Chern class}.
 
\item
The \textit{Berry curvature} $\Omega$ is  evaluated over a closed manifold of quantum states defined on $S$.

\item
The Chern index  $C_n$  isn't  determined by the genus of the surface $S$, as in \eqnoeq{gauss--bonnetEq}, but rather by {\em the  topology of a manifold of quantum states (the Hilbert space)} defined over $S$.
 
\end{enumerate}

\noindent 
 For 2D manifolds the index $C_n$ is a  {\em Chern number of the first kind} $C_1$, which takes values from the set of integers $Z$.  General implications of \eq{28.berryHall1.4}  will be termed the {\em Chern theorem.}

\graybox{
{\bf\em Chern Theorem:} The integral of Berry curvature over a closed  manifold of quantum states is quantized in terms of  topological Chern numbers that take integer values.
}

\noindent
By analogy with the Gauss--Bonnett equation \eqnoeq{gauss--bonnetEq} and \fig{genusInvariance},  \eq{28.berryHall1.4} is topologically protected under smooth deformations that do not change the Chern number on its right side; the Chern number can be changed only by a \textit{phase transition} to a state having a different topology. As discussed in Section \ref{topoProtectPlateau}, this topological protection accounts for the the remarkable flatness of the plateaus in \fig{IQHEsteps}. 

Readable general introductions to topological properties  of the quantum Hall effect may be found in Ref.\ \cite{tong2016} and in Ch.\ 28 of Ref.\ \cite{guid2022}. For a more mathematical discussion of Chern classes in physics, see Frankel  \cite{fran2012} or Nakahara \cite{nak90}.

\section{\label{so8u4su4}U(4) and SU(4) subgroups of SO(8)}

\noindent
From  \eq{20algebraTotal-b}, the 16  operators $B_{ab}$ defined in \eq{20algebra1.2} are closed under commutation and form a subalgebra of \SO8 corresponding to the chain (see  \fig{dynamicalChains_graphene_so8})
\begin{align}
\so8 \supset \unitary4 \supset \unitary1\tsub c \times \su4,
\mwgtag{so8-u4-u1xsu4}
\end{align}
 where the $\unitary1\tsub c$ factor corresponding to conservation of electron number (charge) is generated by $S_0$ defined in \eq{20numberOp1.2}, and the \su4 factor is generated by the 15 operators in Eqs.\ \eqnoeq{algebra1.8}-\eqnoeq{algebra1.11}, but expressed in a different basis than for \eq{20algebra1.2}. 
Because of the direct product $\unitary1\tsub c \times \su4$ in \eq{so8-u4-u1xsu4},  the $\unitary1\tsub c$ and \su4 factors are independent and it is common to keep track of the particle number associated with the $\unitary1\tsub c$ factor separately and refer loosely to the symmetry associated with the generators in \eq{20algebra1.2} as \su4, rather than the more precise \unitary4 or $\unitary1 \times \su4$.

\end{appendix}

\bigskip

\acknowledgments{
We wish to thank Yang Sun for useful discussions and Matthew Murphy for help 
with some of the calculations and illustrations that were used here. L.-A. W. is supported by the Basque Country Government (Grant No.\ IT1470- 22) and Grant No.\ PGC2018-101355-B-I00 funded by MCIN/AEI/10.13039/501100011033. F. W.  was supported partially by LightCone Interactive LLC during the completion of this work.
}

\vfill

\bibliography{referencesGrapheneMWG.bib}



\end{document}